\documentclass[prb,aps,nofootinbib,amssymb,twocolumn,superscriptaddress,10pt,floatfix]{revtex4-2}
\usepackage[T1]{fontenc}
\usepackage{graphicx}
\usepackage{xcolor}
\usepackage{dcolumn}
\usepackage{comment}
\usepackage{amsmath,amssymb,bbold,bm}
\usepackage{hyperref}
\usepackage{amsfonts}
\usepackage{float,latexsym}
\usepackage{multirow}
\usepackage{color}
\usepackage{cleveref}
\usepackage{physics}
\usepackage{tabularx}
\usepackage{array}
\usepackage[caption=false]{subfig}
\usepackage{siunitx}
\usepackage{gensymb}
\usepackage{pifont}
\usepackage{placeins}
\usepackage{mathtools}
\usepackage{longtable}
\usepackage{multirow}
\usepackage{nicematrix}
\usepackage{tikz}
\usepackage{xstring}
\usepackage{makecell}
\usepackage{xr}
\usepackage{ifthen}
\usetikzlibrary{decorations.markings}
\usetikzlibrary{positioning}
\usetikzlibrary{arrows.meta}
\usepackage{tikz-feynman}
\usetikzlibrary{calc}
\tikzfeynmanset{warn luatex=false}
\let\vec\mathbf

\usepackage{chemformula}
\usepackage{pdfpages}

\makeatletter
\def\maketag@@@#1{\hbox{\m@th\normalfont\normalsize#1}}
\makeatother

\crefname{appendix}{Appendix}{Appendices}
\crefname{equation}{Eq.}{Eqs.}
\crefname{figure}{Figure}{Figs.}
\crefname{table}{Table}{Tables}
\crefname{section}{Section}{Sections}
\crefname{enumi}{Point}{Points}
\creflabelformat{appendix}{[#2#1#3]}
\crefmultiformat{figure}{Figures.~#2#1\xdef\mycreffirstarg{#1}#3}
{ and~#2#1#3}{, #2#1#3}{ and~#2#1#3}

\makeatletter     
\renewcommand\onecolumngrid{% <<<<<<
	\do@columngrid{one}{\@ne}%
	\def\set@footnotewidth{\onecolumngrid}% <<<<<<<<<<<<<<<<
	\def\footnoterule{\kern-6pt\hrule width 1.5in\kern6pt}%
}

\makeatletter
\AtBeginDocument{\let\LS@rot\@undefined}
\makeatother

\RequirePackage[normalem]{ulem} %DIF PREAMBLE
\RequirePackage{color}\definecolor{RED}{rgb}{1,0,0}\definecolor{BLUE}{rgb}{0,0,1} %DIF PREAMBLE
\RequirePackage{listings} %DIF PREAMBLE
\RequirePackage{color} %DIF PREAMBLE
\lstdefinelanguage{DIFcode}{ %DIF PREAMBLE
	moredelim=[il][\color{red}\sout]{\%DIF\ <\ }, %DIF PREAMBLE
	moredelim=[il][\color{blue}\uwave]{\%DIF\ >\ } %DIF PREAMBLE
} %DIF PREAMBLE
\lstdefinestyle{DIFverbatimstyle}{ %DIF PREAMBLE
	language=DIFcode, %DIF PREAMBLE
	basicstyle=\ttfamily, %DIF PREAMBLE
	columns=fullflexible, %DIF PREAMBLE
	keepspaces=true %DIF PREAMBLE
} %DIF PREAMBLE
\lstnewenvironment{DIFverbatim}{\lstset{style=DIFverbatimstyle}}{} %DIF PREAMBLE
\lstnewenvironment{DIFverbatim*}{\lstset{style=DIFverbatimstyle,showspaces=true}}{} %DIF PREAMBLE
\newcommand{\ie}{{\it i.e.}}

\usepackage{enumitem}

\newcommand{\vk}{\vec{k}}

\newcommand{\vp}{\vec{p}}

\newcommand{\vQ}{\vec{Q}}
\newcommand{\vq}{\vec{q}}

\newcommand{\twodtqclink}[1]{\href{https://topologicalquantumchemistry.com/topo2d/index.html\#/?serialid=#1}{#1}}

\newcommand{\twodtqclinkmatname}[2][1234]{\href{https://topologicalquantumchemistry.com/topo2d/index.html\#/?serialid=#1}{#2}}

\newcommand{\mctwodlink}[1]{\href{https://mc2d.materialscloud.org/details/#1}{MC2D}}

\newcommand{\ctwodblink}[1]{\href{https://c2db.fysik.dtu.dk/material/#1}{C2DB}}

\newcommand{\mctwodbare}{\href{https://mc2d.materialscloud.org/}{MC2D}}

\newcommand{\ctwodbbare}{\href{https://c2db.fysik.dtu.dk/}{C2DB}}

\newcommand{\TotalNumTwistSemi}{61 }
\newcommand{\TotalNumTwistInsu}{1568 }

\newcommand{\NumSemiElements}{4 }
\newcommand{\NumSemiTMD}{2 }
\newcommand{\NumSemiChalcogenide}{0 }
\newcommand{\NumSemiHalide}{0 }
\newcommand{\NumSemiMixed}{0 }
\newcommand{\NumSemiOther}{0 }
\newcommand{\NumInsuElements}{6 }
\newcommand{\NumInsuTMD}{14 }
\newcommand{\NumInsuChalcogenide}{24 }
\newcommand{\NumInsuHalide}{3 }
\newcommand{\NumInsuMixed}{7 }
\newcommand{\NumInsuOther}{6 }

\newcommand{\ThetaRef}{$5\degree$}

\newcommand{\webBCS}{\href{https://cryst.ehu.es/}{Bilbao Crystallographic Server}}

\newcommand{\webtwoDTQC}{\href{https://topologicalquantumchemistry.com/topo2d/index.html}{Topological 2D Materials Database}}

\newcommand{\lgsymb}[1]{\ifnum#1=1
$p1$\else
\ifnum#1=2
$p\bar{1}$\else
\ifnum#1=3
$p112$\else
\ifnum#1=4
$p11m$\else
\ifnum#1=5
$p11a$\else
\ifnum#1=6
$p112/m$\else
\ifnum#1=7
$p112/a$\else
\ifnum#1=8
$p211$\else
\ifnum#1=9
$p2_111$\else
\ifnum#1=10
$c211$\else
\ifnum#1=11
$pm11$\else
\ifnum#1=12
$pb11$\else
\ifnum#1=13
$cm11$\else
\ifnum#1=14
$p2/m11$\else
\ifnum#1=15
$p2_1/m11$\else
\ifnum#1=16
$p2/b11$\else
\ifnum#1=17
$p2_1/b11$\else
\ifnum#1=18
$c2/m11$\else
\ifnum#1=19
$p222$\else
\ifnum#1=20
$p2_122$\else
\ifnum#1=21
$p2_12_12$\else
\ifnum#1=22
$c222$\else
\ifnum#1=23
$pmm2$\else
\ifnum#1=24
$pma2$\else
\ifnum#1=25
$pba2$\else
\ifnum#1=26
$cmm2$\else
\ifnum#1=27
$pm2m$\else
\ifnum#1=28
$pm2_1b$\else
\ifnum#1=29
$pb2_1m$\else
\ifnum#1=30
$pb2b$\else
\ifnum#1=31
$pm2a$\else
\ifnum#1=32
$pm2_1n$\else
\ifnum#1=33
$pb2_1a$\else
\ifnum#1=34
$pb2n$\else
\ifnum#1=35
$cm2m$\else
\ifnum#1=36
$cm2e$\else
\ifnum#1=37
$pmmm$\else
\ifnum#1=38
$pmaa$\else
\ifnum#1=39
$pban$\else
\ifnum#1=40
$pmam$\else
\ifnum#1=41
$pmma$\else
\ifnum#1=42
$pman$\else
\ifnum#1=43
$pbaa$\else
\ifnum#1=44
$pbam$\else
\ifnum#1=45
$pbma$\else
\ifnum#1=46
$pmmn$\else
\ifnum#1=47
$cmmm$\else
\ifnum#1=48
$cmme$\else
\ifnum#1=49
$p4$\else
\ifnum#1=50
$p\bar{4}$\else
\ifnum#1=51
$p4/m$\else
\ifnum#1=52
$p4/n$\else
\ifnum#1=53
$p422$\else
\ifnum#1=54
$p42_12$\else
\ifnum#1=55
$p4mm$\else
\ifnum#1=56
$p4bm$\else
\ifnum#1=57
$p\bar{4}2m$\else
\ifnum#1=58
$p\bar{4}2_1m$\else
\ifnum#1=59
$p\bar{4}m2$\else
\ifnum#1=60
$p\bar{4}b2$\else
\ifnum#1=61
$p4/mmm$\else
\ifnum#1=62
$p4/nbm$\else
\ifnum#1=63
$p4/mbm$\else
\ifnum#1=64
$p4/nmm$\else
\ifnum#1=65
$p3$\else
\ifnum#1=66
$p\bar{3}$\else
\ifnum#1=67
$p312$\else
\ifnum#1=68
$p321$\else
\ifnum#1=69
$p3m1$\else
\ifnum#1=70
$p31m$\else
\ifnum#1=71
$p\bar{3}1m$\else
\ifnum#1=72
$p\bar{3}m1$\else
\ifnum#1=73
$p6$\else
\ifnum#1=74
$p\bar{6}$\else
\ifnum#1=75
$p6/m$\else
\ifnum#1=76
$p622$\else
\ifnum#1=77
$p6mm$\else
\ifnum#1=78
$p\bar{6}m2$\else
\ifnum#1=79
$p\bar{6}2m$\else
\ifnum#1=80
$p6/mmm$\else
{\color{red}{Invalid LG number}}
\fi
\fi
\fi
\fi
\fi
\fi
\fi
\fi
\fi
\fi
\fi
\fi
\fi
\fi
\fi
\fi
\fi
\fi
\fi
\fi
\fi
\fi
\fi
\fi
\fi
\fi
\fi
\fi
\fi
\fi
\fi
\fi
\fi
\fi
\fi
\fi
\fi
\fi
\fi
\fi
\fi
\fi
\fi
\fi
\fi
\fi
\fi
\fi
\fi
\fi
\fi
\fi
\fi
\fi
\fi
\fi
\fi
\fi
\fi
\fi
\fi
\fi
\fi
\fi
\fi
\fi
\fi
\fi
\fi
\fi
\fi
\fi
\fi
\fi
\fi
\fi
\fi
\fi
\fi
\fi}

\newcommand{\lgsymbnum}[1]{LG #1 (\lgsymb{#1})}

\newcommand{\titlePaper}{
	2D Theoretically Twistable Material Database}

\newcommand{\paperAuthors}{
	\author{Yi Jiang}
	\thanks{These authors contributed equally to this work.}
	\affiliation{Donostia International Physics Center (DIPC), Paseo Manuel de Lardizábal. 20018, San Sebastián, Spain}
	\author{Urko Petralanda}
	\thanks{These authors contributed equally to this work.}
	\affiliation{Department of Physics, University of the Basque Country UPV/EHU, Apartado 644, 48080 Bilbao, Spain}
	\affiliation{Department of Applied Mathematics, University of the Basque Country (UPV/EHU), 01006 Vitoria-Gasteiz, Spain.}
	\author{Hanqi Pi}
	\thanks{These authors contributed equally to this work.}
	\affiliation{Donostia International Physics Center (DIPC), Paseo Manuel de Lardizábal. 20018, San Sebastián, Spain}
	\author{Grigorii Skorupskii}
	\thanks{These authors contributed equally to this work.}
	\affiliation{Department of Chemistry, Princeton University, Princeton, New Jersey 08544, USA}
	\author{Qiaoling Xu}
	\affiliation{College of Physics and Electronic Engineering, Center for Computational Sciences, Sichuan Normal University, Chengdu 610068, China}
	\affiliation{Tsientang Institute for Advanced Study, Zhejiang 310024, China}
	\author{Dumitru C\u{a}lug\u{a}ru}
	\affiliation{Department of Physics, Princeton University, Princeton, NJ 08544, USA}
	\affiliation{Rudolf Peierls Centre for Theoretical Physics, University of Oxford, Oxford OX1 3PU, United Kingdom}
	\author{Haoyu Hu}
	\affiliation{Department of Physics, Princeton University, Princeton, NJ 08544, USA}
	\affiliation{Department of Physics, University of Science and Technology of China, Hefei, Anhui 230026, China}
	\author{Jiaze Xie}
	\affiliation{Department of Chemistry, Princeton University, Princeton, New Jersey 08544, USA}
	\author{Rose Albu Mustaf}
	\affiliation{Department of Physics and Astronomy, Rice University, Houston, TX 77005, USA}
	\affiliation{Rice Center for Quantum Materials (RCQM), Rice University, Houston, TX 77005, USA}
	\affiliation{Smalley-Curl Institute, Rice University, Houston, TX 77005, USA}
	\author{Peter Höhn}
	\affiliation{Max Planck Institute for Chemical Physics of Solids, N\"{o}thnitzer Str. 40, Dresden 01187, Germany}
	\author{Vicky Haase}
	\affiliation{Max Planck Institute for Chemical Physics of Solids, N\"{o}thnitzer Str. 40, Dresden 01187, Germany}
	\author{Abdelmajid Ouahchi}
	\affiliation{Department of Physics, TUM School of Natural Sciences, Technical University of Munich, Garching, Germany}
	\affiliation{Faculty of Physics, Ludwig-Maximilians-University Munich, Munich 80799, Germany}
	\author{Soumyajit Samal}
	\affiliation{Faculty of Physics, Ludwig-Maximilians-University Munich, Munich 80799, Germany}
	\affiliation{Munich Center for Quantum Science and Technology (MCQST), Munich 80799, Germany}
	\author{Jiacheng Zhu}
	\affiliation{School of Applied and Engineering Physics, Cornell University, Ithaca, NY 14850, USA}
	\author{Dongyang Yang}
	\affiliation{Department of Physics, Cornell University, Ithaca, NY 14850, USA}
	\affiliation{Max Planck Institute for the Structure and Dynamics of Matter, Luruper Chaussee 149, 22761 Hamburg, Germany}
	\author{Zuhan Geng}
	\affiliation{Department of Physics, Cornell University, Ithaca, NY 14850, USA}
	\affiliation{Max Planck Institute for the Structure and Dynamics of Matter, Luruper Chaussee 149, 22761 Hamburg, Germany}
	\author{Garen Avedissian}
	\affiliation{Donostia International Physics Center (DIPC), Paseo Manuel de Lardizábal. 20018, San Sebastián, Spain}
	\author{Yongsong Wang}
	\affiliation{Donostia International Physics Center (DIPC), Paseo Manuel de Lardizábal. 20018, San Sebastián, Spain}
	\author{Maia G.~Vergniory}
	\affiliation{Département de Physique et Institut Quantique, Université de Sherbrooke, Sherbrooke, J1K 2R1 Québec, Canada}
	\affiliation{Donostia International Physics Center (DIPC), Paseo Manuel de Lardizábal. 20018, San Sebastián, Spain}
	\author{Martin Claassen}
	\affiliation{Department of Physics and Astronomy, University of Pennsylvania, Philadelphia, PA 19104}
	\author{Luis Elcoro}
	\affiliation{Department of Physics, University of the Basque Country UPV/EHU, Apartado 644, 48080 Bilbao, Spain}
	\author{Nicolas Regnault}
	\affiliation{Center for Computational Quantum Physics, Flatiron Institute, 162 5th Avenue, New York, NY 10010, USA}
	\affiliation{Department of Physics, Princeton University, Princeton, NJ 08544, USA}
	\affiliation{Laboratoire de Physique de l'Ecole normale sup\'{e}rieure, ENS, Universit\'{e} PSL, CNRS, Sorbonne Universit\'{e}, Universit\'{e} Paris-Diderot, Sorbonne Paris Cit\'{e}, 75005 Paris, France}
	\author{Miguel M.~Ugeda}
	\affiliation{Donostia International Physics Center (DIPC), Paseo Manuel de Lardizábal. 20018, San Sebastián, Spain}
	\affiliation{IKERBASQUE, Basque Foundation for Science, Bilbao, Spain}
	\author{Jie Shan}
	\affiliation{School of Applied and Engineering Physics, Cornell University, Ithaca, NY 14850, USA}
	\affiliation{Department of Physics, Cornell University, Ithaca, NY 14850, USA}
	\affiliation{Kavli Institute at Cornell for Nanoscale Science, Ithaca, NY 14850, USA}
	\affiliation{Max Planck Institute for the Structure and Dynamics of Matter, Luruper Chaussee 149, 22761 Hamburg, Germany}
	\author{Kin Fai Mak}
	\affiliation{School of Applied and Engineering Physics, Cornell University, Ithaca, NY 14850, USA}
	\affiliation{Department of Physics, Cornell University, Ithaca, NY 14850, USA}
	\affiliation{Kavli Institute at Cornell for Nanoscale Science, Ithaca, NY 14850, USA}
	\affiliation{Max Planck Institute for the Structure and Dynamics of Matter, Luruper Chaussee 149, 22761 Hamburg, Germany}
	\author{Dmitri K.~Efetov}
	\affiliation{Faculty of Physics, Ludwig-Maximilians-University Munich, Munich 80799, Germany}
	\affiliation{Munich Center for Quantum Science and Technology (MCQST), Munich 80799, Germany}
	\author{Emilia Morosan}
	\affiliation{Department of Physics and Astronomy, Rice University, Houston, TX 77005, USA}
	\affiliation{Rice Center for Quantum Materials (RCQM), Rice University, Houston, TX 77005, USA}
	\affiliation{Smalley-Curl Institute, Rice University, Houston, TX 77005, USA}
	\author{Dante M. Kennes}
	\affiliation{Institut f\"ur Theorie der Statistischen Physik, RWTH Aachen University and JARA-Fundamentals of Future Information Technology, 52056 Aachen, Germany}
	\affiliation{Max Planck Institute for the Structure and Dynamics of Matter, Luruper Chaussee 149, 22761 Hamburg, Germany}
	\author{Angel Rubio}
	\affiliation{Max Planck Institute for the Structure and Dynamics of Matter, Luruper Chaussee 149, 22761 Hamburg, Germany}
	\affiliation{Center for Computational Quantum Physics, Flatiron Institute, 162 5th Avenue, New York, NY 10010, USA}
	\affiliation{Nano-Bio Spectroscopy Group and ETSF, Universidad del País Vasco UPV/EHU- 20018 San Sebastián, Spain}
	\author{Lede Xian}
	\affiliation{Tsientang Institute for Advanced Study, Zhejiang 310024, China}
	\affiliation{Songshan-Lake Materials Laboratory, Dongguan, Guangdong 523808, China}
	\affiliation{Max Planck Institute for the Structure and Dynamics of Matter, Luruper Chaussee 149, 22761 Hamburg, Germany}
	\author{Claudia Felser}
	\affiliation{Max Planck Institute for Chemical Physics of Solids, N\"{o}thnitzer Str. 40, Dresden 01187, Germany}
	\author{Leslie M.~Schoop}
	\affiliation{Department of Chemistry, Princeton University, Princeton, New Jersey 08544, USA}
	\author{B.~Andrei Bernevig}
	\email{bernevig@princeton.edu}
	\affiliation{Department of Physics, Princeton University, Princeton, NJ 08544, USA}
	\affiliation{Donostia International Physics Center (DIPC), Paseo Manuel de Lardizábal. 20018, San Sebastián, Spain}
	\affiliation{IKERBASQUE, Basque Foundation for Science, Bilbao, Spain}
}

\newcommand{\tAA}{\text{AA}}
\newcommand{\tAB}{\text{AB}}

\newcommand{\vg}{\vec{g}}

\crefname{appendix}{Appendix}{Appendices}
\crefname{equation}{Eq.}{Eqs.}
\crefname{figure}{Figure}{Figures}
\crefname{table}{Table}{Tables}
\crefname{section}{Section}{Sections}
\creflabelformat{appendix}{[#2#1#3]}

\makeatletter
\AtBeginDocument{\let\LS@rot\@undefined}
\makeatother

\makeatletter
\renewcommand\onecolumngrid{% <<<<<<
\do@columngrid{one}{\@ne}%
\def\set@footnotewidth{\onecolumngrid}% <<<<<<<<<<<<<<<<
\def\footnoterule{\kern-6pt\hrule width 1.5in\kern6pt}%
}

\allowdisplaybreaks
\begin{document}
\title{\titlePaper}
\paperAuthors
\let\oldaddcontentsline\addcontentsline
\begin{abstract}
	The study of twisted two-dimensional (2D) materials, where twisting layers create moir\'e superlattices, has created opportunities for investigating topological phases and strongly correlated physics. The broader potential of a seemingly infinite set of twistable 2D materials remains largely unexplored. Here, we define ``theoretically twistable materials'' as single- or multi-layer structures that allow for the construction of simple continuum models of their moir\'e structures. Our high-throughput algorithm systematically searches for theoretically twistable semimetals and insulators based on the \textit{Topological 2D Materials Database} (2D-TQCDB). By analyzing key electronic properties, we identify thousands of candidate materials, propose representative twistable materials, and provide examples of crystal growth and exfoliation for several of them. Our results provide a resource for future experimental and theoretical studies of moir\'e systems. 
\end{abstract}
\maketitle

The discovery and study of two-dimensional (2D) materials have revolutionized condensed matter physics, opening avenues for exploring quantum phenomena in reduced dimensions. One of the most exciting developments in the field is the concept of twistronics~\cite{carr2017twistronics, AND21,KEN21, mak2022semiconductor}, where a small twist between layers of 2D materials leads to the formation of moir\'e superlattices, fundamentally altering the electronic structure and interaction landscape, giving rise to novel phenomena.

Twisting introduces periodic modulations and creates a moir\'e pattern that can greatly affect the low-energy electronic properties. The resulting moir\'e bands can become extremely flat owing to the enlarged moir\'e unit cell, quenching the kinetic energy of electrons and thus enhancing the role of electron-electron interactions. These interactions can drive a wide range of topological and correlated phases, such as correlated insulating states~\cite{CAO18}, unconventional superconductivity~\cite{CAO18a}, and fractional Chern insulators~\cite{regnault2011fractional, neupert2011, sheng2011}. The prototypical example is twisted bilayer graphene (TBG)~\cite{BIS11}, where a ``magic angle'' twist between two graphene layers leads to the emergence of flat bands~\cite{CAO18a, CAO18, SHA19,SER20,CHE20b,SON22, KER19, wu2019phonon, LU19, WU20, chou2023kondo, SAI21a, LU24,DON23,DON24,HER24b, ZHA22, SON19, zhu2020twisted, TAR19, HAD20}. Beyond graphene, twisted transition metal dichalcogenides (TMDs), such as twisted \ch{MoTe2} and \ch{WSe2}, have also been shown to exhibit many different types of strongly correlated phases~\cite{WU18c,TAN20, WAN22,LI21d,XU23, PAR23,ZEN23,CAI23,WAN24a,JIA24,YU24a, xia2024superconductivity, guo2024superconductivity, DEV21,WU19b, CLA22a, ANG21, li2024contrasting, foutty2024mapping, sheng2024quantum, zhang2024polarization, wang2020correlated, anderson2023programming}.

Despite the intense focus on TBG and TMD systems, many other potential moir\'e systems remain unexplored. Given the vast number of 2D materials~\cite{Haastrup_2018, Gjerding_2021, mc2d, campi2023expansion}.  with different crystal symmetries and electronic structures, there is a much larger space of twistable materials that could host rich correlated and topological physics~\cite{song2021direct, KEN20, klebl2022moire, xian2021realization, xu2024engineering, crepel2024efficient, ye2023kekule, xu2021moire, liu2025symmetry, xu2026organizing, li2026catalogue}. The systematic exploration of this space is crucial to identifying new systems with properties that go beyond what has been observed in TBG and TMD moir\'e superlattices.

However, systematic progress is limited by the lack of theoretical guidance. Many monolayer, bilayer, and multilayer systems have complex band structures that cannot be reduced to simple effective models, such as the Dirac cone in graphene or the quadratic band edge in TMDs. Twisting such systems produces moir\'e bands that are difficult to model, obscuring the prediction of ``magic'' angles~\cite{BIS11}. In TBG, such a theoretical prediction was essential for identifying the superconducting regime. We therefore define ``theoretically twistable'' materials as exfoliable 2D materials with clean low-energy band structures. For metals, this requires point- or line-like Fermi surfaces amenable to effective modeling. For insulators, it requires isolated band extrema that support simple low-energy descriptions. We also impose experimentally motivated criteria, such as a band gap below \SI{3}{eV}, since larger gaps make electrical contacting difficult. Further criteria are described in the text.

\begin{figure*}[tbp]
	\centering
	\includegraphics[width=\textwidth]{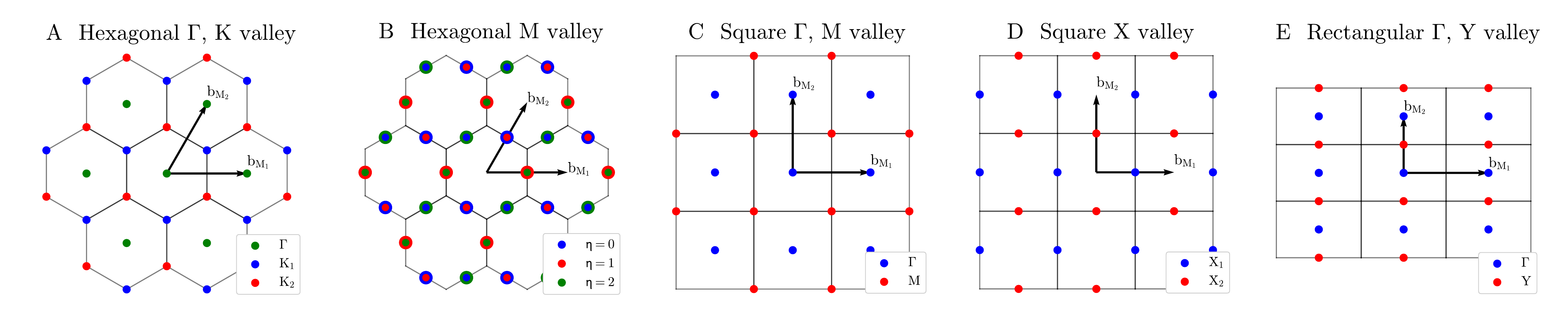}
	\caption{\textbf{Moir\'e $\mathbf{Q}$ lattices derived from different Bravais lattices and valleys.} (A) shows the $\mathbf{Q}$ lattices corresponding to the $\Gamma$ and K valley in the hexagonal lattice. The K valley generates two sets of $\mathbf{Q}$ lattices from the two time-reversal symmetry (TRS)-related K points, $\text{K}_{1}$ and $\text{K}_2$, forming a honeycomb lattice in momentum space. 
		(B) depicts the $\mathbf{Q}$ lattice from the M valley~\cite{calugaru2024mtwist} in the hexagonal lattice, which results in a kagome lattice in momentum space, with three sublattices corresponding to the three $C_3$-symmetry-related M valleys labeled by $\eta=0,1,2$. For each valley, solid (hollow) circles of the same color correspond to the $\vec{Q}$ lattice of the top (bottom) layers. 
		(C) presents the $\Gamma$ and M valleys from the square lattice, both forming a square lattice in momentum space. 
		(D) illustrates the $\mathbf{Q}$ lattice generated by the X valley from the square lattice, forming two nested square lattices in momentum space given by the two $C_4$-related valleys X$_1$ and X$_2$. 
		(E) shows the $\mathbf{Q}$ lattices for the $\Gamma$ and Y valleys from the rectangular lattice. 
	}
	\label{Fig: all-valleys} 
\end{figure*}

Based on these criteria, we develop a high-throughput algorithm to screen the \webtwoDTQC~(2D-TQCDB)~\cite{urko2024two, 2DTQCDB_ref} for theoretically twistable materials. The algorithm identifies semimetals and insulators with clean electronic structures suitable for predictive moir\'e engineering. Applying it, we find \TotalNumTwistSemi twistable semimetals and \TotalNumTwistInsu twistable insulators, which we classify by Bravais lattice, valley type, and spin-orbit-coupling (SOC) strength. 
For representative materials in each class, we present twisted bilayer band structures and corresponding simplified moir\'e continuum models. We also discuss experimental feasibility. The full dataset, including twistability metrics, effective masses, and vacuum-aligned band-edge energies, is available through 2D-TQCDB~\cite{2DTQCDB_ref}. This work expands the materials basis for twistronics and provides a practical guide for future experimental and theoretical studies of 2D moir\'e systems.

\section{Classification of twistable materials}

In twisted systems, the moir\'e potential breaks the translation symmetry of the monolayer and couples the electron states connected by the moir\'e reciprocal lattice vectors $\vec{G}\in \mathcal{Q}=\mathbb{Z}\vec{b}_{M_1} + \mathbb{Z}\vec{b}_{M_2}$, where $\vec{b}_{M_{i}}= 2 \sin \left( \frac{\theta}{2} \right) \vec{b}_i\cross \hat{\vec{z}}$ are moir\'e reciprocal vectors, defined based on the monolayer reciprocal vectors $\vec{b}_{1,2}$ and twist angle $\theta$. The single-particle Hamiltonian for moir\'e systems takes the form
\begin{equation}\mathcal{H}=\sum_{\vec{k},\vec{Q},\vec{Q}',i,j} [h_{\vec{Q},\vec{Q}'}(\vk)]_{ij} \hat{c}^\dagger_{\vec{k},\vec{Q},i} \hat{c}_{\vec{k},\vec{Q}',j}
\end{equation}
where $\vec{k}$ takes value in the first moir\'e Brillouin zone (BZ), $i, j$ are composite indices for orbital, spin, and other degrees of freedom. 
$\vec{Q}$ takes values in the moir\'e $\vQ$ lattice, \ie, $\vec{K}_{\text{valley}}-\mathcal{Q}$, which is formed by the moir\'e plane wave components expanded around the valley momentum $\vec{K}_{\text{valley}}$. Depending on the Bravais lattice and valley types, moir\'e systems exhibit a diverse range of properties.

We classify twistable materials into two types. The first type includes semimetals with (approximately) zero-dimensional (0D) Fermi surfaces (FSs), such as graphene with a linear Dirac crossing. These semimetals may feature various crossing types at the Fermi level, including linear and quadratic degenerate points. The second type includes insulators with clean valence band maxima (VBM) or conduction band minima (CBM), exemplified by TMD materials. We refer to both the crossing points in semimetals and the VBM/CBM in insulators as the ``twisting points'', which are not necessarily located at high-symmetry momentum points (HSPs). Each of the two types of twistable materials is further classified based on their (i) Bravais lattice, (ii) the momentum of the twisting point, and (iii) SOC splitting at the twisting point.

The four Bravais lattices corresponding to 80 layer groups (LGs) are hexagonal, square, rectangular (including centered rectangular), and oblique. The valleys are labeled according to their momenta as follows: for hexagonal lattices, $\Gamma$, M, and K; for square lattices, $\Gamma$, X, and M; and for rectangular and oblique lattices, only HSPs for simplicity. Non-HSPs apply to all lattices. The HSP labeling follows the conventions of the \webBCS~\cite{aroyo2011crystallography}. 
For SOC classification, a numerical threshold of \SI{50}{meV} energy splitting near the twisting point is used to identify materials with strong SOC.

In \cref{Fig: all-valleys}, we illustrate the moir\'e $\vec{Q}$ lattices generated by different Bravais lattices and valley configurations. For a hexagonal lattice, the $\Gamma$ valley gives a triangular $\vec{Q}$ lattice, while the K valleys form a honeycomb lattice whose two sublattices correspond to the TRS-related valleys K$_{1,2}$, as shown in \cref{Fig: all-valleys}A. The M valleys instead generate a kagome $\vec{Q}$ lattice in \cref{Fig: all-valleys}B. This case was introduced in Ref.~\cite{calugaru2024mtwist} for \ch{SnSe2} and \ch{ZrS2}, which are representative materials from the current high-throughput results and can be exfoliated into single layers in experiments. They were shown to host emergent non-symmorphic symmetries. 
For a square lattice, the $\Gamma$ and M valleys generate square $\vec{Q}$ lattices, whereas the X valleys form two nested square lattices associated with the $C_4$-related valleys X$_{1,2}$, as shown in \cref{Fig: all-valleys} C and D. For a rectangular lattice, both the $\Gamma$ and Y valleys generate rectangular $\vec{Q}$ lattices in \cref{Fig: all-valleys}E. Thus, the Bravais lattice and valley configuration determine the geometry of the moir\'e $\vec{Q}$ lattice. 
We have already described the non-trivial physics of twisting the M point in the hexagonal lattice with negligible SOC~\cite{calugaru2024mtwist}. This represents just one of the universality classes of twisted materials, with each class presented here likely to host previously unexplored physics.

\begin{figure*}[htbp]
	\centering
	\includegraphics[width=\textwidth]{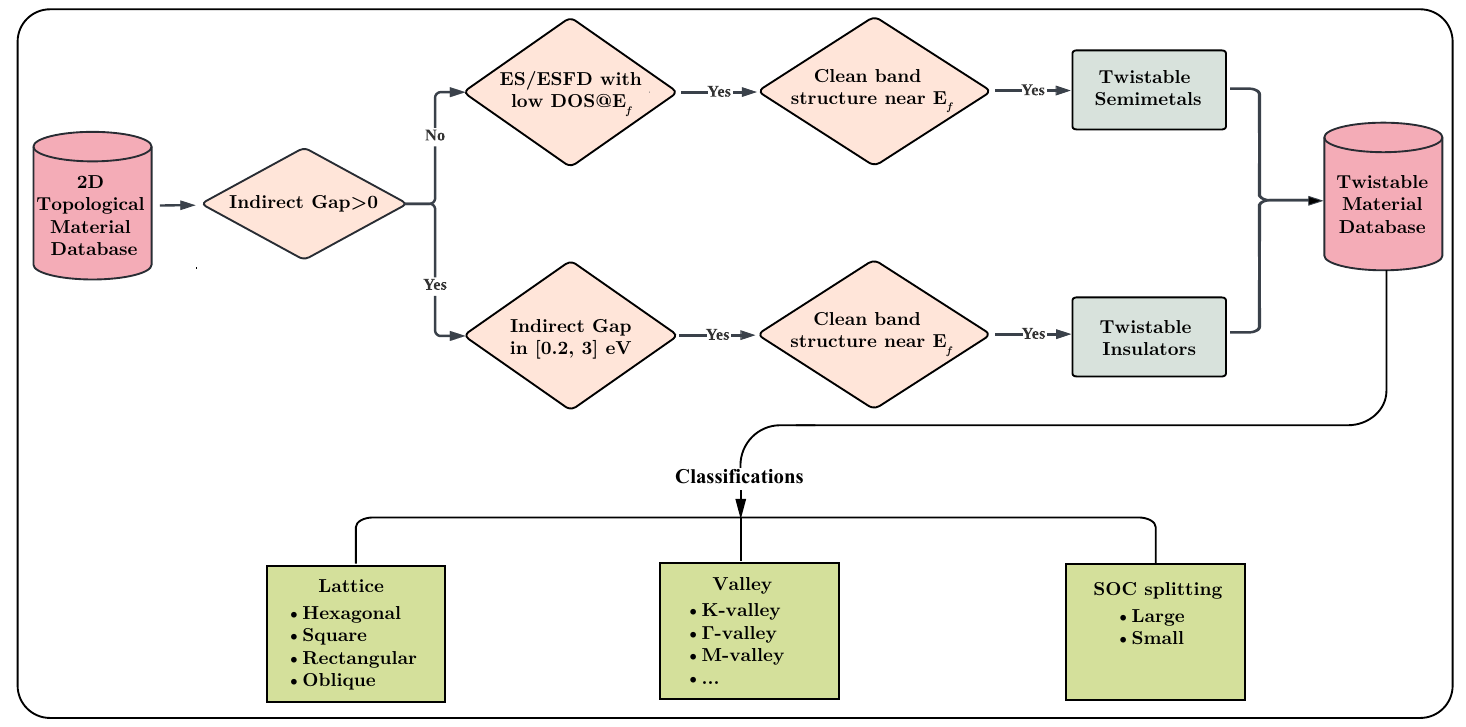}
	\caption{\textbf{The high-throughput algorithm used to search for 2D twistable materials.} We start from all 2D materials in the 2D-TQCDB~\cite{urko2024two}. 
		For a given material, if it has a zero indirect gap, we then determine whether it has a topological classification of ES or ESFD, a low DOS, and a clean band structure at the Fermi level (see \cref{sec:algorithm} High-throughput algorithm for the definition of cleanness). If these criteria are met, the material is classified as a theoretically twistable semimetal with symmetry-protected degeneracy at the Fermi level. Conversely, if the material has an indirect gap within the range of $[0.2, 3]$ \SI{}{eV} and a clean band edge in either the valence or conduction bands, it is considered a theoretically twistable insulator. The identified twistable semimetals and insulators are further categorized based on their Bravais lattice, valley momentum, and SOC splitting strength.
	}
	\label{Fig: searching-algorithm} 
\end{figure*}

\section{High-throughput algorithm}\label{sec:algorithm}

We now outline the high-throughput screening algorithm for identifying theoretically twistable materials, illustrated in \cref{Fig: searching-algorithm}. We first exclude conventional metals with an odd number of valence electrons, which generally cannot form 0D Fermi surfaces or insulating states. We then restrict the search to monolayer unit cells with at most 12 atoms and four chemical elements.

To quantify the isolation of low-energy bands, we define a ``clean energy range'': the energy window around the twisting point containing only one band without SOC, or one Kramers pair with SOC. This condition ensures that the band edge admits a simple $\vk\cdot\vp$ description and is likely to yield clean moir\'e bands after twisting. The detailed criteria for twistable semimetals and insulators are given below.

For semimetals, the criteria are: 
\begin{enumerate}[label=(\roman*)]
	\item Enforced-Semimetal Condition: the topological classification of the material, either with or without SOC, must be either an enforced semimetal with Fermi degeneracy (ESFD), featuring crossings at high-symmetry points, or an enforced semimetal (ES) with crossings along high-symmetry lines, thereby ensuring a symmetry-protected crossing point between valence and conduction bands. Semimetals with a small SOC gap at the crossing points are also included.
	
	\item Density-of-States (DOS) Condition: the DOS at the Fermi level $E_f$ must be $\leq 2$ states per unit cell per eV. This threshold takes an empirical value and aims to account for significant DOS contributions from quadratic crossing points, small Fermi surface pockets from other bands, or multiple crossing points. For instance, \twodtqclinkmatname[1.3.341]{\ch{AuTe}} features several crossing points at $E_f$, resulting in a relatively large DOS of about 1 state per unit cell per \SI{}{eV}. Nevertheless, the bands near $E_f$ remain relatively clean, making it a viable twistable candidate.
	
	\item Clean-Energy-Range Condition: both the VBM and CBM must exhibit a clean energy range of at least \SI{200}{meV}. This energy range is on the order of the interlayer hopping in moir\'e systems, within which a reliable low-energy description can be achieved.
	
	\item Band-Edge Condition: the highest valence band edge and lowest conduction band edge should be within \SI{0.1}{eV} from $E_f$. Even if the semimetal is gapped by SOC or other mechanisms, the gap is typically not large enough to fully separate the conduction and valence bands. As a result, the theoretical description must account for both bands together. We also allow small FS pockets within \SI{0.1}{eV} from $E_f$, as they may be eliminated upon twisting. 
\end{enumerate}

For insulators, we apply the following criteria:
\begin{enumerate}[label=(\roman*)]
	\item Gap Condition: a global band gap between \SI{200}{meV} and \SI{3}{eV};
	
	\item Clean-Energy-Range Condition: a clean energy range of at least \SI{200}{meV} in either the highest valence band or the lowest conduction band. 
	
	\item DOS condition: we require the maximal DOS within \SI{200}{meV} to the twisting point should be $\leq 4$ states per unit cell per \SI{}{eV}. This criterion also takes an empirical threshold used to exclude extremely flat bands at the twisting point, which are likely to produce spaghetti-like bands after twisting.
\end{enumerate}
These conditions are applied to ensure well-isolated band structures. A maximum gap of \SI{3}{eV} is chosen because a larger gap would be experimentally challenging for making electrical contacts and performing effective gating. Given that ab initio calculations generally underestimate the band gap, compounds with large gaps will be ranked with low scores defined in the following. We note that multiple, symmetry-unrelated valleys may exist within the clean energy range. In practice, these valleys are typically either energetically distinct or effectively decoupled at the single-particle level after twisting. Therefore, they are included in our algorithm.

To rank the twistable materials identified by the algorithm, we have established a scoring system to quantify their suitability. The scores for semimetals and insulators are defined as follows:
\begin{equation}
	\begin{aligned}
		\mathcal{S}_\text{semimetal} &= \frac{1}{5}(\mathcal{S}_\text{DOS} + \mathcal{S}_\text{clean VB range} + \mathcal{S}_\text{clean CB range}\\ 
		&+ \mathcal{S}_\text{symmetry} + \mathcal{S}_\text{atom number}) \\
		\mathcal{S}_\text{insulator}^{\text{VB/CB}} &= \frac{1}{4}(\mathcal{S}_\text{gap} + \mathcal{S}_\text{clean VB/CB range} \\
		&+ \mathcal{S}_\text{symmetry} + \mathcal{S}_\text{atom number})
	\end{aligned}
	\label{Eq: score_twisting}
\end{equation}
Here, the score for twistable insulators, \ie, $\mathcal{S}_\text{insulator}^{\text{VB/CB}}$, is defined for the twisting point at VBM and CBM separately. The criteria for each component are as follows:
\begin{itemize}
	\item Gap Score ($\mathcal{S}_{\text{gap}}$): This score aims for an ideal indirect band gap $\Delta$ range of 1 to 1.5  \SI{}{eV}. The score decreases linearly with deviation from this range, \ie, $\mathcal{S}_{\text{gap}} = 1$ for $\Delta \in [1, 1.5]$  \SI{}{eV},
	$\mathcal{S}_{\text{gap}} = 1 - |\Delta - 1|/0.8$ for $\Delta \in [0.2, 1]$  \SI{}{eV},
	and $\mathcal{S}_{\text{gap}} = 1 - |\Delta - 1.5|/1.5$ for $\Delta \in [1.5, 3]$  \SI{}{eV}.
	
	\item Clean-Range Score ($\mathcal{S}_{\text{clean VB/CB range}}$): Measures the cleanliness of the energy range around the VBM or CBM, defined as $\mathcal{S}_{\text{clean CB/VB range}} = 1 - |\max(\Delta E_{\text{VB/CB}}, 1) - 1|/0.8$, where $\Delta E_{\text{CB/VB}}$ is the clean range at the VB/CM, truncated at \SI{1}{eV} because an excessively large energy range is not critical for twisting. 
	
	\item DOS Score ($\mathcal{S}_{\text{DOS}}$):  $\mathcal{S}_{\text{DOS}} = 1 - D(E_f)/2$, where $D(E_f)$ is the density of states at the Fermi level in states per unit cell per \SI{}{eV}. A lower density of states results in a higher score.
	
	\item Symmetry Score ($\mathcal{S}_{\text{symmetry}}$): Defined as the number of point group (PG) symmetries divided by 24, which is the maximum number of PG operations in layer groups.  
	\item Atom Number Score ($\mathcal{S}_{\text{atom number}}$): Inversely related to the number of atoms in the unit cell to favor simpler and more symmetrical compounds.
\end{itemize}
Each component takes value in $[0,1]$. The total scores for both semimetals and insulators range from 0 to 1, where higher scores indicate better candidates. Although these scores may involve a degree of arbitrariness, they are based on both theoretical and experimental facts. 
We intentionally applied less strict criteria to capture a broader range of potentially twistable materials. Although adjusting the definition of the score would alter their rankings, the materials are already categorized into sub-classes of lattices, valleys, and SOC strength, and there are relatively few in each. This allows for the manual selection of candidates from each sub-class for further study, with the score serving as a reference rather than a strict determinant.

\section{Applying the algorithm}\label{sec:results}

We applied the algorithm to all materials in the \webtwoDTQC~\cite{urko2024two} and identified \TotalNumTwistSemi candidates as twistable semimetals and \TotalNumTwistInsu as twistable insulators, with materials further categorized as experimental, computationally exfoliable, computationally stable, and computationally unstable (see Section S5 of \cite{seeSI}
% Supplementary materials Section S5 
% \cref{app:sec:table_intro} 
for more details). Computationally unstable materials are also included to ensure that potential twistable candidates are not overlooked. 
The significantly smaller number of twistable semimetals compared to insulators is a consequence of the strict requirement for clean, symmetry-protected crossings at the Fermi level—a condition that is difficult to meet despite the relatively large number of 2D topological semimetals~\cite{urko2024two}, which typically exhibit large FSs. 
A comprehensive list of these materials is provided in Sections S6 and S7 of \cite{seeSI}. 
% Supplementary materials Section S6 and S7. 
% \cref{app:sec:table_twistable_semimetal} and \cref{app:sec:table_twistable_insulator}. 
In the following, we discuss representative twistable semimetals and insulators across different classes.

\begin{table}[htbp]
	\centering
	\caption{\textbf{Representative twistable semimetals.} They are classified into different classes according to Bravais lattices, valleys, and SOC strength. }
	\label{table:twistable-semimetal}
	\begin{tabular}{c|c|c|c|c}
		\hline\hline
		Lattice & SOC & $\Gamma$ & $\mathrm{K}$ & nHSP \\ \hline
		\multirow{2}{*}{Hexagonal} & Strong & 
		\twodtqclinkmatname[1.2.107]{\ch{ZrBr}}
		& \twodtqclinkmatname[1.3.321]{\ch{Ta2Te2S}}
		& \twodtqclinkmatname[1.3.333]{\ch{ZrTe}}
		\\ \cline{2-5} 
		& Weak 
		& \twodtqclinkmatname[1.4.326]{\ch{Cu2Se}}
		& \twodtqclinkmatname[1.1.15]{\ch{Ge}}
		& \twodtqclinkmatname[1.3.186]{\ch{IrPS3}}
		\\ \hline\hline
		&  & HSP & nHSP &   \\ \hline
		\multirow{2}{*}{\makecell{Rectangular}} & Strong 
		& / 
		& \twodtqclinkmatname[1.1.1]{\ch{MoS2}}
		&  \\ \cline{2-5} 
		& Weak 
		& /
		& \twodtqclinkmatname[1.2.29]{\ch{Hg3S2}} 
		&  \\ \hline\hline
	\end{tabular}
\end{table}

\section{Representative Twistable Semimetals}

Twistable semimetals possess symmetry-protected crossing points near the Fermi level, which can be gapped in the presence of SOC. 
\cref{table:twistable-semimetal} presents representative twistable semimetals, from which three examples are selected and discussed in the following.

\begin{figure}[tbp]
	\centering
	\includegraphics[width=\linewidth]{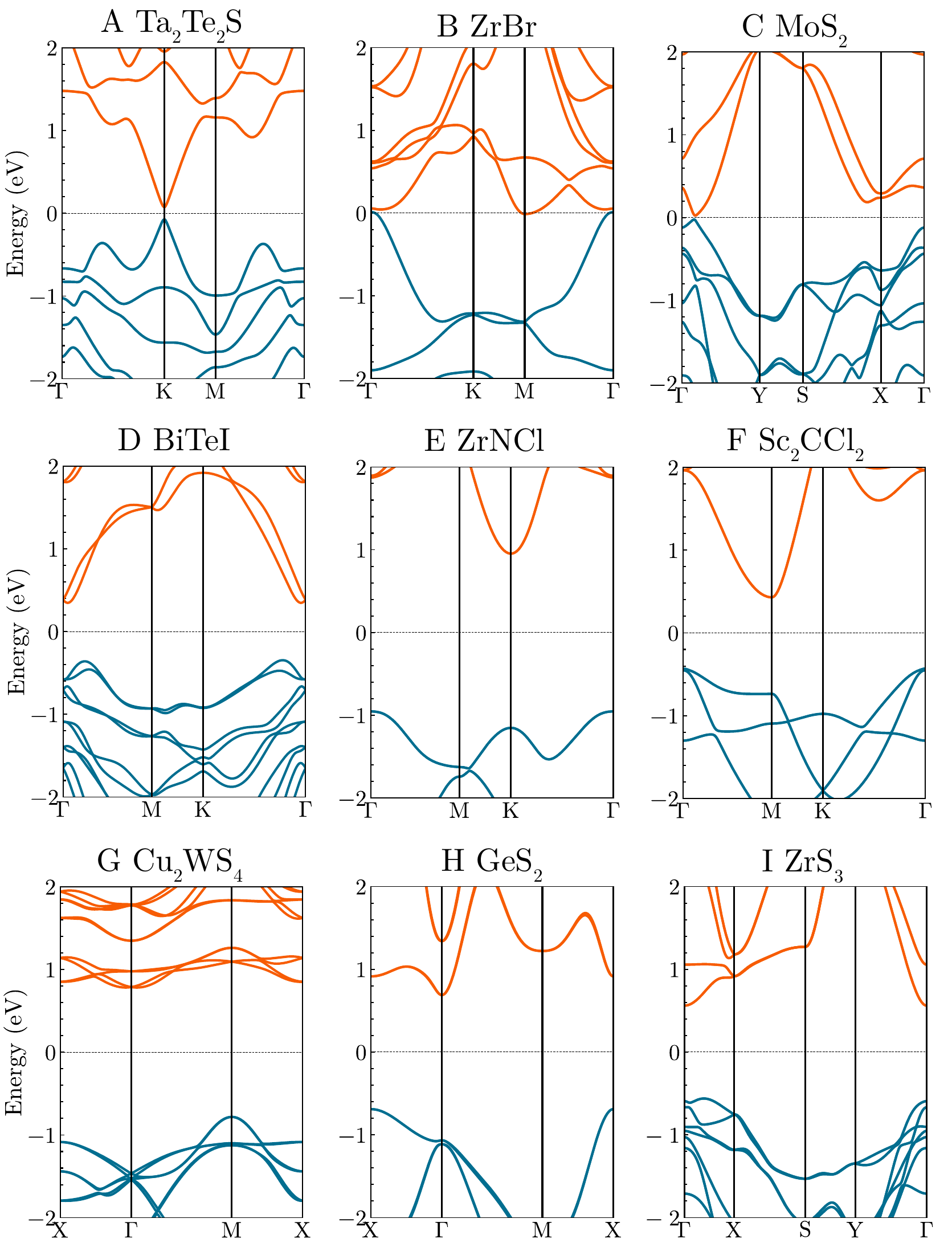}
	\caption{\textbf{The monolayer band structures of representative twistable materials.} (A) Hexagonal \twodtqclinkmatname[1.3.321]{\ch{Ta2Te2S}} exhibits a Dirac cone at the K point, which is gapped by SOC with a gap of approximately \SI{100}{meV}, placing it in a distinctly different universal class from TBG. 
		(B) Hexagonal \twodtqclinkmatname[1.2.107]{\ch{ZrBr}} features a quadratic crossing at the $\Gamma$ point, further gapped by SOC with a gap of about \SI{50}{meV}. (C) Rectangular \twodtqclinkmatname[1.1.1]{\ch{MoS2}} displays a linear crossing along the $\Gamma-Y$ line, which opens a SOC gap of about \SI{50}{meV}. 
		(D) Hexagonal \twodtqclinkmatname[6.1.13]{BiTeI} hosts the CBM at the $\Gamma$ point with strong Rashba SOC splitting. (E) Hexagonal \twodtqclinkmatname[6.3.2408]{ZrNCl} features a clean K valley at the CBM with negligible SOC splitting. (F) Hexagonal \twodtqclinkmatname[6.2.1200]{\ch{Sc2CCl2}}  has the CBM at the M valley with negligible SOC splitting. (G) Square \twodtqclinkmatname[6.3.1620]{\ch{Cu2WS4}} has the VBM at the M valley with negligible SOC splitting. (H) Square \twodtqclinkmatname[6.2.941]{\ch{GeS2}} has the VBM at the X valley with negligible SOC splitting. (I) Rectangular \twodtqclinkmatname[6.2.903]{\ch{ZrS3}} hosts the CBM at the $\Gamma$ valley with negligible SOC splitting.
	}
	\label{Fig: representative-twist-material} 
\end{figure}

\cref{Fig: representative-twist-material}A shows the monolayer band structure of a predicted stable material \twodtqclinkmatname[1.3.321]{\ch{Ta2Te2S}} in \lgsymbnum{72}, which features a linear Dirac crossing at the K point without SOC. This crossing is gapped by SOC, resulting in a sizable gap of approximately \SI{100}{meV}, much larger than the negligible SOC gap in graphene. The large SOC in \twodtqclinkmatname[1.3.321]{\ch{Ta2Te2S}} leads to the formation of quantum spin Hall (QSH) states in both the valence and conduction bands, which are absent in graphene. After twisting, the moir\'e band structure is expected to be markedly different from that of graphene and show distinct moir\'e physics.

\cref{Fig: representative-twist-material}B is for the computationally exfoliable material \twodtqclinkmatname[1.2.107]{\ch{ZrBr}} in \lgsymbnum{72}, where a quadratic crossing is observed at $\Gamma$ with an SOC gap of about \SI{50}{meV}. The quadratic dispersion arises from the higher symmetry of the $D_{3d}$ point group at $\Gamma$, which includes three in-plane $C_2$ rotations and the TRS in addition to the $C_{3z}$ rotation. 
The effective $\vec{k}\cdot\vec{p}$ Hamiltonian for the quadratic dispersion at $\Gamma$ has the form:
\begin{equation}
	h(\delta\vec{k}) \approx v_1 \delta\vec{k}^2 \sigma_0 + v_2
	\begin{bmatrix}
		0 & e^{i\frac{\pi}{3}} \delta k_+^2 \\
		e^{-i\frac{\pi}{3}} \delta k_-^2 & 0
	\end{bmatrix},
\end{equation}
where $\delta k_{\pm}= \delta k_x\pm i \delta k_y$, and $v_{1,2}$ are the parameters for the two terms. 
The quadratic dispersion leads to a quasi-flat segment of band near $\Gamma$ and a large DOS at the Fermi level, which can result in flatter moir\'e bands after twisting. These flatter bands reduce the kinetic energy of electrons, enhancing correlation effects. As a result, \twodtqclinkmatname[1.2.107]{\ch{ZrBr}} could exhibit strong interactions in its twisted form.

\cref{Fig: representative-twist-material}C shows the experimental material \twodtqclinkmatname[1.1.1]{\ch{MoS2}}~\cite{yin2017tunable} in \lgsymbnum{15}, featuring a linear crossing along the $\Gamma-\mathrm{Y}$ high-symmetry line. \twodtqclinkmatname[1.1.1]{\ch{MoS2}} adopts the $1T'$ structure, similar to $1T'$-\ch{WTe2}~\cite{wu2018observation, he2021giant, zhang2023symmetry, yuan2023atomic}. Note that $1T'$-\ch{MoS2} is a meta-stable phase and is different from the common 2H-\twodtqclinkmatname[3.1.39]{\ch{MoS2}}~\cite{backes2015functionalization}. In $1T'$-\twodtqclinkmatname[1.1.1]{\ch{MoS2}}, the linear crossing along the $\Gamma-\mathrm{Y}$ line without SOC is protected by a $C_2$ rotation symmetry. When SOC is introduced, this crossing point is gapped by approximately \SI{50}{meV}, and both the valence and conduction bands exhibit quantum spin Hall (QSH) states. Unlike \ch{WTe2}, where small Fermi surface pockets remain even with SOC, \twodtqclinkmatname[1.1.1]{\ch{MoS2}} has a fully gapped band structure, potentially leading to much simpler moir\'e band structures and offering a cleaner platform for exploring correlated moir\'e phases. We note that the \twodtqclinkmatname[1.3.24]{\ch{WS2}} in the same group exhibits similar gapped band structures and can be synthesized in this structure starting from \ch{K_xWS2}~\cite{song2023acid}.

Among all twistable semimetals, only four candidates have a square lattice: \ch{CuCl2}, \ch{PtS}, \ch{SnS}, and \ch{RuCl2}. Of these, \ch{CuCl2} is computationally exfoliable and has a twisting point at the X point, which is gapped by SOC with small magnetic moments developed on Cu. The others, however, are computationally unstable. For the hexagonal lattice, we note that the M valley has only 1D irreducible representations (IRREPs) in the absence of SOC, and therefore cannot host twistable semimetals.

\section{Representative Twistable Insulators}

\begin{table}[tbp]
	\centering
	\caption{\textbf{Representative twistable insulators.} They are classified into different types of lattices, valleys, and SOC strength. }
	\label{table:twistable-insulator}
	\begin{tabular}{c|c|c|c|c|c}
		\hline\hline
		Lattice & SOC & $\Gamma$ & $\mathrm{M}$ & $\mathrm{K}$ & nHSP \\ \hline
		\multirow{2}{*}{Hexagonal} & Strong 
		& \twodtqclinkmatname[6.1.13]{BiTeI}
		& \twodtqclinkmatname[3.1.35]{GaTe}
		& \twodtqclinkmatname[3.1.33]{\ch{GaSe}}
		& \twodtqclinkmatname[6.2.1090]{\ch{AsSb}} 
		\\ \cline{2-6} 
		& Weak 
		& \twodtqclinkmatname[3.1.37]{InSe} 
		& \twodtqclinkmatname[6.2.1200]{\ch{Sc2CCl2}} 
		& \twodtqclinkmatname[6.3.2408]{ZrNCl} 
		& \twodtqclinkmatname[6.1.58]{\ch{PtSe2}} 
		\\ \hline\hline
		&  & $\Gamma$ & $\mathrm{M}$ & $\mathrm{X}$ & nHSP \\ \hline
		\multirow{2}{*}{Square} & Strong 
		& \twodtqclinkmatname[6.2.982]{\ch{BiIO}} 
		& \twodtqclinkmatname[6.4.385]{\ch{HgH2}}
		& / 
		& \twodtqclinkmatname[6.3.1672]{\ch{SnI2}} 
		\\ \cline{2-6} 
		& Weak 
		& \twodtqclinkmatname[6.3.1642]{\ch{GeCl2}} 
		& \twodtqclinkmatname[6.3.1620]{\ch{Cu2WS4}}
		& \twodtqclinkmatname[6.2.941]{\ch{GeS2}} 
		& \twodtqclinkmatname[6.3.1768]{\ch{GeI}} \\ 
		\hline\hline
		&  & HSP & nHSP &  &  \\ \hline
		\multirow{2}{*}{\makecell{Rectangular/ \\ Oblique}} & Strong 
		& \twodtqclinkmatname[6.3.398]{\ch{Sb2Te2O}}
		& \twodtqclinkmatname[6.1.4]{SnSe}
		&  &  \\ \cline{2-6} 
		& Weak 
		& \twodtqclinkmatname[6.2.903]{\ch{ZrS3}} 
		& \twodtqclinkmatname[3.2.113]{\ch{SnPS3}} 
		&  &  \\ \hline\hline
	\end{tabular}
\end{table}

\cref{table:twistable-insulator} presents representative twistable insulators. From this list, we select six compounds with distinct properties, with their band structures shown in \cref{Fig: representative-twist-material}.

The first three compounds, \twodtqclinkmatname[6.1.13]{BiTeI}, \twodtqclinkmatname[6.2.1200]{\ch{Sc2CCl2}}, and \twodtqclinkmatname[6.3.2408]{ZrNCl}, all possess a hexagonal lattice but have different valleys and SOC splitting strength. 

\cref{Fig: representative-twist-material}D shows the experimental material \twodtqclinkmatname[6.1.13]{BiTeI} in \lgsymbnum{69}, with the CBM near the $\Gamma$ point, characterized by strong Rashba-type SOC splitting. To model the spin splitting, we write down a $\vec{k}\cdot\vec{p}$ Hamiltonian at $\Gamma$ point with the first and second order terms:
\begin{equation}
	h(\delta\vec{k}) \approx v \delta \vec{k}\cdot \bm{\sigma} + \frac{\delta k_x^2 + \delta k_y^2}{2m} \sigma_0,
\end{equation}
where $v$ characterizes the strength of the linear SOC term and $m$ is the effective mass. 
As the $\Gamma$ valley has the time-reversal symmetry (TRS), the strong SOC effects could potentially induce QSH states in the moir\'e bands. At small twist angles where interactions become dominant, this non-trivial topology may give rise to fascinating correlated physical phenomena, including fractionalized topological states at non-integer fillings.

\cref{Fig: representative-twist-material}E shows \twodtqclinkmatname[6.3.2408]{ZrNCl} in \lgsymbnum{72}, with the CBM at the K valley and the VBM at the $\Gamma$ valley, both of which exhibit negligible SOC splitting. The weak SOC splitting at the K valley contrasts sharply with that of the well-studied TMD materials such as \ch{MoTe2} and \ch{WSe2}. 
In the literature, bulk \ch{ZrNCl} and its Hf counterpart, \ch{HfNCl}, have been shown to exhibit superconductivity at approximately \SI{15}{K} and \SI{25}{K}, respectively, upon doping into the conduction band ~\cite{yamanaka1998superconductivity, taguchi2006increase}. Therefore, twisted \twodtqclinkmatname[6.3.2408]{ZrNCl} may similarly exhibit superconducting properties upon doping.

\cref{Fig: representative-twist-material}F features a predicted exfoliable material \twodtqclinkmatname[6.2.1200]{\ch{Sc2CCl2}}  in \lgsymbnum{72}, displaying a CBM at the $\mathrm{M}$ point.  In contrast to the $\mathrm{K}$ valley in TMDs,  there are three $C_3$-related $\mathrm{M}$ valleys in the monolayer BZ. When mapped onto the moir\'e BZ, these M valleys form a kagome lattice in momentum space, which could lead to novel correlated physics enriched by the rich valley degrees of freedom~\cite{calugaru2024mtwist}.

The next two compounds, \twodtqclinkmatname[6.3.1620]{\ch{Cu2WS4}} and \twodtqclinkmatname[6.2.941]{\ch{GeS2}}, both have a square lattice. Although moir\'e systems with a square lattice have been theoretically proposed~\cite{kariyado2019flat, xu2024engineering, eugenio2024tunable}, they have not yet been realized in experiments. Here, we propose two twistable square lattice insulators \twodtqclinkmatname[6.3.1620]{\ch{Cu2WS4}} and \twodtqclinkmatname[6.2.941]{\ch{GeS2}}, both of which are computationally exfoliable~\cite{mc2d, campi2023expansion} with bulk structures and exhibit different valley properties.

\cref{Fig: representative-twist-material}G shows \twodtqclinkmatname[6.3.1620]{\ch{Cu2WS4}} in \lgsymbnum{57}, featuring a VBM at the $\mathrm{M}$ point with negligible SOC splitting, and a flat CBM in the vicinity of $\Gamma$ with strong SOC-splitting. Both the $\Gamma$ and $\mathrm{M}$ valleys exhibit $C_4$ symmetry and form a square lattice in the moir\'e BZ upon twisting. 
\cref{Fig: representative-twist-material}H presents \twodtqclinkmatname[6.2.941]{\ch{GeS2}} in \lgsymbnum{59}, with the VBM at the $\mathrm{X}$ point and the CBM at the $\Gamma$ point. The $\mathrm{X}$ point differs from the $\Gamma$ and $\mathrm{M}$ points, as it lacks $C_4$ symmetry. Upon twisting, the $\mathrm{X}$ valley forms two sets of staggered square lattices in the moir\'e BZ, related by the $C_4$ symmetry, as shown in \cref{Fig: all-valleys}D. 

We note that the $\Gamma$, $\mathrm{M}$, and $\mathrm{X}$ valleys from the square lattice all respect TRS, and could potentially give rise to quantum spin Hall (QSH) states in the presence of strong SOC. Moir\'e systems with a square lattice could serve as a platform for simulating the Hubbard model, particularly in relation to high-temperature SC in cuprates~\cite{bednorz1986possible, orenstein2000advances}, and may also exhibit unconventional superconductivity.

Lastly, \cref{Fig: representative-twist-material}I shows the band structure of \twodtqclinkmatname[6.2.903]{\ch{ZrS3}} in \lgsymbnum{46}, which has a rectangular lattice and is computationally exfoliable~\cite{mc2d}. The CBM appears at $\Gamma$ and VBM is located along the $\Gamma-X$ line. In the rectangular lattice, the two in-plane lattice constants differ, and this disparity is further amplified in the moir\'e unit cell at small twist angles~\cite{WAN22, yu2023evidence}. As a result, rectangular moir\'e systems are expected to exhibit quasi-1D characteristics, potentially acting as Luttinger liquid simulators~\cite{KEN20}. In \twodtqclinkmatname[6.2.903]{\ch{ZrS3}}, the two in-plane lattice constants have a ratio of approximately 1.5. Additionally, the bands near the VBM are relatively flat along the $k_x$ direction, which could potentially lead to flat moir\'e bands and further enhance the potential for strongly correlated physics.

\subsection{Band structures for twisted bilayer materials}

The database of theoretically twistable 2D materials developed in this work identifies promising candidates for twist engineering from monolayer properties. These candidates have clean band structures, making them suitable for simple continuum modeling and likely to yield clean moir\'e bands after twisting. We demonstrate the predictive power of this approach by comparing database-based expectations with full density-functional calculations of selected twisted bilayers, including structural relaxation. Because small twist angles lead to moir\'e cells with many thousands of atoms, fully ab initio calculations are computationally demanding. We therefore study representative candidates at moderate twist angles. AA-stacked bilayers are constructed by directly aligning the two layers, while AB-stacked bilayers are obtained by rotating the top layer by 180$\degree$ before stacking.

\begin{figure*}[tbp]
	\centering
	\includegraphics[width=\textwidth]{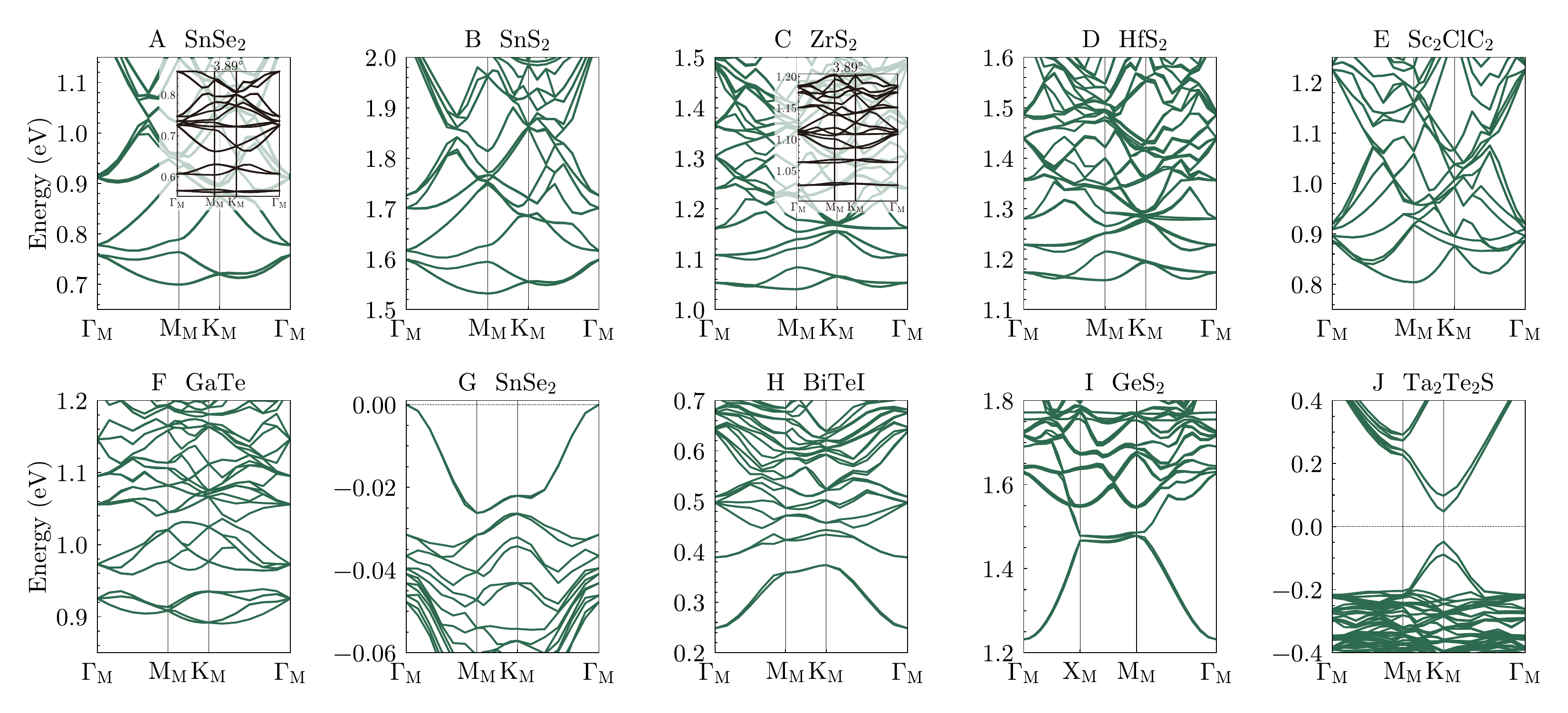}
	\caption{\textbf{Band structures for twisted bilayer materials.} Conduction bands of (A) \twodtqclinkmatname[6.1.71]{\ch{SnSe2}}, (B) \twodtqclinkmatname[6.1.75]{\ch{SnS2}}, (C) \twodtqclinkmatname[6.2.1299]{\ch{ZrS2}}, (D) \twodtqclinkmatname[6.2.1259]{\ch{HfS2}}, and (E) \twodtqclinkmatname[6.2.1200]{\ch{Sc2CCl2}} at twist angle $\theta=7.34^\circ$, representing hexagonal lattice M-valley systems with negligible SOC. In the inset of (A) and (C), we show the conduction band of \ch{SnSe2} and \ch{ZrS2} at $\theta=3.89\degree$~\cite{calugaru2024mtwist}. (F) Conduction bands of \ch{GaTe} at $\theta = 5.09^\circ$, illustrating a hexagonal lattice M-valley system with strong SOC. (G) shows the valence bands of \twodtqclinkmatname[6.1.71]{\ch{SnSe2}} and (H) shows the conduction bands of \ch{BiTeI} at $\theta = 5.09^\circ$, exemplifying hexagonal lattice $\Gamma$-valley systems with strong SOC. (I) is the conduction bands of \ch{GeS2} at $\theta = 12.68^\circ$, demonstrating the square-lattice $\Gamma$-valley type. (J) shows \ch{Ta2Te2S} at $\theta = 7.34^\circ$, a hexagonal lattice K-valley twistable semimetal resembling TBG but with significantly stronger SOC than TBG. The high-energy entangled valence bands come from other valleys. 
	}
	\label{Fig: twisted-bands} 
\end{figure*}

We first focus on the universality class of hexagonal-lattice M-valley with weak SOC. \cref{Fig: twisted-bands} A-E summarizes the band structures at twist angle $\theta=7.34^\circ$, for AA-stacked \twodtqclinkmatname[6.1.71]{\ch{SnSe2}}, \twodtqclinkmatname[6.1.75]{\ch{SnS2}}, \twodtqclinkmatname[6.2.1299]{\ch{ZrS2}}, \twodtqclinkmatname[6.2.1259]{\ch{HfS2}}, and \twodtqclinkmatname[6.2.1200]{\ch{Sc2CCl2}}, respectively. All five materials demonstrate the emergence of isolated (or nearly isolated) moir\'e flat bands, confirming the utility of our twistable materials database. The insets complement these results with moir\'e bands obtained at a smaller twist angle of 3.89$\degree$~\cite{calugaru2024mtwist}, for which we have already built continuum models. This comparison illustrates that although twist angle variation enables bandwidth control of the few-flat-band physics, the essential characteristics are already established at moderate angles.

We next extend our analysis of moir\'e bands to additional universality classes, highlighting distinct electronic properties emerging from different valley configurations and SOC strengths. 
\cref{Fig: twisted-bands}F depicts the conduction bands of twisted  AB-stacked \twodtqclinkmatname[3.1.35]{\ch{GaTe}}, a hexagonal lattice material with the valley at the M point, showing pronounced spin splitting in its moir\'e bands owing to the strong Rashba SOC in the monolayer, offering potential for spin-dependent phenomena. 
\cref{Fig: twisted-bands} G and H present the valence bands of AB-stacked \twodtqclinkmatname[6.1.71]{\ch{SnSe2}} and conduction bands of AB-stacked \twodtqclinkmatname[6.1.13]{\ch{BiTeI}}, respectively. These hexagonal-lattice $\Gamma$-valley systems with strong SOC provide a promising platform to host non-trivial QSH topology, making them promising candidates for topological moir\'e physics, including fractionalized topological states. 
\cref{Fig: twisted-bands}I presents the conduction bands of \twodtqclinkmatname[6.3.1660]{\ch{GeS2}}, a square-lattice material with $\Gamma$ valley and weak SOC. The two lowest spinful conduction bands are isolated from other bands, which could be flatter at a lower twisting angle and modeled by a square-lattice Hubbard model. 
Finally, \cref{Fig: twisted-bands}J shows the moir\'e bands of twistable semimetal \twodtqclinkmatname[1.3.321]{\ch{Ta2Te2S}}, originating from the monolayer Dirac cone with a large SOC gap. Compared with TBG, several distinct features emerge in \twodtqclinkmatname[1.3.321]{\ch{Ta2Te2S}}: Strong SOC induces a substantial gap between the conduction and valence bands, and significant spin splitting reduces the 4-fold valley-spin degeneracy in TBG to 2-fold valley-only degeneracy in \twodtqclinkmatname[1.3.321]{\ch{Ta2Te2S}}. Note that the densely packed valence bands below \SI{-0.2}{eV} originate from other valleys at nHSP, as seen in monolayer bands in \cref{Fig: representative-twist-material}A. 

These diverse case studies of moir\'e band structures highlight the predictive power of our database. By leveraging monolayer descriptors alone, we can efficiently identify materials poised to host clean, twist-induced quasi-flat bands. In Section S6 of \cite{seeSI}, 
% Supplementary materials Section S6 
% \cref{app:sec:table_intro}, 
we provide the complete list of theoretically twistable materials in each universality class. 
A high-throughput algorithm to establish a database of moderate-twist-angle materials is presented in \cite{xu2026organizing}. Combining the two databases will provide an even more holistic guide to experiments on twisted two-dimensional materials and will hone in on the next generation of twist-engineering of correlated and topological 2D quantum materials.

\section{Simple moir\'e continuum model}
Different $\vQ$-lattices constrained by the intra-valley symmetries produce different moir\'e potentials, which control both the single-particle bands and the interaction-driven many-body physics. Below, we present simple continuum models for several typical cases and summarize their main features, including quasi-one-dimensional dispersion and non-trivial band topology.

We start with the M-valley on the hexagonal lattice with negligible SOC. The M-valley of the hexagonal lattice has three $C_3$-related subvalleys, forming a kagome $\vQ$ lattice as shown in \cref{Fig: all-valleys}B. 
In valley $\eta = 0$, a simplified moir\'e Hamiltonian has the form
{\small
	\begin{equation}
		\begin{aligned}
			\left[ h_{\vQ,\vQ'} \left( \vk \right) \right]_{s l; s' l'} &=  \delta_{\vQ, \vQ'} \delta_{s s'} \delta_{l l'} \left[ \frac{\left( k_{x} - Q_{x} \right)^2}{2 m_x} + \frac{\left( k_{y} - Q_{y} \right)^2}{2 m_y} \right] \\
			&+ \left[ T_{\vQ,\vQ'} \right]_{s l; s' l'}, \qq{for} \vQ^{(\prime)} \in \mathcal{Q}_{l^{(\prime)}}, \label{eqn:single_particle_hamiltonian}
		\end{aligned}    
\end{equation}}
where $m_x$ and $m_y$ are two effective masses along the $k_x$ and $k_y$ directions, respectively, and $s$ ($l$) is the spin (layer) index. The interlayer hopping terms have the form:  $\left[ T^{\tAA}_{\vQ,\vQ'} \right]_{ls;(-l)s} = \left( \pm i w^{\tAA}_1 + w^{\tAA}_2 \right) \delta_{\vQ \pm \vq_{0}, \vQ'} + w^{\prime \tAA}_3 \delta_{\vQ \pm \left( \vq_1 - \vq_2 \right), \vQ'}$ and $\left[ T^{\tAB}_{\vQ,\vQ'} \right]_{ls;(-l)s} = w^{\tAB}_2 \delta_{\vQ \pm \vq_{0}, \vQ'} + w^{\prime \tAB}_4 \delta_{\vQ \pm \left( \vq_1 - \vq_2 \right), \vQ'}$. Here, $\vq_i=C_{3z}^{i}\frac{\vec{b}_{M1}}{2}$ is the first harmonic vectors for the Fourier expansion of moir\'e potential in M-valley systems. The superscript AA (AB) is for AA (AB) stacked bilayer, which has space group 149 (150) symmetry. 
Owing to the negligible SOC, the model presents spin $\text{SU}(2)$ symmetry. 
In Ref.\cite{calugaru2024mtwist}, we use this simplified M-valley model to fit the moir\'e bands of \ch{SnSe2} and \ch{ZrS2}, which gives good agreement in both dispersion and wavefunction. 
One key feature of this simplified M-valley moir\'e Hamiltonian is the emergence of an effective $\tilde{M}_z$ symmetry, which acts non-symmorphically in momentum space by mapping $\vk$ to $\vk + \vq_0$. 
The momentum-space non-symmorphic $\tilde{M}_z$ symmetry, which exists without applied magnetic fields, enforces perfect nesting at wavevector $\vq_0$ and induces an effective 1D character~\cite{calugaru2024mtwist}. Although the three M-valleys remain nearly decoupled at the single-particle level, strong electron-electron interactions arising from the flat moir\'e bands can couple them. Combining with the underlying $C_3$ symmetry relating the three valleys, an exotic crossed sliding Luttinger liquid phase may be stabilized. These unique characteristics make M-valley moir\'e systems promising platforms for realizing and studying correlated Luttinger liquid physics in 2D materials.

We then discuss a continuum model of the $\Gamma$ valley in hexagonal lattice systems. Under the symmetry constraints of $C_{3z}$ and $C_{2x}$ (for AA-stacking) or $C_{2y}$ (for AB stacking) and $\mathcal{T}$, the simplified general model has the form 
{\small
	\begin{equation}
		\begin{aligned}
			\left[ h_{\vQ,\vQ'} \left( \vk \right) \right]_{s l; s' l'} &=  \delta_{\vQ, \vQ'} \delta_{s s'} \left[\delta_{l l'}  \frac{\left(\vk - \vQ \right)^2}{2 m_1}+ \delta_{-l l'}  \frac{\left(\vk - \vQ \right)^2}{2 m_2}\right]\\
			&+ \left[ T_{\vQ,\vQ'} \right]_{s l; s' l'}. \label{eqn:single_particle_hamiltonian_Gamma}
		\end{aligned}    
\end{equation}}
The specific form of $\left[ T^{\tAA}_{\vQ,\vQ'} \right]$ is provided in Section S2 of \cite{seeSI}. 
% Supplementary materials Section S2. 
% \cref{app:sec:details}. 
Because the top and bottom layer share the same hexagonal $\vQ$-lattice, the layer coupling is particularly strong in $\Gamma$ valley, making layer bonding and antibonding states the natural bases. 
To account for the potentially different effective masses of these states, we introduce two mass parameters, $m_1$ and $m_2$. To realize QSH states, strong SOC terms are required in the continuum model, which can be introduced through either the second harmonic SOC term or the on-site Rashba term, depending on the ab initio band structures of the system. 
As detailed in \cite{pi2026engineering}, the QSH bands observed in \ch{SnSe2} require strong SOC terms in the continuum model. They can be introduced through either the second-harmonic SOC term or the on-site Rashba term, depending on the ab initio band structures of the system.

Lastly, we examine the continuum model for a general square lattice $\Gamma$-valley system with $C_{4z}$, $C_{2y}$, and $\mathcal{T}$ symmetries. 
Although the Hamiltonian retains the general form of \cref{eqn:single_particle_hamiltonian_Gamma}, the distinct $\vQ$-lattice geometry produces characteristically different moir\'e potentials and resulting band structures. 
When the SOC is negligible, we simplify the model by retaining only the onsite term for interlayer coupling $\left[ T_{\vQ,\vQ'} \right]_{ls;(-l)s} =w_0\delta_{\vQ , \vQ'}$, and keep the first harmonic terms for the intralayer part, \ie, $\left[ T_{\vQ,\vQ'} \right]_{ls;ls} =V_1 \delta_{\vQ + \vg_{1}, \vQ'}$. Here, $\boldsymbol{g}_{1,i}=C_{4z}^{i}\vec{b}_{M_1}$ represents the first harmonic vectors. 
When the lowest set of moir\'e bands is topologically trivial, it can be described by a single-orbital square-lattice Hubbard model, which may provide insight into the cuprate physics.

We also briefly examine the emergence of ``magic angles'' in moir\'e systems. 
For twistable semimetals with a Dirac crossing at the Fermi level, magic angles occur when the Fermi velocity in the moir\'e bands vanishes at the Dirac point, a condition that can be analytically determined from the monolayer Fermi velocity and interlayer coupling. 
For twistable insulators, moir\'e bands generally flatten at smaller angles, though bandwidths may vary non-monotonically owing to the strong moir\'e potential. The magic angle can be defined as the critical angle where band extrema shift away from the valley momentum in the moir\'e Brillouin zone~\cite{DEV21, morales2024magic}. 
Our 2D-TQCDB database enables preliminary estimates through the band-edge effective masses, where larger masses generally indicate flatter moir\'e bands and potentially larger magic angles. However, precise predictions require material-specific modeling of the full moir\'e Hamiltonian, including interlayer coupling strength, which is an important direction for future study.

\section{Experimental validation}\label{sec:experiment_observation}

A central goal of this work is to identify twistable materials suitable for immediate experimental studies. Since moir\'e devices are typically fabricated from exfoliated mono- or few-layer crystals~\cite{huang2020universal}, we focus on systems that satisfy three practical criteria: (i) the structures have bulk crystal analogs, (ii) bulk crystals can be grown with sufficient size and quality, and (iii) the bulk crystals contain van der Waals gaps or have other viable exfoliation pathways. For these materials, we further consider the following factors: a) stability of monolayers to brief air exposure; b) how clean the systems are expected to be in terms of defects, impurities, or disorder; c) reports of experimental exfoliation down to monolayers. 

Based on these criteria, the experimentally relevant twistable candidates can be grouped into six broad classes:
\begin{enumerate}
	\item \textbf{Elements:} Exfoliation of graphene to monolayers kickstarted the field of 2D electronics~\cite{geim_rise_2007}. Yet, to this day, we have relatively few experimentally exfoliated elements to create moir\'e structures with. 
	
	\item \textbf{TMDs:} TMDs have had an overwhelming influence on  moir\'e research. Despite this, only a small subset of transition metals has been used to produce twisted moir\'e systems experimentally, or even theoretically. Materials including \ch{SnS2}, \ch{HfS2}, \ch{ZrS2}, 
	and their selenium analogs, have all been exfoliated down to monolayers \cite{huang2020universal}, however no twistronic devices with them are reported to date. What makes them even more interesting is that they are all hexagonal M-valley systems, distinct from the well-studied moir\'e TMDs, which primarily involve the K valley. This distinction leads to significantly different physics~\cite{calugaru2024mtwist}.
	
	\item \textbf{Other chalcogenides :} Chalcogenides are particularly likely to create van der Waals layered materials thanks to the chalcogen atoms' lone pairs. It is not surprising then that many candidate twistable materials belong to this class. Some standout materials here include group IIIA monochalcogenides, such as \ch{GaSe}. These are well established as exfoliable van der Waals systems, however few studies have considered using them in heterostructure or twistronic devices. Group IVB trichalcogenides such as \ch{ZrSe3} and \ch{TiS3} are another promising family here, representing well-established exfoliable systems. Lastly, tetradymite-type and related chalcogenides of antimony, bismuth, and group IVA metals comprise the last important family of twistable experimental chalcogenides, including \ch{Bi2Te3}.
	
	\item \textbf{Halides:} Similar to chalcogenides, van der Waals halides are fairly common. Not all of them, however, are stable to air or moisture, complicating exfoliation in some cases. The most promising systems according to our methodology include \ch{PbI2}, \ch{BiI3}, and \ch{CdI2}. More details can be found in Section S3 of \cite{seeSI}. 
	
	\item \textbf{Mixed pnictides, chalogenides, and halides:} Both chalcogenides and halides are especially suitable to produce exfoliable materials; it is therefore unsurprising that combinations of halide and chalcogenide ions, or either of those with pnictides, also can produce materials good for exfoliation. Some key examples of materials in this class include \ch{ZrNCl}, which has been exfoliated to monolayers~\cite{nong_layer-dependent_2022}, and is reported to be a superconductor when gated~\cite{saito_metallic_2015}; and bismuth chalcohalides such as \twodtqclinkmatname[6.1.13]{BiTeI}, which likewise form stable monolayers, so far unexplored in moir\'e research. Thiophosphates such as \ch{AgInP2S6}~\cite{gao_vacancy-defect_2021} have also recently been exfoliated experimentally, and can potentially produce clean twistable systems.
	
	\item \textbf{Other materials:} Most experimental twistable materials fit into the five classes listed above.
	\ch{GeH} is one of the candidates that are outside of this classification. Several oxides also show promise for twisting theoretically, although experimental work may be somewhat challenging, including \ch{TiO2} and the MXene \ch{Ti2CO2}. Lastly, one of the highest twist scores for experimentally obtained materials we found in 2D-\ch{GaN}. This phase of \ch{GaN} was recently obtained by encapsulation between sheets of graphene \cite{al_balushi_two-dimensional_2016}. Experimental twisting of this system does not appear possible, but it could become an interesting material in the future if synthetic strategies affording freestanding monolayers can be devised.
\end{enumerate}

Among the experimentally twistable materials, we have successfully synthesized several of the most promising ones in bulk form, including TMDs \ch{SnSe2}, \ch{SnS2}, and \ch{HfS2}, chalcogenide GaTe, and pnicto-halide ZrNCl. 
We also demonstrate the experimental exfoliation of several twistable materials into high-quality monolayers, including \twodtqclinkmatname[6.1.71]{\ch{SnSe2}}, 
\twodtqclinkmatname[6.1.45]{\ch{HfSe2}}, 
\twodtqclinkmatname[6.2.1299]{\ch{ZrS2}}, and
\twodtqclinkmatname[1.3.24]{1T'-\ch{WS2}}. 
We employ an ultra-clean gold-assisted exfoliation technique~\cite{wu2024gold}, which enables the deterministic exfoliation of a monolayer in each attempt, ensuring reproducibility and near-unity yield in monolayer production. We apply this technique to several twistable materials, including \twodtqclinkmatname[6.1.75]{\ch{SnS2}}, \twodtqclinkmatname[6.2.1299]{\ch{ZrS2}}, \twodtqclinkmatname[6.1.43]{\ch{HfS2}}, and \twodtqclinkmatname[6.3.2408]{ZrNCl}. The exfoliated monolayer flakes have significantly larger areas compared to conventional scotch-tape exfoliation. 
The exfoliated monolayer samples can be readily used to make twist bilayer devices and perform characterizations. As an example, we realized dual-gated devices of homobilayer \twodtqclinkmatname[6.1.71]{\ch{SnSe2}}, which shows a comparable mobility as \ch{MoTe2}. 
As a further demonstration, we also fabricated a twisted homobilayer \twodtqclinkmatname[6.1.43]{\ch{HfS2}} device for millikelvin scanning tunneling microscopy/spectroscopy (STM/STS) measurements, confirming the suitability of these exfoliated monolayers for twist-device assembly and local spectroscopy. More details are provided in Section S3 of \cite{seeSI}.

\section{Discussion and outlook}\label{sec:conclusion}

This work provides a practical starting point for systematic moir\'e materials design beyond graphene and TMDs. By identifying theoretically twistable semimetals and insulators with clean low-energy electronic structures, the resulting database highlights materials for which predictive continuum modeling is feasible and experimentally meaningful. These candidates offer diverse lattice symmetries, valley structures, and SOC strengths, opening routes to correlated states and topological phases in previously unexplored moir\'e systems. Going forward, the database can guide material synthesis and device fabrication, while experimental outcomes can in turn refine the twistability criteria. Coupling this resource to high-throughput moir\'e calculation and automated model construction would transform the present screening into an iterative platform for discovering and engineering new moir\'e quantum matter.

\section{Methods}

The electronic properties calculations for the monolayer and twisted bilayer system were performed using density functional theory with the Vienna ab initio software package (VASP) \cite{kresse_1996}. The Perdew-Burke-Ernzerhof (PBE) exchange-correlation functionals \cite{perdew1996generalized} were employed. More details are left in Section S4 in \cite{seeSI}.

Bulk crystals of selected twistable candidates were synthesized to assess their experimental feasibility, including \ch{SnSe2}, \ch{SnS2}, \ch{HfS2}, GaTe, and ZrNCl. Among them, \ch{SnSe2}, \ch{SnS2}, and \ch{HfS2} were grown by chemical vapor transport methods, GaTe was grown by a self-flux method, and ZrNCl by a two-step solid-state synthesis route. Crystal quality, phase purity, and composition were characterized by powder X-ray diffraction. 
Monolayers of several representative materials, including \ch{SnSe2}, \ch{HfSe2}, \ch{ZrS2}, 1T$'$-\ch{WS2}, \ch{SnS2}, \ch{HfS2}, and ZrNCl, were obtained by mechanical exfoliation onto \ch{SiO2/Si} or Au-coated substrates. For materials with low conventional exfoliation yield, we used gold-assisted exfoliation, which produces large-area, clean monolayers suitable for device fabrication. As demonstrations, we fabricated dual-gated homobilayer \ch{SnSe2} devices and twisted homobilayer \ch{HfS2} devices, confirming the compatibility of these materials with standard moir\'e-device assembly. Further synthesis, exfoliation, characterization, and device-fabrication details are provided in Section S3 of \cite{seeSI}.

% \begin{acknowledgments}
\section*{Acknowledgments}

\paragraph*{Funding:} 
We thank for technical support provided by the staff at SGIker (UPV/EHU/ ERDF, EU), and the Donostia International Physics Center Supercomputing Center. 
The simulations presented in this article were partially performed on computational resources managed and supported by Princeton Research Computing, a consortium of groups including the Princeton Institute for Computational Science and Engineering (PICSciE) and the Office of Information Technology's High Performance Computing Center and Visualization Laboratory at Princeton University. 
Y.J. and H.H. were supported by the European Research Council (ERC) under the European Union’s Horizon 2020 research and innovation program (Grant Agreement No. 101020833), as well as by the IKUR Strategy under the collaboration agreement between Ikerbasque Foundation and DIPC on behalf of the Department of Education of the Basque Government. 
U.P. acknowledges funding from the European Union’s Next Generation EU plan through the María Zambrano Programme. 
U.P. and L.E. were supported by the Government of the Basque Country (Project No. IT1458-22).
G.S. was supported by the Arnold and Mabel Beckman Foundation via an AOB postdoctoral fellowship (dx.doi.org/10.13039/100000997). 
D.C. acknowledges support from the DOE Grant No. DE-SC0016239 and the hospitality of the Donostia International Physics Center, at which this work was carried out. 
M.G.V and H.P. were supported by the Ministry for Digital Transformation and of Civil Service of the Spanish Government through the QUANTUM ENIA project call - Quantum Spain project, and by the European Union through the Recovery, Transformation and Resilience Plan - NextGenerationEU within the framework of the Digital Spain 2026 Agenda. 
M.G.V. thanks support to the Deutsche Forschungsgemeinschaft (DFG, German Research Foundation) GA 3314/1-1 – FOR 5249 (QUAST), the Spanish Ministerio de Ciencia e Innovacion (PID2022-142008NB-I00) and the Canada Excellence Research Chairs Program for Topological Quantum Matter.
L.M.S. was supported by the Gordon and Betty Moore Foundation's EPiQS Initiative (grant GBMF9064), the David and Lucile Packard Foundation, the National Science Foundation (NSF) MRSEC program through the Princeton Center for Complex Materials (grant DMR-2011750), the NSF through the AI Research Institutes program (award DMR-2433348), the Princeton Catalysis Initiative (PCI), and the Schmidt Transformative Technology Fund at Princeton University. 
D.K.E. acknowledges funding from the European Research Council (ERC) under the European Union’s Horizon 2020 research and innovation program (grant agreement No. 852927), the German Research Foundation (DFG) under the priority program SPP2244 (project No. 535146365), the EU EIC Pathfinder Grant “FLATS” (grant agreement No. 101099139) and the Keele Foundation.
R.A.M. and E.M. and acknowledge support from the Robert A. Welch Foundation Grant C-2114 and the Department of Defense, Air Force Office of Scientific Research under Grant No. FA9550-21-1-0343.
D.M.K. acknowledges support by the DFG via the Priority Program SPP 2244 ``2DMP'' --- 443274199. 
L.X. and Q.X. acknowledge the support by the National Key Research and Development Program of China (Grant No. 2022YFA1403501), the National Natural Science Foundation of China (Grants No. 12474174 and No. 12504218), Hangzhou Tsientang Education Foundation and the Max Planck Partner group programme.  
A.R. acknowledges the support from the Max Planck-New York City Center for Non-Equilibrium Quantum Phenomena, the Cluster of Excellence 'CUI: Advanced Imaging of Matter'- EXC 2056 - project ID 390715994, SFB-925 "Light induced dynamics and control of correlated quantum systems" – project 170620586  of the Deutsche Forschungsgemeinschaft (DFG)  and Grupos Consolidados (IT1453-22). The Flatiron Institute is a division of the Simons Foundation.
M.M.U. acknowledges support from the European Union ERC Starting grant LINKSPM (Grant \#758558) and by the grant PID2023-153277NBI00 funded by MCIN/AEI/10.13039/501100011033. M.M.U. and B.A.B. acknowledge support by the Gipuzkoa Quantum 2025 grant (ref: 2025-QUAN-000009-01). 
B.A.B. was supported by the Gordon and Betty Moore Foundation through Grant No. GBMF8685 towards the Princeton theory program, the Gordon and Betty Moore Foundation’s EPiQS Initiative (Grant No. GBMF11070), the Global Collaborative Network Grant at Princeton University, the Simons Investigator Grant No. 404513, the NSF MRSEC (grant DMR-2011750), the Simons Collaboration on New Frontiers in Superconductivity (Grant No. SFI-MPS-NFS-00006741-01), Princeton Catalysis Initiative (PCI), the Schmidt Transformative Technology Fund at Princeton University and the National Science Foundation through the AI Research Institutes program Award No. DMR-2433348. 

\paragraph*{Author contributions:} B.A.B. and Y.J. conceived of the study. Y.J., U.P., H.P., D.C., and H.H developed the code and performed the high-throughput calculations. 
J.X., R.A.M., P.H., V.H., E.M., C.F., and L.M.S. synthesized the samples.
A.O., S.S., J.Z., D.Y., Z.G., J.S., K.F.M., D.K.E., G.A., Y.W., and M.M.U. exfoliated and characterized the monolayer and bilayer samples. 
N.R. built the Topological 2D materials Database with input data from U.P. and Y.J.. 
Q.X., H.P., M.C., D.M.K., A.R., and L.X. performed calculations and provided DFT results for moderate twist angles.
Y.J., G.S., H.P., and B.A.B. wrote the original draft and supplementary materials. 
All authors contributed to the review and editing of the final draft. 

\paragraph*{Competing interests:}
The authors declare that they have no competing interests.

\paragraph*{Data and materials availability:}
The crystal structures analyzed in this work are available in the C2DB (https://c2db.fysik.dtu.dk) and MC2D (https://mc2d.materialscloud.org). Programs DLCOMPREL, LBANDREP, Check Topological Layer Mat., RSI-layer, and TopologicalIndices can be accessed from https://cryst.ehu.es/. First-principles calculations were performed using version 5.4.4 of the Vienna Ab-initio Simulation Package (VASP), a commercial software package available at https://www.vasp.at/. 
All data are available in the supplementary materials, through our public website 2D-TQCDB~\cite{2DTQCDB_ref}, and Zenodo~\cite{zenodo_doi}. Details for the synthesis and characterization of all compounds/materials are in the methods section and \cite{seeSI}.

\renewcommand{\addcontentsline}[3]{}
\bibliographystyle{apsrev4-2}

\let\addcontentsline\oldaddcontentsline

\renewcommand{\thetable}{S\arabic{table}}
\renewcommand{\thefigure}{S\arabic{figure}}
\renewcommand{\theequation}{S\arabic{section}.\arabic{equation}}

\onecolumngrid
\pagebreak
\thispagestyle{empty}

\clearpage
\begin{center}
	\textbf{\large Supplementary Information for ''\titlePaper{}``}\\[.2cm]
\end{center}

\appendix
\renewcommand{\thesection}{\Roman{section}}
\tableofcontents
\let\oldaddcontentsline\addcontentsline

\newpage
% !TEX root = ./main.tex

\section{Dispersion of Twistable Semimetals}

In this section, we analyze the electronic dispersion characteristics of the twisting points of 2D twistable semimetals. These twistable semimetals host different types of dispersion. The nature of the dispersion, \ie, whether linear or quadratic, at these crossing points is determined by the symmetry of the associated valley in the Brillouin zone (BZ). 
In the presence of certain symmetries, linear terms in the  $\vec{k} \cdot \vec{p}$  Hamiltonian may be either allowed or forbidden, leading to distinct types of dispersion~\cite{bradlyn2016beyond, jiang2021k, tang2021exhaustive, zhan2021programmable, yu2022encyclopedia, zhang2023magnetickp, zhang2023vasp2kp}. 
A linear dispersion appears when symmetry permits first-order terms, creating Dirac-like cones in the band structure. Conversely, quadratic dispersion emerges when the symmetry constraints eliminate the linear terms, resulting in parabolic band crossings.
Note that crystalline symmetries and time-reversal symmetry (TRS) do not forbid all first- and second-order terms, meaning that cubic dispersion can only occur along specific directions, which we omit here for simplicity. Higher-dimensional ($>2$) crossings are generally not allowed in the 80 layer groups (LGs) without SOC, except in rare cases involving non-symmorphic LGs, such as the $S=(\frac{1}{2}, \frac{1}{2})$ point in \lgsymbnum{33}, which hosts a 4D irreducible representation (IRREP) $S_1S_1$. However, in practice, we do not find twistable semimetals with such high-dimensional crossings. In the presence of SOC, crossing points are typically gapped.
The high-symmetry points (HSPs) and IRREPs labeling follows the conventions of the \webBCS~\cite{aroyo2011crystallography, aroyo2006bilbao1, aroyo2006bilbao2}.

We proceed by discussing the dispersion of semimetals based on the symmetries of the valley.

\subsubsection{Hexagonal lattice}
For semimetals in hexagonal lattices, quadratic dispersion can only appear at the $\Gamma$ point. 
$\Gamma$ valley of the hexagonal lattice exhibits $C_3$ and TRS symmetry, which lead to the 2D irreducible representation (IRREP) with $C_3$ eigenvalues $e^{\pm i\frac{2\pi}{3}}$ (without SOC). In this case, the presence of TRS symmetry forbids the linear $\vec{k}\cdot\vec{p}$ terms and leads to the quadratic dispersion.

The $K$ valley, however, can only host linear crossings, \ie, the Dirac cone. The $K$ point possesses $C_3$ symmetry but lacks TRS. To host a Dirac cone at the $K$ point, the system must also exhibit inversion or $C_{2z}$ symmetry. In this case, a 2D IRREP with $C_3$ eigenvalues $e^{\pm i\frac{2\pi}{3}}$ (without SOC) is protected by $C_3$ and the space-time inversion symmetry $\mathcal{P} \cdot \mathcal{T}$ (or $C_{2z}\cdot \mathcal{T}$) with a linear dispersion. 

The $M$ valley of the hexagonal lattice can only host 1D IRREPs (without SOC) and is therefore omitted from consideration. 
Non-high-symmetry points (nHSPs), however, can host 2D degenerate points if the momentum possesses a twofold rotation $C_2$ or mirror symmetry. When two bands with opposite $C_2$ or mirror eigenvalues intersect, the crossing point is symmetry-protected, resulting in linear dispersion.

We note that in the presence of SOC, these degenerate points generally become gapped, leading to two quadratic band edges. However, since SOC is generally not large, a semimetal, even after the SOC-induced gap opens, will still require a model that includes both the valence and conduction bands around the formerly gapless point.

\subsubsection{Square lattice}

In the square lattice, quadratic dispersion at a crossing point can appear at the $\Gamma$ and $M$ points. These points possess $C_4$ symmetry and TRS, with TRS enforcing a two-dimensional irreducible representation (IRREP) characterized by $C_4$ eigenvalues of $\pm i$. In this case, TRS forbids linear terms in the $\vec{k} \cdot \vec{p}$ Hamiltonian, resulting in quadratic dispersion.

The $X$ valley in the square lattice lacks $C_4$ symmetry but can still host 2D IREREPs (without SOC) when non-symmorphic symmetries are present. For example, in \lgsymbnum{63}, there are two such IRREPs, $X_1$ and $X_2$. However, linear $\vec{k} \cdot \vec{p}$ terms are allowed in this case. Similarly, symmetry-protected crossings at non-HSPs exhibit linear dispersion.

\subsubsection{Rectangular and Oblique lattice}

For rectangular and oblique lattices, the situation is similar to that of the $X$ valley or non-HSPs in square lattices, which only have $C_2$ rotation, mirror symmetries, inversion, and TRS symmetries, but have no $C_{n}$ $(n\leq 3)$ symmetries. As a result, linear dispersion is generally allowed.

One semimetal with a rectangular lattice worth mentioning is 
\href{https://c2db.fysik.dtu.dk/material/2AuCrO4-1}{\ch{AuCrO4}} in \lgsymbnum{41}. In the absence of SOC, it exhibits a twofold (or fourfold, if spin is considered) degenerate nodal line (NL)~\cite{fang2016topological, yu2017topological, ahn2018band, song2018diagnosis} along the $X-S$ line near the Fermi level. Upon introducing SOC, the nodal line is expected to be gapped, but the fourfold degenerate points at $X$ and $S$ should persist. However, our calculations show that \ch{AuCrO4} develops spontaneous magnetic moments, breaking the fourfold degeneracy at $X$ and $S$. As \webtwoDTQC ~ does not include magnetic materials, \ch{AuCrO4} is thus not included in the current twistable material list. 
In contrast, a similar compound, \twodtqclinkmatname[2.3.285]{\ch{AgCrO4}}, does not develop magnetic order and retains the fourfold degeneracies at $X$ and $S$. However, the NL in \ch{AgCrO4} leaves a large density of states (DOS) at the Fermi level, disqualifying it as a twistable semimetal.

\section{Additional Details for the high-throughput algorithm and continuum models}\label{app:sec:details}

\paragraph{SOC classification}
We give more details on the high-throughput algorithm for searching theoretically twistable materials. 
For the SOC classification, a numerical threshold of \SI{50}{meV} energy splitting near the twisting point is used to identify materials with strong SOC. 
In practice, we evaluate the maximal SOC splitting within a specified momentum range relative to the twisting point—specifically, within the first moir\'e BZ at a chosen twist angle of \ThetaRef, an angle that covers the relevant momenta after twisting. 
This momentum range is particularly useful for time-reversal-invariant momenta (TRIM) such as the $\Gamma$ and M points in the hexagonal lattice, where SOC splitting is zero (for a single band with two spins) at the TRIMs themselves but can be significant in their vicinity. 
The same moir\'e BZ is also used as the momentum resolution to determine whether a valley is located at an HSP. If the distance from the valley to an HSP falls within this first moir\'e BZ, we define the valley as being at the corresponding HSP.

\paragraph{Hexagonal lattice $\Gamma$-valley continuum model}
We discuss a continuum model of the $\Gamma$ valley in hexagonal lattice systems. Under the symmetry constraints of $C_{3z}$, $C_{2x}/C_{2y}$ (assuming twisted AA and AB stacking have $C_{2x}$ and $C_{2y}$, respectively) and $\mathcal{T}$, the simplified general model has the form 
\begin{equation}
\begin{aligned}
\left[ h_{\vQ,\vQ'} \left( \vk \right) \right]_{s l; s' l'} &=  \delta_{\vQ, \vQ'} \delta_{s s'} \left[\delta_{l l'}  \frac{\left(\vk - \vQ \right)^2}{2 m_1}+ \delta_{-l l'}  \frac{\left(\vk - \vQ \right)^2}{2 m_2}\right] + \left[ T_{\vQ,\vQ'} \right]_{s l; s' l'}.
\end{aligned}    
\end{equation}
Because the top and bottom layer share the same hexagonal $\vQ$-lattice, the layer coupling is particularly strong in $\Gamma$ valley, making layer bonding and antibonding states the natural bases. 
To account for the potentially different effective masses of these states, we introduce two mass parameters, $m_1$ and $m_2$. Without SOC taken into consideration, the intralayer hopping terms have the form 
$\left[ T^{\tAA}_{\vQ,\vQ'} \right]_{ls;ls} = \left(\pm i V^{\tAA}_1 + V^{\tAA}_2 \right)\delta_{\vQ \pm \vg_{1}, \vQ'}$ 
and 
$\left[ T^{\tAB}_{\vQ,\vQ'} \right]_{ls;ls} = \left[\pm i (-)^lV^{\tAB}_1 + V^{\tAB}_2 \right] \delta_{\vQ \pm \vg_{1}, \vQ'}$,
while the interlayer hopping terms are
$\left[ T^{\tAA}_{\vQ,\vQ'} \right]_{ls;(-l)s} = w^{\tAA}_1 \delta_{\vQ, \vQ'} + \left(\pm iw^{ \tAB}_2+w^{ \tAB}_3 \right)\delta_{\vQ \pm \vg_{1}, \vQ'}$ 
and
$\left[ T^{\tAB}_{\vQ,\vQ'} \right]_{ls;(-l)s} = w^{\tAB}_1 \delta_{\vQ, \vQ'} + w^{ \tAB}_2 \delta_{\vQ \pm \vg_{1}, \vQ'} $. 
Here $(-)^{\text{top}}=1$, $(-)^{\text{bottom}}=-1$, and $\boldsymbol{g}_{1,i}=C_{3z}^{i}\vec{b}_{M1}$ represent the first harmonic vectors in $\Gamma$-valley system. 
As detailed in our forthcoming work, the QSH bands observed in \twodtqclinkmatname[6.1.71]{\ch{SnSe2}} and \ch{BiTeI} require strong SOC terms in the continuum model. They can be introduced through either second harmonic SOC term $\left[ T^{\tAB}_{\vQ,\vQ'} \right]_{ls;(-l)s} = \pm (-)^{l}(-)^s w^{ \tAB}_3 \delta_{\vQ \pm \vg_{2}, \vQ'}$ or on-site Rashba term $\left[ T^{\tAA,\tAB}_{\vQ,\vQ'} \right]_{ls;(-l)(-s)} =  (-)^{l} V^{ \tAA,\tAB}_3\left[ i(k_x-Q_x)-(k_y-Q_y)\right]\delta_{\vQ , \vQ'}$, depending on the ab initio band structures of the system. Here $\boldsymbol{g}_{2,i}=C_{3z}^{i} (\vec{b}_{M1}+\vec{b}_{M2})$ represents the second harmonic vectors, and we define $(-)^{\uparrow}=1$, $(-)^{\downarrow}=-1$ for spin states.

\section{Experimental twistable materials}

\subsection{Table of experimental twistable materials}

\begin{table}[htbp]
\centering
\caption{\textbf{Statistics of experimental twistable materials.} In the table, Ch denotes other chalcogenides than TMD, and Mixed P/C/H denotes mixed pnictides, chalcogenides, and halides.}
\label{table: material-class} 
\begin{tabular}{c|c|c|c|c|c|c}
\hline\hline
Class & Element & TMD & Ch & Halide & Mixed P/C/H & Other \\ \hline
Semimetal & \NumSemiElements & \NumSemiTMD & \NumSemiChalcogenide & \NumSemiHalide & \NumSemiMixed & \NumSemiOther \\ \hline
Insulator & \NumInsuElements & \NumInsuTMD & \NumInsuChalcogenide & \NumInsuHalide & \NumInsuMixed & \NumInsuOther \\ \hline\hline
\end{tabular}
\end{table}

In this section, we give more details on the experimental twistable materials, extending the discussion in the main text. 
\begin{itemize}
    \item \textbf{Elements:} Besides graphene, relatively few experimentally exfoliated elements to create moir\'e structures with. To the best of our knowledge, only phosphorus has been exfoliated to produce large crystalline flakes. Some studies report other elemental monolayers~\cite{shah2020experimental}, but those either are produced as nanoscale particles or are grown epitaxially on substrates~\cite{hussain2017ultrathin, yang20182d, lu2024recent}. Of promise is also elemental bismuth, which was recently obtained as high-quality 2D sheets by crystallization molded by van der Waals materials. Although monolayers have not been obtained so far, transport devices have been made with few-layer bismuth grown in this fashion \cite{chen_exceptional_2024}.
    
    \item \textbf{Halides:} Similar to chalcogenides, van der Waals halides are fairly common. Not all of them, however, are stable to air or moisture, complicating exfoliation in some cases. The most promising systems according to our methodology are: \ch{PbI2}, which has a 2H structure similar to many TMDs, and can be exfoliated mechanically \cite{wangyang_mechanical_2016}; \ch{BiI3}, which has been exfoliated to yield small, but few-layer or even monolayer, flakes \cite{wang_green_2020}; or, alternatively, can be grown on \ch{SiO2/Si} substrates to produce nanoplates with large lateral areas \cite{li_synthesis_2017}; and \ch{CdI2}, which can also be grown as nanoplates on substrates \cite{ai_growth_2017}, but has no reports of experimental exfoliation to the best of our knowledge. 
    Twisted magnetic \ch{CrI3} has also been studied extensively~\cite{huang2017layer, xu2022coexisting, akram2021moire, xie2022twist, wang2020stacking}, but it is not included in the current work as we do not consider magnetic orderings.
    
    \item \textbf{Other materials:} Most experimental twistable materials fit into the five classes listed above.
    \ch{GeH} is one of the candidates that are outside of this classification. Although bulk exfoliable crystals cannot be obtained directly, few- and monolayer samples can be obtained through topochemical reactions of \ch{CaGe2} \cite{bianco_stability_2013}. Its silicon analog has also been made, but is less pure, with terminal hydrogens partially replaced by hydroxyl groups \cite{ryan_silicene_2020}. Several oxides also show promise for twisting theoretically, although experimental work may be somewhat challenging: \ch{TiO2} and the MXene \ch{Ti2CO2} have been obtained as nanosheets, but only of small area \cite{sun_generalized_2014, melchior_high-voltage_2018} No large monolayers have been reported so far; neither of these materials have exfoliable bulk crystals, and thus, similar to \ch{GeH}, they rely on wet chemistry to produce flakes, complicating the synthesis. Lastly, one of the highest twist scores for experimentally obtained materials we found in 2D-\ch{GaN}. This phase of \ch{GaN} was recently obtained by encapsulation between sheets of graphene \cite{al_balushi_two-dimensional_2016}. Experimental twisting of this system does not appear possible, but it could become an interesting material in the future if synthetic strategies affording freestanding monolayers can be devised.
\end{itemize}

We list the most promising materials that have either been experimentally exfoliated to mono- or few-layer flakes, or have been grown as mono- or few-layer thin films epitaxially on substrates. They are shown in \cref{semimetal_exp_full} (semimetals) and \cref{app:table_insulator_exp_full} (insulators), sorted by the material class and twist score. 
The “Class” column designates each material as Element, TMD, Chalcogenide, Halide, Mixed P/C/H (mixed pnictides, chalcogenides, and halides), or Other, as introduced in the main text. A statistic of these experimental material classes is given in \cref{table: material-class}.  
The “Mat. Type” column reflects the experimental synthesis method, marked as “Exp.M.Exfo” for mechanical exfoliation, “Exp.W.Exfo” for wet exfoliation methods, “Exp.Substr” for materials grown on substrates, or “Other” for unique methods specific to the system or study. 
For insulators, the score is taken as the maximum of the VBM and CBM scores for simplicity.

    \begin{longtable}{c|c|c|c|c|c|c|c|c|c|c|c}
    \caption{\label{semimetal_exp_full} Experimental twistable semimetals.}\\
    \centering
    Formula & LG & ID & Gap & Database & Lattice & valley & SOC gap & Twist score & Topology & Mat. type & Class\\
    \endfirsthead
    \hline
    \ch{C} & 80 & \twodtqclink{5.1.3} & 0.00 & \ctwodblink{2C-1} & hexagonal & $K$ & 0.00 & 0.90 & AccidentalFermi & \href{https://doi.org/10.1126/science.1102896}{Exp.M.Exfo} & Elements \\\hline
\ch{Si} & 72 & \twodtqclink{5.1.2} & 0.00 & \ctwodblink{2Si-1} & hexagonal & $K$ & 0.00 & 0.80 & SEBR & \href{https://onlinelibrary.wiley.com/doi/full/10.1002/adma.201903013}{Exp.W.Exfo} & Elements \\\hline
\ch{Ge} & 72 & \twodtqclink{1.1.15} & 0.02 & \ctwodblink{2Ge-1} & hexagonal & $K$ & 0.02 & 0.68 & SEBR & \href{https://pubs.rsc.org/en/content/articlehtml/2020/tc/d0tc03892j}{Exp.W.Exfo} & Elements \\\hline
\ch{Sn} & 72 & \twodtqclink{1.1.17} & 0.07 & \ctwodblink{2Sn-1} & hexagonal & $K$-SOC & 0.07 & 0.66 & SEBR & \href{https://www.sciencedirect.com/science/article/pii/S1005030223000415}{Exp.W.Exfo} & Elements \\\hline
\ch{WS2} & 15 & \twodtqclink{1.3.24} & 0.04 & \ctwodblink{2WS2-1} & rectangular & nHSP-SOC & 0.15 & 0.29 & NLC & \href{https://doi.org/10.1063/1.5091997}{Exp.W.Exfo} & TMD \\\hline
\ch{MoS2} & 15 & \twodtqclink{1.1.1} & 0.05 & \ctwodblink{2MoS2-1} & rectangular & nHSP-SOC & 0.05 & 0.28 & NLC & \href{https://www.nature.com/articles/s41467-017-00640-2}{Exp.M.Exfo} & TMD \\\hline

    \hline
    \end{longtable}
    
    \begin{longtable}{c|c|c|c|c|c|c|c|c|c|c|c}
    \caption{\label{app:table_insulator_exp_full} Experimental twistable insulators.}\\
    \centering
    Formula & LG & ID & Gap & Database & Lattice & VBM valley & CBM valley & Twist score & Topology & Mat. type & Class\\
    \endfirsthead
    \hline
    \ch{Sb} & 72 & \twodtqclink{3.1.25} & 1.01 & \ctwodblink{2Sb-1} & hexagonal & $\Gamma$-SOC & nHSP & 0.75 & OAI & \href{https://pubs.rsc.org/en/content/articlelanding/2019/ta/c9ta06072c}{Exp.W.Exfo} & Elements \\\hline
\ch{As} & 72 & \twodtqclink{3.1.11} & 1.48 & \ctwodblink{2As-1} & hexagonal & $\Gamma$-SOC & nHSP & 0.75 & OAI & \href{https://onlinelibrary.wiley.com/doi/full/10.1002/adfm.201807004}{Exp.W.Exfo} & Elements \\\hline
\ch{P} & 72 & \twodtqclink{3.1.23} & 1.95 & \ctwodblink{2P-1} & hexagonal & / & nHSP & 0.68 & OAI & \href{https://onlinelibrary.wiley.com/doi/10.1002/smll.201804066}{Exp.Substr} & Elements \\\hline
\ch{P} & 42 & \twodtqclink{3.1.7} & 0.91 & \ctwodblink{4P-1} & rectangular & $\Gamma$ & $\Gamma$ & 0.62 & OAI & \href{https://doi.org/10.1038/nnano.2014.35}{Exp.M.Exfo} & Elements \\\hline
\ch{Bi} & 72 & \twodtqclink{1.1.10} & 0.53 & \mctwodlink{Bi} & hexagonal & $\Gamma$-SOC & $\Gamma$ & 0.60 & SEBR & \href{https://www.nature.com/articles/s41563-024-01894-0}{Exp.Other} & Elements \\\hline
\ch{Bi} & 72 & \twodtqclink{1.1.9} & 0.49 & \ctwodblink{2Bi-1} & hexagonal & / & $\Gamma$ & 0.59 & SEBR & \href{https://www.nature.com/articles/s41563-024-01894-0}{Exp.Other} & Elements \\\hline
\ch{WSe2} & 78 & \twodtqclink{3.1.46} & 1.26 & \ctwodblink{1WSe2-1} & hexagonal & $K$-SOC & $K$-SOC & 0.71 & OAI & \href{https://journals.aps.org/prl/abstract/10.1103/PhysRevLett.113.026803}{Exp.M.Exfo} & TMD \\\hline
\ch{WS2} & 78 & \twodtqclink{3.1.45} & 1.55 & \ctwodblink{1WS2-1} & hexagonal & $K$-SOC & $K$-SOC & 0.70 & OAI & \href{https://www.science.org/doi/10.1126/science.aao3503}{Exp.M.Exfo} & TMD \\\hline
\ch{WTe2} & 78 & \twodtqclink{3.1.47} & 0.75 & \ctwodblink{1WTe2-1} & hexagonal & $K$-SOC & $K$-SOC & 0.61 & OAI & \href{https://pubs.acs.org/doi/10.1021/acs.inorgchem.0c02049}{Exp.W.Exfo} & TMD \\\hline
\ch{MoS2} & 78 & \twodtqclink{3.1.39} & 1.60 & \ctwodblink{1MoS2-1} & hexagonal & $K$-SOC & $K$ & 0.68 & OAI & \href{https://onlinelibrary.wiley.com/doi/full/10.1002/advs.202301243}{Exp.M.Exfo} & TMD \\\hline
\ch{MoSe2} & 78 & \twodtqclink{3.1.41} & 1.34 & \ctwodblink{1MoSe2-1} & hexagonal & $K$-SOC & $K$ & 0.64 & OAI & \href{https://doi.org/10.1364/OE.21.004908}{Exp.M.Exfo} & TMD \\\hline
\ch{MoSSe} & 69 & \twodtqclink{3.1.10} & 1.48 & \ctwodblink{1MoSSe-1} & hexagonal & $K$-SOC & $K$ & 0.59 & OAI & \href{https://www.nature.com/articles/nnano.2017.100}{Exp.Substr} & TMD \\\hline
\ch{MoTe2} & 78 & \twodtqclink{3.1.43} & 0.96 & \ctwodblink{1MoTe2-1} & hexagonal & $K$-SOC & $K$ & 0.60 & OAI & \href{https://doi.org/10.1038/nature24043}{Exp.M.Exfo} & TMD \\\hline
\ch{SnS2} & 72 & \twodtqclink{6.1.75} & 1.58 & \ctwodblink{SnS2-42a44e4e7298} & hexagonal & / & $M$ & 0.70 & LCEBR & \href{https://pubs.acs.org/doi/10.1021/acsaelm.2c01010}{Exp.M.Exfo} & TMD \\\hline
\ch{SnSe2} & 72 & \twodtqclink{6.1.71} & 0.76 & \ctwodblink{1SnSe2-1} & hexagonal & $\Gamma$-SOC & $M$ & 0.63 & LCEBR & \href{https://www.nature.com/articles/s41467-020-16266-w}{Exp.M.Exfo} & TMD \\\hline
\ch{HfS2} & 72 & \twodtqclink{6.1.43} & 1.24 & \ctwodblink{1HfS2-1} & hexagonal & / & $M$ & 0.64 & LCEBR & \href{https://doi.org/10.1002/adma.201503864}{Exp.M.Exfo} & TMD \\\hline
\ch{PtSe2} & 72 & \twodtqclink{6.1.58} & 1.18 & \ctwodblink{1PtSe2-1} & hexagonal & $\Gamma$-SOC & nHSP & 0.53 & LCEBR & \href{https://www.nature.com/articles/s41467-020-16266-w}{Exp.M.Exfo} & TMD \\\hline
\ch{HfSe2} & 72 & \twodtqclink{6.1.44} & 0.45 & \ctwodblink{1HfSe2-1} & hexagonal & $\Gamma$-SOC & $M$ & 0.49 & LCEBR & \href{https://doi.org/10.1063/1.4917458}{Exp.M.Exfo} & TMD \\\hline
\ch{ZrS2} & 72 & \twodtqclink{6.1.61} & 1.16 & \ctwodblink{1ZrS2-1} & hexagonal & / & $M$ & 0.55 & LCEBR & \href{https://onlinelibrary.wiley.com/doi/full/10.1002/smll.202205763}{Exp.M.Exfo} & TMD \\\hline
\ch{ZrSe2} & 72 & \twodtqclink{6.1.73} & 0.34 & \ctwodblink{1ZrSe2-1} & hexagonal & $\Gamma$-SOC & $M$ & 0.39 & LCEBR & \href{https://doi.org/10.3390/app6090264}{Exp.M.Exfo} & TMD \\\hline
\ch{InSe} & 78 & \twodtqclink{3.1.37} & 1.40 & \ctwodblink{2InSe-1} & hexagonal & nHSP & $\Gamma$ & 0.64 & OAI & \href{https://onlinelibrary.wiley.com/doi/10.1002/adma.201302616}{Exp.M.Exfo} & Chalcogenide \\\hline
\ch{GaSe} & 78 & \twodtqclink{3.1.32} & 1.74 & \ctwodblink{2GaSe-1} & hexagonal & nHSP & $\Gamma$ & 0.51 & OAI & \href{https://onlinelibrary.wiley.com/doi/full/10.1002/admt.201600197}{Exp.M.Exfo} & Chalcogenide \\\hline
\ch{GaTe} & 78 & \twodtqclink{3.1.35} & 1.29 & \ctwodblink{2GaTe-1} & hexagonal & / & $M$-SOC & 0.58 & OAI & \href{https://doi.org/10.1039/C6CP01963C}{Exp.M.Exfo} & Chalcogenide \\\hline
\ch{GaS} & 78 & \twodtqclink{3.1.30} & 2.30 & \ctwodblink{2GaS-1} & hexagonal & nHSP & $\Gamma$ & 0.48 & OAI & \href{https://www.nature.com/articles/s41467-020-16266-w}{Exp.M.Exfo} & Chalcogenide \\\hline
\ch{Bi2Se2Te} & 72 & \twodtqclink{6.1.23} & 0.37 & \ctwodblink{1TeBi2Se2-1} & hexagonal & / & $\Gamma$ & 0.48 & LCEBR & \href{https://pubs.acs.org/doi/10.1021/nl3019802}{Exp.Substr} & Chalcogenide \\\hline
\ch{Bi2Se2Te} & 72 & \twodtqclink{6.1.24} & 0.33 & \mctwodlink{Bi2TeSe2} & hexagonal & $\Gamma$-SOC & $\Gamma$ & 0.46 & LCEBR & \href{https://pubs.acs.org/doi/10.1021/nl3019802}{Exp.Substr} & Chalcogenide \\\hline
\ch{Bi2Se3} & 72 & \twodtqclink{6.1.25} & 0.47 & \ctwodblink{1Bi2Se3-1} & hexagonal & / & $\Gamma$ & 0.51 & LCEBR & \href{https://www.nature.com/articles/s41467-020-16266-w}{Exp.M.Exfo} & Chalcogenide \\\hline
\ch{Bi2SeTe2} & 72 & \twodtqclink{6.1.27} & 0.31 & \ctwodblink{1SeBi2Te2-1} & hexagonal & / & $\Gamma$ & 0.46 & LCEBR & \href{https://pubs.acs.org/doi/10.1021/acsaelm.3c01195}{Exp.M.Exfo} & Chalcogenide \\\hline
\ch{Sb2Se2Te} & 72 & \twodtqclink{6.1.63} & 0.61 & \ctwodblink{1TeSb2Se2-1} & hexagonal & / & $\Gamma$ & 0.44 & LCEBR & \href{https://www.nature.com/articles/s41598-017-05369-y.pdf}{Exp.M.Exfo} & Chalcogenide \\\hline
\ch{SnS} & 32 & \twodtqclink{6.1.2} & 1.43 & \ctwodblink{2SSn-1} & rectangular & nHSP & nHSP-SOC & 0.40 & LCEBR & \href{https://www.nature.com/articles/s41598-023-46092-1}{Exp.M.Exfo} & Chalcogenide \\\hline
\ch{Sb2SeTe2} & 72 & \twodtqclink{6.1.67} & 0.44 & \ctwodblink{1SeSb2Te2-1} & hexagonal & nHSP & $\Gamma$ & 0.41 & LCEBR & \href{https://www.nature.com/articles/srep45413}{Exp.M.Exfo} & Chalcogenide \\\hline
\ch{GeSe} & 32 & \twodtqclink{6.1.1} & 1.12 & \ctwodblink{2GeSe-1} & rectangular & nHSP & / & 0.41 & LCEBR & \href{https://www.nature.com/articles/s41598-023-46092-1}{Exp.M.Exfo} & Chalcogenide \\\hline
\ch{SnSb2Se4} & 72 & \twodtqclink{6.1.66} & 0.39 & \ctwodblink{1SnSb2Se4-1} & hexagonal & / & $\Gamma$ & 0.38 & LCEBR & \href{https://onlinelibrary.wiley.com/doi/10.1002/adfm.202316849}{Exp.Substr} & Chalcogenide \\\hline
\ch{SnSe} & 32 & \twodtqclink{6.1.4} & 0.89 & \ctwodblink{2SeSn-1} & rectangular & nHSP & nHSP-SOC & 0.37 & LCEBR & \href{https://www.nature.com/articles/s41598-023-46092-1}{Exp.M.Exfo} & Chalcogenide \\\hline
\ch{Bi2PbSe4} & 72 & \twodtqclink{6.1.18} & 0.51 & \ctwodblink{1PbBi2Se4-1} & hexagonal & / & $\Gamma$ & 0.42 & LCEBR & \href{https://pubs.rsc.org/en/content/articlelanding/2014/cp/c4cp01885k}{Exp.Substr} & Chalcogenide \\\hline
\ch{Bi2Te3} & 72 & \twodtqclink{6.1.30} & 0.27 & \ctwodblink{1Bi2Te3-1} & hexagonal & / & $\Gamma$ & 0.41 & LCEBR & \href{https://www.nature.com/articles/s41467-020-16266-w}{Exp.M.Exfo} & Chalcogenide \\\hline
\ch{GaGeTe} & 72 & \twodtqclink{3.1.16} & 0.67 & \ctwodblink{2GaGeTe-1} & hexagonal & / & $\Gamma$ & 0.44 & OAI & \href{https://pubs.aip.org/aip/apl/article/111/20/203504/34427}{Exp.M.Exfo} & Chalcogenide \\\hline
\ch{ZrS3} & 46 & \twodtqclink{6.1.6} & 1.18 & \ctwodblink{2ZrS3-1} & rectangular & / & $\Gamma$ & 0.41 & LCEBR & \href{https://pubs.rsc.org/en/content/articlelanding/2014/qi/c4qi00127c}{Exp.W.Exfo} & Chalcogenide \\\hline
\ch{Sb2Te3} & 72 & \twodtqclink{6.1.70} & 0.41 & \mctwodlink{Sb2Te3} & hexagonal & / & $\Gamma$ & 0.36 & LCEBR & \href{https://pubs.aip.org/aip/jap/article/111/5/054305/345973/Micro-Raman-spectroscopy-of-mechanically}{Exp.M.Exfo} & Chalcogenide \\\hline
\ch{SnSb2Te4} & 72 & \twodtqclink{6.3.2582} & 0.42 & \ctwodblink{1SnSb2Te4-1} & hexagonal & / & $\Gamma$ & 0.31 & LCEBR & \href{https://onlinelibrary.wiley.com/doi/10.1002/adfm.202316849}{Exp.Substr} & Chalcogenide \\\hline
\ch{Bi2SnTe4} & 72 & \twodtqclink{6.1.29} & 0.20 & \ctwodblink{1SnBi2Te4-1} & hexagonal & / & $\Gamma$ & 0.25 & LCEBR & \href{https://link.springer.com/article/10.1007/s12274-017-1679-z}{Exp.Substr} & Chalcogenide \\\hline
\ch{Bi2PbTe4} & 72 & \twodtqclink{6.1.19} & 0.34 & \ctwodblink{1PbBi2Te4-1} & hexagonal & / & $\Gamma$ & 0.30 & LCEBR & \href{https://link.springer.com/article/10.1007/s10854-018-9114-0}{Exp.Substr} & Chalcogenide \\\hline
\ch{TiS3} & 46 & \twodtqclink{6.1.8} & 0.29 & \ctwodblink{2TiS3-1} & rectangular & / & $\Gamma$ & 0.18 & LCEBR & \href{https://doi.org/10.1002/adom.201400043}{Exp.M.Exfo} & Chalcogenide \\\hline
\ch{ZrSe3} & 46 & \twodtqclink{6.1.7} & 0.41 & \ctwodblink{2ZrSe3-1} & rectangular & $\Gamma$ & / & 0.18 & LCEBR & \href{https://onlinelibrary.wiley.com/doi/abs/10.1002/adma.202103571}{Exp.M.Exfo} & Chalcogenide \\\hline
\ch{PbI2} & 72 & \twodtqclink{6.1.54} & 1.87 & \ctwodblink{1PbI2-1} & hexagonal & / & $\Gamma$ & 0.65 & LCEBR & \href{https://doi.org/10.1016/j.matlet.2016.01.034}{Exp.M.Exfo} & Halide \\\hline
\ch{CdI2} & 72 & \twodtqclink{6.1.38} & 2.18 & \ctwodblink{1CdI2-2} & hexagonal & $\Gamma$-SOC & $M$ & 0.60 & LCEBR & \href{https://pubs.acs.org/doi/10.1021/acsnano.7b01507}{Exp.Substr} & Halide \\\hline
\ch{BiI3} & 71 & \twodtqclink{6.1.15} & 1.64 & \ctwodblink{2BiI3-1} & hexagonal & / & $\Gamma$ & 0.49 & LCEBR & \href{https://doi.org/10.1002/smll.201701034}{Exp.Substr} & Halide \\\hline
\ch{ZrClN} & 72 & \twodtqclink{6.3.2408} & 1.91 & \ctwodblink{2ClNZr-1} & hexagonal & $\Gamma$ & $K$ & 0.58 & LCEBR & \href{https://doi.org/10.1002/smll.202107490}{Exp.M.Exfo} & Mixed-P/C/H \\\hline
\ch{BiClTe} & 69 & \twodtqclink{6.3.1963} & 0.95 & \ctwodblink{1BiClTe-2} & hexagonal & $\Gamma$-SOC & $\Gamma$ & 0.63 & LCEBR & \href{https://analyticalsciencejournals.onlinelibrary.wiley.com/doi/full/10.1002/jrs.5253}{Exp.M.Exfo} & Mixed-P/C/H \\\hline
\ch{ZrBrN} & 72 & \twodtqclink{6.3.2323} & 1.62 & \ctwodblink{2BrNZr-1} & hexagonal & nHSP & $K$ & 0.65 & LCEBR & \href{https://doi.org/10.1016/j.chemphys.2024.112487}{Exp.W.Exfo} & Mixed-P/C/H \\\hline
\ch{BiBrTe} & 69 & \twodtqclink{6.1.10} & 0.92 & \ctwodblink{1BiBrTe-2} & hexagonal & $\Gamma$-SOC & $\Gamma$-SOC & 0.62 & LCEBR & \href{https://www.nature.com/articles/s41467-023-44439-w}{Exp.M.Exfo} & Mixed-P/C/H \\\hline
\ch{BiClTe} & 69 & \twodtqclink{6.1.11} & 0.94 & \ctwodblink{1BiClTe-1} & hexagonal & / & $\Gamma$ & 0.63 & LCEBR & \href{https://analyticalsciencejournals.onlinelibrary.wiley.com/doi/full/10.1002/jrs.5253}{Exp.M.Exfo} & Mixed-P/C/H \\\hline
\ch{AgInP2S6} & 67 & \twodtqclink{3.3.222} & 1.33 & \ctwodblink{1AgInP2S6-1} & hexagonal & / & $\Gamma$ & 0.59 & OAI & \href{https://www.nature.com/articles/s41467-021-25068-7}{Exp.W.Exfo} & Mixed-P/C/H \\\hline
\ch{BiITe} & 69 & \twodtqclink{6.1.13} & 0.70 & \ctwodblink{1BiITe-1} & hexagonal & nHSP-SOC & $\Gamma$-SOC & 0.55 & LCEBR & \href{https://doi.org/10.1088/2053-1583/aac652}{Exp.M.Exfo} & Mixed-P/C/H \\\hline
\ch{GaN} & 78 & \twodtqclink{6.1.78} & 1.82 & \ctwodblink{1GaN-1} & hexagonal & $K$ & $\Gamma$ & 0.70 & LCEBR & \href{https://www.nature.com/articles/nmat4742}{Exp.Other} & Other \\\hline
\ch{GeH} & 72 & \twodtqclink{3.1.18} & 0.90 & \ctwodblink{2GeH-1} & hexagonal & / & $\Gamma$ & 0.66 & OAI & \href{https://doi.org/10.1021/nn4009406}{Exp.M.Exfo} & Other \\\hline
\ch{C3N} & 80 & \twodtqclink{3.1.48} & 0.39 & \ctwodblink{2NC3-1} & hexagonal & $M$ & / & 0.59 & OAI & \href{https://doi.org/10.1002/adma.201605625}{Exp.Other} & Other \\\hline
\ch{HSi} & 72 & \twodtqclink{3.1.19} & 2.18 & \ctwodblink{2HSi-1} & hexagonal & / & $M$ & 0.57 & OAI & \href{https://pubs.acs.org/doi/10.1021/acs.chemmater.9b04180}{Exp.Other} & Other \\\hline
\ch{TiO2} & 72 & \twodtqclink{6.1.49} & 2.66 & \ctwodblink{1TiO2-1} & hexagonal & nHSP & / & 0.27 & LCEBR & \href{https://www.nature.com/articles/ncomms4813}{Exp.Other} & Other \\\hline
\ch{Ti2CO2} & 72 & \twodtqclink{6.1.34} & 0.31 & \ctwodblink{1CO2Ti2-1} & hexagonal & / & $M$ & 0.24 & LCEBR & \href{https://doi.org/10.1149/2.0401803jes}{Exp.Other} & Other \\\hline

    \hline
    \end{longtable}

\FloatBarrier

\subsection{Short-list of promising unexplored experimental twistable materials.}

Based on the criteria specified in the main text, here we provide several families of materials that are especially likely to be immediately available for fabrication of moir\'e devices.

\begin{enumerate}
    \item M-valley TMDs, such as \ch{SnSe2} show much promise, and are well known to be exfoliable as large and high-quality monolayers~\cite{huang_universal_2020}.
    \item Group IIIa monochalcogenides such as \ch{GaS} likewise have been obtained as high-quality monolayer crystals~\cite{huang_universal_2020}, and feature fairly high twist scores according to our methodology.
    \item Tetradymite-type chalcogenides such as \ch{Bi2Te3} have well-established procedures for crystal growth and exfoliation~\cite{huang_universal_2020}, and are some of the highest-scoring experimental twistable insulators.
    \item \ch{ZrNCl} has been exfoliated to monolayers~\cite{nong_layer-dependent_2022}, and can be gated to reach a superconducting state~\cite{saito_metallic_2015}. It remains to be seen what states can be achieved in twistronic devices featuring the material.
\end{enumerate}

\subsection{Experimental growth of twistable materials}\label{app:sec:sample-growth}

\begin{figure}[tbp]
    \centering
    \includegraphics[width=0.6\linewidth]{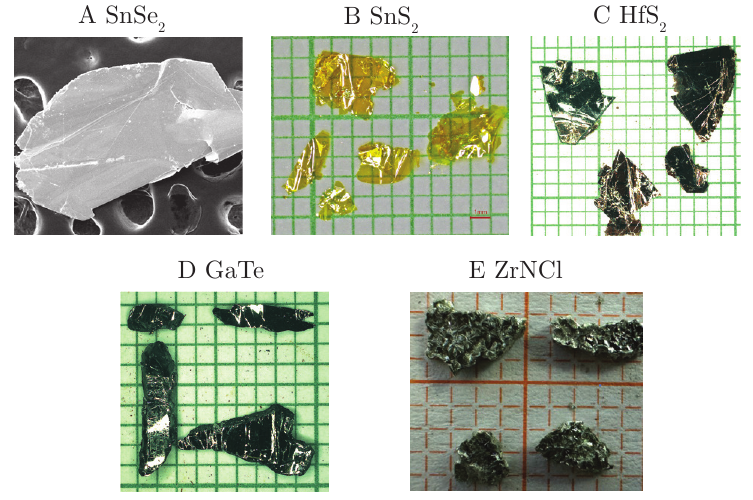}
    \caption{\textbf{Samples of twistable materials.} 
    (A) \twodtqclinkmatname[6.1.71]{\ch{SnSe2}}, (B) \twodtqclinkmatname[6.1.75]{\ch{SnS2}}, (C) \twodtqclinkmatname[6.2.1259]{\ch{HfS2}}
   (D) \href{https://materials.springer.com/isp/crystallographic/docs/sd_1940259}{GaTe}, and (E) \twodtqclinkmatname[6.3.2408]{ZrNCl}. 
    }
    \label{Fig: samples} 
\end{figure}

Among the experimentally twistable materials, we have successfully synthesized several of the most promising ones in bulk form, including TMDs \twodtqclinkmatname[6.1.71]{\ch{SnSe2}}, \ch{SnS2}, and \twodtqclinkmatname[6.1.43]{\ch{HfS2}}, chalcogenide GaTe~\cite{jacobsen2013high}, and pnicto-halide ZrNCl, with images of the samples shown in \cref{Fig: samples}. In the following, we give a more detailed description of the growth procedures for these materials. 

\subsubsection{\ch{SnSe2} and \ch{SnS2}}

Both \ch{SnS2} and \ch{SnSe2} crystals were synthesized using the iodine vapor transport technique, with commercially available starting materials: tin (Sn, Thermo Scientific Chemicals, 99.999\%), sulfur (S, Sigma-Aldrich, 99.999\%), selenium (Se, Sigma-Aldrich, 99.99\%), and iodine (I2, Sigma-Aldrich, 99.99\%). \ch{SnS2} was synthesized following literature conditions (5 mg/cm$^3$ \ch{I2}, 850–650 \SI{}{\degreeCelsius}, 12 days)~\cite{burton2013synthesis}.
For \ch{SnSe2}, Sn and Se (1:2 molar ratio) and 120 mg \ch{I2} were sealed in a 14 cm quartz ampule (ice bath used during sealing) and heated to 850 \SI{}{\degreeCelsius} over 12 h, held for 96 h, then cooled to room temperature over 12 h. Grey metallic-looking crystals formed in the hot and middle zones. Phase purity and composition were confirmed by powder X-ray diffraction (PXRD), inductively coupled plasma optical emission spectroscopy (ICP-OES) and scanning electron microscopy (SEM) with energy-dispersive X-ray spectroscopy (EDX).

The crystallinity and phase purity of bulk crystals were checked using pXRD patterns obtained from a STOE STADI P powder diffractometer with Mo $K\alpha 1$ radiation and a Dectris Mythen 2R 1K detector, in the $2\theta$ range from $1\degree$ to $45\degree$. SEM and EDX spectra were taken with a Verios 460 extreme high-resolution scanning electron microscope with an Oxford energy dispersive X-ray spectrometer. ICP-OES analysis was performed with Agilent 5800 ICP-OES spectrometer.

\subsubsection{\ch{HfS2}}

HfS$_2$ single crystals were synthesized using chemical vapor transport (CVT).  Stoichiometric amounts of Hafnium (Alfa Aesar, 99.6\%) and Sulfur (Alfa Aesar, 99.5\%) powders were sealed in a quartz tube under vacuum along with iodine as a transport agent. The tube was heated with a temperature gradient of 800-900 \SI{}{\degreeCelsius} and kept at that gradient for 10 days. The as-grown crystals were reddish and transparent layered plates with a typical dimension of $5\times 5\times 0.1$ mm$^3$. Room-temperature powder X-ray diffraction (XRD) measurements were carried out in a Bruker diffractometer with Cu K$\alpha$ radiation and revealed the hexagonal crystal structure with lattice parameters a = 3.632 {\AA} and c = 5.847 {\AA}.

\subsubsection{\ch{GaTe}}

GaTe single crystals were synthesized by a self-flux growth technique, using an excess amount of Ga. Elemental Ga (Sigma-Aldrich, 99.995\%) and Te (Thermo Scientific, 99.999\%) pieces in an atomic ratio of 0.57 : 0.43 were packed in an alumina crucible and sealed in a quartz ampoule under vacuum. The ampoule was heated up to 900 \SI{}{\degreeCelsius}, kept at that temperature for 4 h, then slowly cooled down to 760 \SI{}{\degreeCelsius} over 100 h, after which the excess flux was removed in a centrifuge. The as-grown crystals were layered plates with a typical dimension of $5\times 2\times 0.2$ mm$^3$. Room-temperature powder X-ray diffraction (XRD) measurements were carried out in a Bruker diffractometer with Cu K$\alpha$ radiation. Rietveld analysis was performed and showed the monoclinic crystal structure with lattice parameters a = 17.404 {\AA}, b =10.456 {\AA}, and c =4.077 {\AA}.

\subsubsection{\ch{ZrNCl}}

Synthesis of ZrNCl was achieved via a two-step synthesis employing an approach slightly modified from literature ~\cite{ohashi1988novel}.

Due to the sensitivity of the starting materials and reaction products to air and moisture, all work is carried out in a glove box or in a closed apparatus in the absence of air. To prepare the precursor, two alumina boats are prepared, one with a mixture of 1 g \ch{ZrH2} (Alfa-Aesar, 99\%) and 1 g \ch{NH4Cl} (Roth, 99\%); a second alumina boat is charged with 1.6 g \ch{NH4Cl}. The two boats are placed in a quartz reaction tube and installed in a horizontal furnace. After purging with argon for five minutes, the apparatus is heated within 90 minutes to 650 \SI{}{\degreeCelsius} in an ammonia stream (~ 60 ml/min, Air Liquide 5.0) and held at this temperature for 30 minutes. It is then cooled down to room temperature in the ammonia stream within 120 minutes. After evacuation, the reaction tube is inserted into the glove box.

At the end of the reaction, the alumina boat originally filled with \ch{NH4Cl} is empty; the sample in the second alumina boat has changed its appearance considerably: with an increase in volume, a voluminous, microcrystalline green product with a mass of 0.88 g has formed in the ammonia/ammonium chloride mist. In the cooler area of the reaction tube towards the gas outlet, there is a clearly separated smaller area of an unidentifiable bright yellow precipitate as well as large, unspecified amounts of white fine crystalline powder, which is identified as \ch{(NH4)2[ZrCl6]} and \ch{NH4Cl}.

In the second step, 710 mg of the above product together with 35 mg \ch{NH4Cl} and 60 mg \ch{PtCl2} in a capillary as transport medium are placed in a quartz tube, evacuated and sealed. The chemical transport is carried out in a two-zone furnace with a temperature gradient of 750 – 850 \SI{}{\degreeCelsius}; the starting material is placed in the temperature range of 750 \SI{}{\degreeCelsius}, the transport takes place towards 850 \SI{}{\degreeCelsius}. After a reaction time of 10 days, millimeter-sized green polycrystalline agglomerates of $\beta$-ZrNCl are obtained in the hot area, whereas the small amount of grey, non-transported residue is identified as a mixture of \ch{Zr7N4O8}~\cite{bredow2007anion} and \ch{ZrO2}~\cite{bohm1925verglimmen}.

\subsection{Experimental exfoliation and characterization of twistable materials}\label{app:sec:exp-exfoliation}

In this section, we demonstrate the exfoliation of several promising twistable materials into high-quality monolayers and perform characterization for homobilayer \ch{SnSe2}.

\begin{figure}[htbp]
	\centering
	\includegraphics[width=0.6\textwidth]{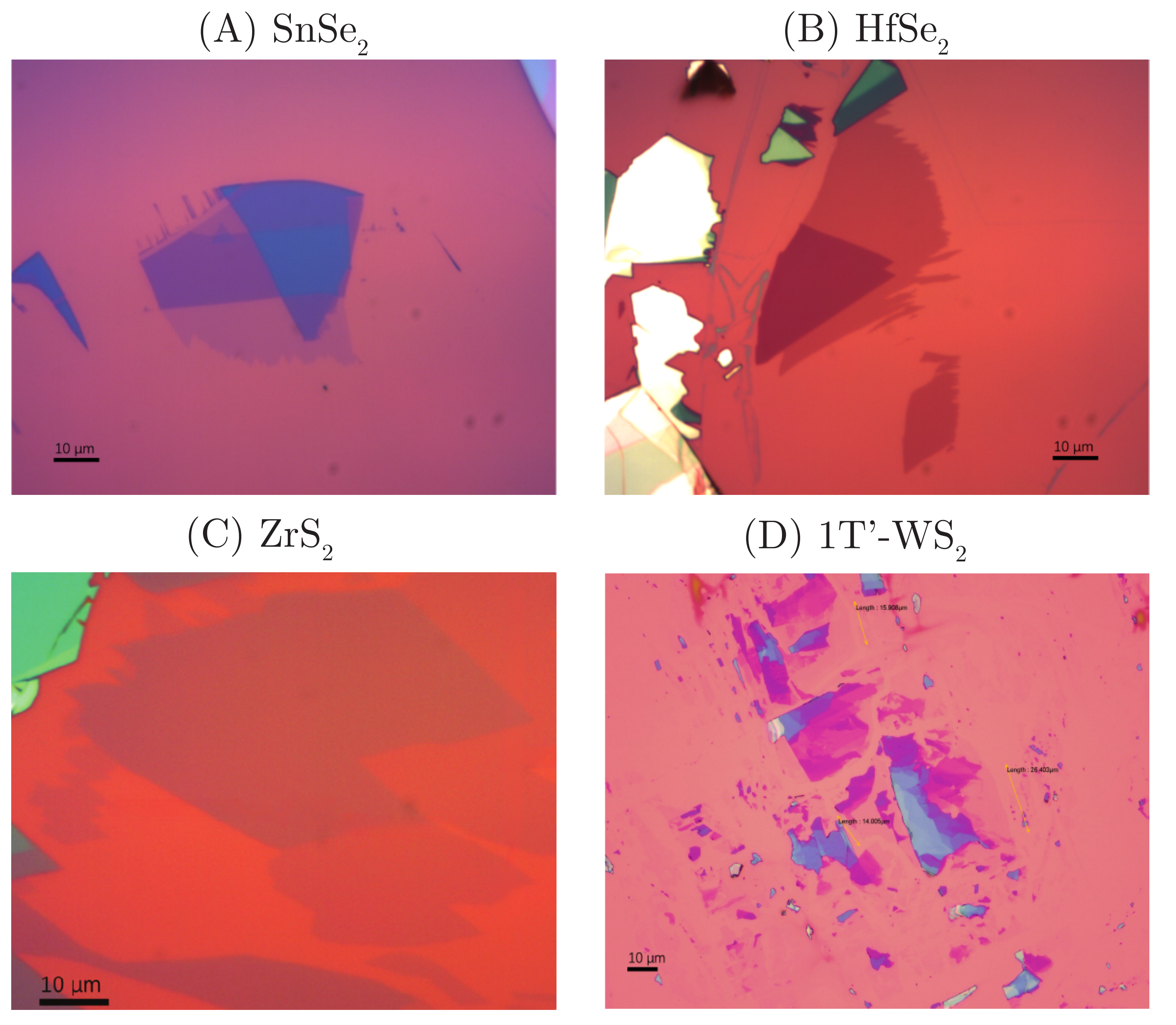}%
	\caption{Optical image of exfoliated monolayer materials, where (A) is \twodtqclinkmatname[6.1.71]{\ch{SnSe2}}, (B) is \twodtqclinkmatname[6.1.45]{\ch{HfSe2}}, (C) is \twodtqclinkmatname[6.2.1299]{\ch{ZrS2}}, and (D) is \twodtqclinkmatname[1.3.24]{1T'-\ch{WS2}}. The first three monolayers are exfoliated from \ch{SiO2/Si} substrates, while the last one is exfoliated from the gold substrates.
    }
    \label{fig:exp_exfoliated_monolayer}
\end{figure}

\cref{fig:exp_exfoliated_monolayer} shows optical images of the exfoliated monolayers, including \twodtqclinkmatname[6.1.71]{\ch{SnSe2}}, 
\twodtqclinkmatname[6.1.45]{\ch{HfSe2}}, 
\twodtqclinkmatname[6.2.1299]{\ch{ZrS2}}, and
\twodtqclinkmatname[1.3.24]{1T'-\ch{WS2}}. 
The four bulk materials were first repeatedly cleaved using Scotch tape. The tape with thinner crystals was then pressed onto a clean \ch{SiO2/Si} or gold substrate. To improve adhesion between the crystal flakes and the substrate, we heated the substrate to \SI{90}{\degreeCelsius} on a hotplate.  Finally, the tape was slowly peeled off, leaving behind atomically thin flakes, as shown in \cref{fig:exp_exfoliated_monolayer}.

\subsubsection{Characterization of homobilayer \ch{SnSe2}}

The exfoliated monolayer samples can be readily used to make twist bilayer devices and perform characterizations. As an example, we realized dual-gated devices of homobilayer \twodtqclinkmatname[6.1.71]{\ch{SnSe2}}, where two- and four-terminal resistances were measured as functions of gate voltage.  
% as shown in \cref{fig:exp_SnSe2_characterization}. 
The measured mobility of \ch{SnSe2} exceeds $\SI{1000}{\centi\meter^{2}\volt^{-1}\second^{-1}}$, which is comparable to well-studied transition metal dichalcogenides (TMDs) such as \ch{WSe2} (\SIrange{100}{250}{\centi\meter^{2}\volt^{-1}\second^{-1}})~\cite{ALL14}, \ch{MoTe2} (\SIrange{100}{200}{\centi\meter^{2}\volt^{-1}\second^{-1}})~\cite{SHA23}, or \ch{MoS2} (\SIrange{250}{1000}{\centi\meter^{2}\volt^{-1}\second^{-1}})~\cite{BAU13,RAD13}. 

This successful device implementation confirms the experimental viability of our identified materials for twistronics applications. We leave detailed studies of the twisted devices to future work. 

% \begin{figure}[htbp]
% 	\centering
% 	\includegraphics[width=0.8\textwidth]{fig/Fig-exp-SnSe2-characterization.pdf}%
% 	\caption{Experimental characterization of a \twodtqclinkmatname[6.1.71]{\ch{SnSe2}} homobilayer heterostructure. (A) Gate voltage dependence of two-terminal (blue) and four-terminal (red) resistance of a bilayer \ch{SnSe2} device encapsulated by hBN and gated by a few-layer graphite gate (at temperature $\SI{4}{\kelvin}$ and zero magnetic field). (B) Conductivity (green) and field-effect mobility (purple) of the device as a function of the gate voltage. A mobility above $\SI{1000}{\centi\meter^{2}\volt^{-1}\second^{-1}}$ is observed.}
% 	\label{fig:exp_SnSe2_characterization}
% \end{figure}

\subsubsection{Exfoliation of monolayers on gold substrates}\label{app:sec:exfoliation_gold}

Conventional exfoliation of two-dimensional (2D) materials typically relies on mechanical cleavage using adhesive tape (e.g., Scotch tape) to transfer flakes onto \ch{SiO2} substrates. However, for certain materials, this method often results in very low monolayer yield, small area, and polymer residues. To overcome these limitations and significantly enhance both the yield and lateral dimensions of exfoliated monolayers, we employ an ultra-clean gold-assisted exfoliation technique.

The gold-assisted approach enables large-area and clean monolayer exfoliation through two key mechanisms: (i) quasi-covalent bonding between gold and the bottommost layer of the 2D crystal~\cite{huang2020universal}, and (ii) lattice-mismatch-induced strain at the gold-2D material interface, which weakens interlayer coupling and promotes monolayer separation~\cite{sun2018theory}. With optimal parameters, this technique enables the deterministic exfoliation of a monolayer in each attempt, thereby ensuring reproducibility and near-unity yield in monolayer production. The gold substrate thus serves as both a mechanical and electronic decoupling layer, dramatically increasing the yield and size of monolayers.

\begin{figure}[htbp]
	\centering
	\includegraphics[width=0.8\textwidth]{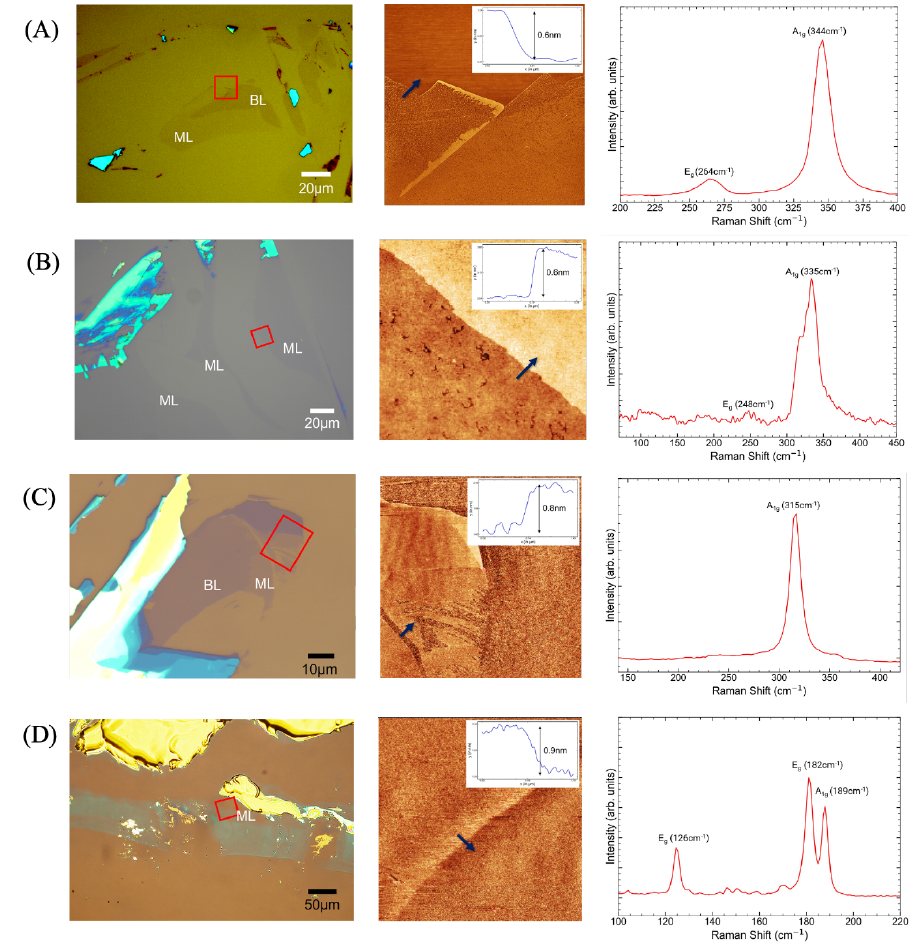}%
	\caption{Exfoliation and characterization of (A) \twodtqclinkmatname[6.1.43]{\ch{HfS2}}, (B) \twodtqclinkmatname[6.2.1299]{\ch{ZrS2}}, (C) \twodtqclinkmatname[6.1.75]{\ch{SnS2}}, and (D) \twodtqclinkmatname[6.3.2408]{ZrNCl} flakes. For each material, the first column presents the optical microscope image of the exfoliated flake, with red-marked regions indicating the areas selected for AFM analysis. The second column displays the corresponding AFM topography scans; the inset shows the height profile measured along the blue-marked line, highlighting the step height between layers. The arrow denotes the direction of the scan. The third column presents the Raman spectra of each material, with the prominent Eg and A1g peaks labeled. Together, these panels confirm the layer number, thickness, and quality of each 2D material.
    }
    \label{fig:exp_exfoliation_on_gold}
\end{figure}

As demonstrated in \cref{fig:exp_exfoliation_on_gold}, we have successfully applied gold-assisted exfoliation on Au(5 nm)/Cr(2 nm)/\ch{SiO2}(285 nm)/Si substrate to several twistable 2D materials, including \twodtqclinkmatname[6.1.75]{\ch{SnS2}}, \twodtqclinkmatname[6.1.43]{\ch{HfS2}}, \twodtqclinkmatname[6.2.1299]{\ch{ZrS2}}, and \twodtqclinkmatname[6.3.2408]{ZrNCl}. The resulting monolayers exhibit dimensions larger than those reported using conventional methods, with \ch{ZrS2} flakes reaching up to 150 $\mu$m $\times$ 30 $\mu$m. \cref{table:experimental_flake_dimensions} indicates a comparison of the sizes of the monolayers in this work versus the reported literature. The high quality of the exfoliated few-layered materials is confirmed by Raman spectroscopy, while the number of layers is independently verified by contact-mode atomic force microscopy (AFM).

\begin{table}[h!]
\centering
\begin{tabular}{l|c|c|c}
\hline
% \rowcolor[gray]{0.85}
\textbf{Crystal} & \textbf{Largest ML Flake Size} & \textbf{Reported in Literature} & \textbf{References} \\
\hline \hline
\twodtqclinkmatname[6.2.1299]{\ch{ZrS2}} & 150$\times$100 $\mu$m & 3$\times$3 $\mu$m & \cite{manas2016raman} \\\hline
\twodtqclinkmatname[6.1.43]{\ch{HfS2}} & 50$\times$20 $\mu$m & 15$\times$2 $\mu$m & \cite{kanazawa2016few,yue2019raman} \\\hline
%\twodtqclinkmatname[6.1.71]{2H-\ch{SnSe2}} & 35$\times$35 $\mu$m & 5$\times$5 $\mu$m & \cite{zhou2017lattice} \\
\twodtqclinkmatname[6.1.75]{\ch{SnS2}} & 20$\times$10 $\mu$m & 5$\times$1 $\mu$m & \cite{sriv2018low} \\
\hline
\twodtqclinkmatname[6.3.2408]{ZrNCl} & 60$\times$10 $\mu$m & 10$\times$5 $\mu$m & \cite{nong_layer-dependent_2022}\\
\hline 
\end{tabular}
\caption{Comparison of the largest flake dimensions of various crystals obtained in this work with the maximum sizes previously reported in the literature. The literature flake sizes and corresponding references are indicated for each material.}
\label{table:experimental_flake_dimensions}
\end{table}

\begin{figure}[htbp]
	\centering
	\includegraphics[width=0.6\textwidth]{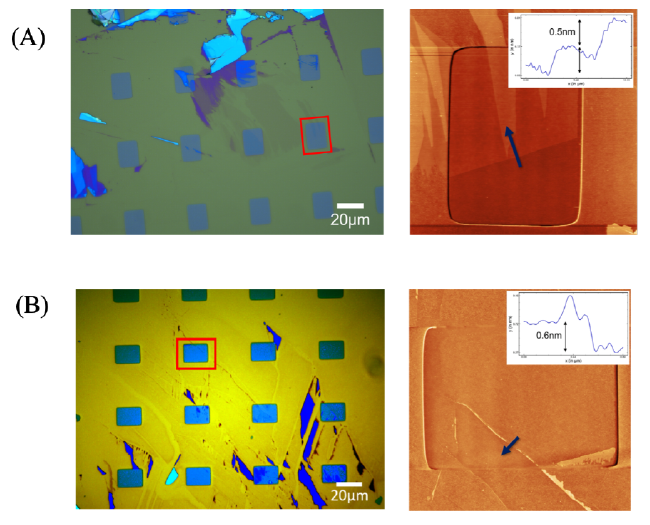}%
	\caption{Exfoliation of (A) \twodtqclinkmatname[6.1.43]{\ch{HfS2}} and (B) \twodtqclinkmatname[6.2.1299]{\ch{ZrS2}} flakes on gold template. Boxes of SiO2 were patterned to allow efficient transfer of a monolayer of the materials. The first column shows the optical image of the gold template. The second column shows the AFM topography image of the flake on the SiO2 boxes, with the inset showing the height difference across the blue-marked regions. 
    }
    \label{fig:exp_exfoliation_on_gold2}
\end{figure}

Furthermore, since the gold-monolayer adhesion is strong, a hybrid substrate solution known as gold template exfoliation is implemented~\cite{wu2024gold}. The gold template substrate consists of an Au-coated area and isolated Au-free \ch{SiO2} surfaces. In the latter, the substrate-flake adhesion is weaker enough to be transferred. Thus, this technique can enable clean transfer of these monolayers onto different substrates, facilitating device fabrication. Results of \ch{HfS2} and \ch{ZrS2} exfoliation using the gold template method are shown in~\cref{fig:exp_exfoliation_on_gold2}. Collectively, these advances establish gold-assisted exfoliation as a robust and scalable platform for producing monolayers on demand and subsequently fabricating twisted heterostructures.

\subsection{Experimental twisted \ch{HfS2} sample}

The \ch{HfS2} heterostructure was assembled using a dry pick-up and drop technique with a convex LCC polymer stamp (nail varnish, \textit{Lakme Color Crush}) deposited on a polydimethylsiloxane (PDMS) support. First, the graphite gate electrode and the thick bottom hBN dielectric were picked up at $T=55\,^\circ\mathrm{C}$. Subsequently, the first \ch{HfS2} monolayer was picked up at $T=70\,^\circ\mathrm{C}$, followed by the second \ch{HfS2} monolayer with a relative twist angle of $5^\circ$ with respect to the first layer. Finally, the top monolayer hBN, used for encapsulation, was picked up at $T=70\,^\circ\mathrm{C}$, followed by the graphite bias electrode, which makes direct contact with the \ch{HfS2} flakes.

\begin{figure}
    \centering
    \includegraphics[width=0.7\linewidth]{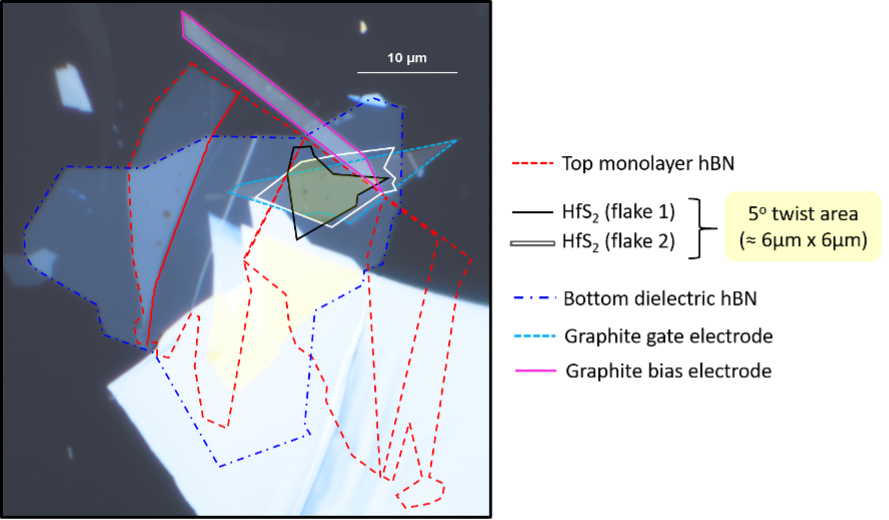}
    \caption{Representative optical micrograph (100$\times$) of a twisted homobilayer \ch{HfS2} (\textit{t}\ch{HfS2}) device for millikelvin STM/STS measurements. The graphite gate and bias electrodes, the bottom hBN dielectric, and the monolayer \ch{HfS2} flakes were mechanically exfoliated onto a \ch{Si/SiO2} substrate with a $300~\mathrm{nm}$ oxide layer, while the top monolayer hBN was exfoliated onto \ch{Si/SiO2} with a $90~\mathrm{nm}$ oxide layer to improve optical contrast. Monolayer hBN and \ch{HfS2} flakes were initially identified by optical microscopy using established contrast criteria and were subsequently confirmed by atomic force microscopy (AFM) and micro-Raman spectroscopy. To prevent oxidation, the \ch{HfS2} flakes were exfoliated in an Ar-filled glovebox with $\mathrm{O_2}<0.1$ ppm and $\mathrm{H_2O}<0.1$ ppm.}
    \label{app:fig:HfS2-twist-sample}
\end{figure}

In the optical micrograph shown in \cref{app:fig:HfS2-twist-sample}, the red, blue, and cyan dashed lines indicate the top monolayer hBN, the bottom hBN dielectric, and the graphite gate electrode, respectively. The purple solid line marks the graphite bias electrode, while the black and white solid lines outline the two individual \ch{HfS2} monolayers twisted by $5^\circ$. The yellow shaded region denotes the twisted \ch{HfS2} (\textit{t}\ch{HfS2}) area of the device, with dimensions $6~\mu\mathrm{m}\times 6~\mu\mathrm{m}$, where the scanning tunneling microscopy (STM) and scanning tunneling spectroscopy (STS) measurements will be performed.

\section{First-principles methods}\label{app:sec:DFT-methods}

The electronic properties calculations for the monolayer and twisted bilayer system were performed using density functional theory with the Vienna ab initio software package (VASP) \cite{kresse_1996}. The Perdew-Burke-Ernzerhof (PBE) exchange-correlation functionals \cite{blochl_1994} were employed. 
We use a 1.3 times larger cutoff value compared to the default value for the plane wave basis set. 
For the monolayer calculations, we use a  $\Gamma$-centered k-mesh such that $n_{k_i}=\frac{75}{a_i}$, where $n_{k_i}$ is the $\mathbf{k}$ point amount in $i$ direction and $\mathbf{a}_i$ is the unit cell length in $i$ direction in \r{A}.
For the twisted bilayer calculations, we use a $\Gamma$-centered k-mesh of 1$\times$1$\times$1  for the geometry optimization and electronic structure calculations, with the Tkatchenko-Scheffler (TS) van der Waals (vdW) corrections \cite{tkatchenko_2009,anatole_2010}, which have been shown to yield results consistent with experimental observations in our previous work \cite{wang2020correlated}. 
For other materials, we choose the vdW functional that best matches the experimental bulk lattice constant: DFT-ulg~\cite{ulg-kim2012} for \ch{GaTe}; Grimme’s D3 with Becke–Johnson damping~\cite{d3bj2011} for \ch{BiTeI}; and Grimme’s D2~\cite{d2grimme2006} for \ch{GeS2}. For \ch{Ta2Te2S}, we choose PBE + rVV10L~\cite{pbervv10l2017}.
The calculation is performed for periodic boundary conditions in all three spatial dimensions, including a vacuum thickness larger than 15$\AA$ for the out-of-plane direction of the two-dimensional material. This is large enough to suppress artificial interactions between the periodic slab images and thus reflects the two-dimensional limit. All atoms were fully relaxed, ensuring a residual force less than 0.02 eV$\AA^{-1}$ per atom. While the internal atomic positions were fully optimized, the lattice constant for the moir\'e supercell was kept fixed at a value corresponding to the optimized lattice constant for a 1$\times$1 unit cell in the monolayer to leverage the computational cost of the superlattice calculations. Spin-orbit coupling is included in every band-structure calculation. 
The twisted bilayer structures are generated by twisting the top (bottom) layers by $+\frac{\theta}{2}$ ($-\frac{\theta}{2}$), with the twist axis centered at the origin of the monolayer unit cell.

\section{Introduction to the tables in the catalogue}\label{app:sec:table_intro}

\begin{table}[tbp]
\centering
\caption{\label{table:statistics} \textbf{Statistics of twistable materials.} The materials are classified by Bravais lattice, valley, and SOC strength. 
The number in parentheses indicates the number of corresponding twistable semimetals (omitted if zero), 
while the number without parentheses represents the number of twistable insulators. 
Valley types are based on the momentum at the twisting point. 
When the SOC splitting near the twisting point is significant, the valley is labeled as ``valley-SOC.'' 
Materials are further classified into experimental (Exp.), computationally exfoliable (Exfo.), and computationally stable (Stab.), 
and computationally unstable (Unstab.).
For insulators, we evaluate the valence-band maximum (VBM) and conduction-band minimum (CBM) separately. Consequently, a single material may appear twice in the table, and the total number of entries can exceed the number of unique twistable materials.}
\scriptsize
\begin{tabular}{c|c|c|c|c|c}
\hline\hline
Lattice & Valley & Exp. & Exfo. & Stab. & Unstab.\\ \hline
\multirow{8}{*}{\makecell{Hexagonal \\ 1072 (48)}}
 & $\Gamma$ & 28 & 28 & 298 & 60 (2) \\ \cline{2-6}
 & $\Gamma$-SOC & 13 & 10 (1) & 124 (10) & 20 (5) \\ \cline{2-6}
 & $K$ & 7 (3) & 8 & 75 & 25 (9) \\ \cline{2-6}
 & $K$-SOC & 7 (1) & 2 (1) & 43 (4) & 12 (4) \\ \cline{2-6}
 & $M$ & 10 & 10 & 54 & 20 \\ \cline{2-6}
 & $M$-SOC & 1 & 0 & 3 & 2 \\ \cline{2-6}
 & nHSP & 10 & 8 & 79 (3) & 18 \\ \cline{2-6}
 & nHSP-SOC & 1 & 3 & 65 (3) & 28 (2) \\ \hline
\multirow{8}{*}{\makecell{Square \\ 167 (3)}}
 & $\Gamma$ & 0 & 23 & 45 & 13 \\ \cline{2-6}
 & $\Gamma$-SOC & 0 & 3 & 14 & 0 (1) \\ \cline{2-6}
 & $M$ & 0 & 2 & 20 & 3 \\ \cline{2-6}
 & $M$-SOC & 0 & 0 & 3 & 1 \\ \cline{2-6}
 & $X$ & 0 & 2 & 8 & 5 \\ \cline{2-6}
 & $X$-SOC & 0 & 0 & 0 & 1 \\ \cline{2-6}
 & nHSP & 0 & 1 & 3 & 9 \\ \cline{2-6}
 & nHSP-SOC & 0 & 0 & 7 & 4 (2) \\ \hline
\multirow{4}{*}{\makecell{Rectangular \\ 710 (10)}}
 & HSP & 4 & 96 & 290 & 85 (1) \\ \cline{2-6}
 & HSP-SOC & 0 & 0 & 17 & 17 \\ \cline{2-6}
 & nHSP & 3 & 13 (1) & 98 (1) & 25 \\ \cline{2-6}
 & nHSP-SOC & 2 (2) & 4 (2) & 38 (2) & 18 (1) \\ \hline
\multirow{4}{*}{\makecell{Oblique \\ 38}}
 & HSP & 0 & 6 & 10 & 3 \\ \cline{2-6}
 & HSP-SOC & 0 & 0 & 3 & 1 \\ \cline{2-6}
 & nHSP & 0 & 4 & 3 & 0 \\ \cline{2-6}
 & nHSP-SOC & 0 & 1 & 3 & 4 \\ \hline

\hline
\end{tabular}
\end{table}

In the following two sections, we tabulate the twistable semimetals (\cref{app:sec:table_twistable_semimetal}) and insulators (\cref{app:sec:table_twistable_insulator}) found in the high-throughput search, as summarized in \cref{table:statistics}. The tables list all materials with nonzero twist scores and contain the relevant properties of each type of material. Materials that fail to meet our criteria—including the gap condition, clear-energy-range condition, and density of states (DOS) condition (see Section III in the main text) —are deemed as not "clean" and theoretically unsuitable for twisting, and are assigned a zero twist score. As such, they are excluded from the tables. 

In each section, the twistable materials are classified based on their Bravais lattice, the valley of the twisting point, and the SOC splitting strength. Within each type, we further separate materials into four groups, \ie,  experimental, computationally exfoliable, computationally stable, and computationally unstable, based on the following: 
\begin{itemize}
\item \emph{Experimentally existing}: materials that are already manually verified in the literature to have been experimentally fabricated, which could be reported as exfoliated (labeled in the table as ``Exp.M.Exfo'' for mechanical exfoliation, and ``Exp.W.Exfo'' for the various wet exfoliation methods), grown on a substrate (``Exp.Substr''), or other methods, unique to the system or study (``Other''). 

\item \emph{Computationally exfoliable (MC2D)}: materials from the \mctwodbare~\cite{mc2d, campi2023expansion}. In this database, all materials were computationally exfoliated from an experimentally existing material in 3D and then relaxed.

\item \emph{Stable (C2DB)}: materials that are marked as highly thermodynamically stable (the energy above the convex hull $<0.2$ eV/atom) or dynamically stable (all phonon frequencies are real) in the \ctwodbbare~\cite{Haastrup_2018,Gjerding_2021}. 

\item \emph{Not-stable (C2DB)}: materials that do not satisfy the conditions defined above from \ctwodbbare. 
\end{itemize}

In each table, we list the following properties of each material:
\begin{itemize}
\item Formula: the chemical formula of the material.
\item LG: the layer group.
\item ID: the serial ID used in \webtwoDTQC, accompanied by the corresponding web link.

\item Gap: the global (indirect) gap, given in eV.
\item Database: the link of the material to the database where we obtain the crystal structure, including \ctwodbbare ~ and \mctwodbare. 

\item Lattice: the Bravais lattice of the material, \ie, hexagonal, square, rectangular (including centered rectangular), and oblique. 

\item Valley: the momentum of the twisting point. For insulators, we give the VBM and CBM valley separately. When the SOC splitting near the valley is stronger than \SI{50}{meV}, we add a ``-SOC'' suffix to the valley. 

\item SOC gap: the SOC splitting gap near the valley. 
In practice, we compute the SOC splitting within a specified momentum range relative to the twisting point — specifically, within the first moir\'e BZ at a chosen twist angle of \ThetaRef, an angle that covers the relevant momenta after twisting. This moir\'e BZ is also used as the momentum resolution to determine whether a valley is located at an HSP. If the distance from the valley to an HSP falls within this first moir\'e BZ, we define the valley as being at the corresponding HSP.

\item Twist Score: the twisting score of the material, defined in \cref{Eq: score_twisting}. For insulators, since each table is for one specific valley type, the score of the corresponding valley is used. In case the VBM and CBM have the same valley, the twist score is taken as the larger one.

\item Dispersion: The type of dispersion at the twisting point for twistable semimetals (without SOC), which can be either linear or quadratic.

\item Topology (SOC): the topological classification of the material with SOC, including LCEBR (linear combination of elementary band representations (EBR)), SEBR (split EBR), AccidentalFermi (accidental crossing point at Fermi level), ES (enforced semimetal), ESFD (enforced semimetal with Fermi degeneracy), OAI (obstructed atomic insulators), and OOAI (orbital-selective OAI)~\cite{bradlyn2017topological}. Among these classifications, the SEBR and NLC are topological insulators, the ES and ESFD are topological semimetals, the LCEBR are trivial insulators, the OAI are trivial insulators but have some electronic Wannier centers at empty sites, and the OOAI has Wannier centers at some atomic sites but can not form the outer-shell orbitals of those atoms. 

\item Bulk: If the material has an existing bulk structure, mark it as ``Yes''; If it does not have or is unknown, mark it as ``/''. Their bulk structures can be found in either \ctwodbbare ~or \mctwodbare. 

\item Mat. type: the type of material, including experimental (Exp), computationally exfoliable (Comp.Exfo), computationally stable (Stable), and computationally unstable (Unstable). 

\end{itemize}

Note that the 2D materials in \webtwoDTQC ~include structures from \ctwodbbare ~and \mctwodbare, which contain some duplicate materials. We remove these duplicates based on the following criteria: (1) same chemical formula, (2) same layer group, (3) same topological classification with and without SOC, (4) same valley type at both the VBM and CBM, and (5) differences in the indirect gap and twist score within 0.2 (in the unit of eV for the gap).

\clearpage

\section{Catalogue of twistable semimetals}\label{app:sec:table_twistable_semimetal}

\subsection{Summary of results}

\begin{table}[hbp]
\centering
\scriptsize
\begin{tabular}{c|c|c|c|c|c|c|c}
\hline\hline
Lattice & Valley & \# of materials & Experimental & Comp. exfoliated & Comp. stable & Comp. unstable & Total \# \\ \hline
\multirow{8}{*}{Hexagonal} & $\Gamma$ & 2 & 0 & 0 & 0 & 2 & \multirow{8}{*}{48} \\ \cline{2-7}
 & $\Gamma$-SOC & 16 & 0 & 1 & 10 & 5 &  \\ \cline{2-7}
 & $K$ & 12 & 3 & 0 & 0 & 9 &  \\ \cline{2-7}
 & $K$-SOC & 10 & 1 & 1 & 4 & 4 &  \\ \cline{2-7}
 & $M$ & 0 & 0 & 0 & 0 & 0 &  \\ \cline{2-7}
 & $M$-SOC & 0 & 0 & 0 & 0 & 0 &  \\ \cline{2-7}
 & nHSP & 3 & 0 & 0 & 3 & 0 &  \\ \cline{2-7}
 & nHSP-SOC & 5 & 0 & 0 & 3 & 2 &  \\ \hline\multirow{8}{*}{Square} & $\Gamma$ & 0 & 0 & 0 & 0 & 0 & \multirow{8}{*}{3} \\ \cline{2-7}
 & $\Gamma$-SOC & 1 & 0 & 0 & 0 & 1 &  \\ \cline{2-7}
 & $M$ & 0 & 0 & 0 & 0 & 0 &  \\ \cline{2-7}
 & $M$-SOC & 0 & 0 & 0 & 0 & 0 &  \\ \cline{2-7}
 & $X$ & 0 & 0 & 0 & 0 & 0 &  \\ \cline{2-7}
 & $X$-SOC & 0 & 0 & 0 & 0 & 0 &  \\ \cline{2-7}
 & nHSP & 0 & 0 & 0 & 0 & 0 &  \\ \cline{2-7}
 & nHSP-SOC & 2 & 0 & 0 & 0 & 2 &  \\ \hline\multirow{4}{*}{Rectangular} & HSP & 1 & 0 & 0 & 0 & 1 & \multirow{4}{*}{10} \\ \cline{2-7}
 & HSP-SOC & 0 & 0 & 0 & 0 & 0 &  \\ \cline{2-7}
 & nHSP & 2 & 0 & 1 & 1 & 0 &  \\ \cline{2-7}
 & nHSP-SOC & 7 & 2 & 2 & 2 & 1 &  \\ \hline\multirow{4}{*}{Oblique} & HSP & 0 & 0 & 0 & 0 & 0 & \multirow{4}{*}{0} \\ \cline{2-7}
 & HSP-SOC & 0 & 0 & 0 & 0 & 0 &  \\ \cline{2-7}
 & nHSP & 0 & 0 & 0 & 0 & 0 &  \\ \cline{2-7}
 & nHSP-SOC & 0 & 0 & 0 & 0 & 0 &  \\ \hline
\hline
\end{tabular}
\caption{\label{app:table:statistic_semimetal} Statistics of twistable semimetal, classified into Bravais lattice, valley, and strength of SOC. 
When the SOC splitting near the twisting point is strong, the valley is labeled as ``valley-SOC'' type. 
The materials are further grouped into four types, \ie, experimental, computationally (Comp.) exfoliable, computationally stable, and computationally unstable. }
\end{table}

\subsection{Hexagonal lattice}

\subsubsection{$\Gamma$}

\begin{longtable}{c|c|c|c|c|c|c|c|c|c|c|c|c}
    \caption{\label{semimetal_hexagonal-GM_unstable} Computationally unstable materials with valley type: hexagonal-$\Gamma$.}\\
    \centering
    Formula & LG & ID & Gap & Database & Lattice & valley & SOC gap & Twist score & Dispersion & Topology & Bulk & Mat. type\\
    \endfirsthead
    \hline
    \ch{Cu2Se} & 80 & \twodtqclink{1.4.326} & 0.00 & \ctwodblink{1SeCu2-1} & hexagonal & $\Gamma$ & 0.05 & 0.81 & quadratic & SEBR & / & Unstable \\\hline
\ch{NiZrH6} & 67 & \twodtqclink{3.4.66} & 0.05 & \ctwodblink{1NiZrH6-2} & hexagonal & $\Gamma$ & 0.05 & 0.40 & quadratic & OAI & / & Unstable \\\hline

\hline
\end{longtable}

\subsubsection{$\Gamma$-SOC}

    \begin{longtable}{c|c|c|c|c|c|c|c|c|c|c|c|c}
    \caption{\label{semimetal_hexagonal-GM-SOC_exfo} Computationally exfoliable materials with valley type: hexagonal-$\Gamma$-SOC.}\\
    \centering
    Formula & LG & ID & Gap & Database & Lattice & valley & SOC gap & Twist score & Dispersion & Topology & Bulk & Mat. type\\
    \endfirsthead
    \hline
    \ch{HfBr} & 72 & \twodtqclink{1.2.104} & 0.06 & \mctwodlink{Hf2Br2} & hexagonal & $\Gamma$-SOC & 0.23 & 0.57 & quadratic & SEBR & Yes & Comp.Exfo \\\hline

\hline
\end{longtable}

    \begin{longtable}{c|c|c|c|c|c|c|c|c|c|c|c|c}
    \caption{\label{semimetal_hexagonal-GM-SOC_stable} Computationally stable materials with valley type: hexagonal-$\Gamma$-SOC.}\\
    \centering
    Formula & LG & ID & Gap & Database & Lattice & valley & SOC gap & Twist score & Dispersion & Topology & Bulk & Mat. type\\
    \endfirsthead
    \hline
    \ch{Ag2Te} & 80 & \twodtqclink{1.3.420} & 0.14 & \ctwodblink{1TeAg2-1} & hexagonal & $\Gamma$-SOC & 0.19 & 0.84 & quadratic & SEBR & / & Stable \\\hline
\ch{BrGe} & 72 & \twodtqclink{1.3.209} & 0.16 & \ctwodblink{2BrGe-1} & hexagonal & $\Gamma$-SOC & 0.25 & 0.57 & quadratic & SEBR & / & Stable \\\hline
\ch{ZrBr} & 72 & \twodtqclink{1.3.213} & 0.00 & \ctwodblink{2BrZr-1} & hexagonal & $\Gamma$-SOC & 0.06 & 0.54 & quadratic & SEBR & Yes & Stable \\\hline
\ch{BiMoAs} & 69 & \twodtqclink{6.3.1923} & 0.00 & \ctwodblink{1AsBiMo-1} & hexagonal & $\Gamma$-SOC & 0.09 & 0.47 & quadratic & LCEBR & / & Stable \\\hline
\ch{Zr3C2Cl2} & 78 & \twodtqclink{1.3.350} & 0.00 & \ctwodblink{1C2Cl2Zr3-1} & hexagonal & $\Gamma$-SOC & 0.61 & 0.47 & quadratic & NLC & / & Stable \\\hline
\ch{HgO} & 72 & \twodtqclink{1.3.288} & 0.15 & \ctwodblink{2HgO-1} & hexagonal & $\Gamma$-SOC & 0.15 & 0.41 & quadratic & SEBR & / & Stable \\\hline
\ch{GaSnTe} & 72 & \twodtqclink{1.3.277} & 0.06 & \ctwodblink{2GaSnTe-1} & hexagonal & $\Gamma$-SOC & 0.44 & 0.40 & quadratic & SEBR & / & Stable \\\hline
\ch{HgTe} & 72 & \twodtqclink{3.3.447} & 0.09 & \ctwodblink{2HgTe-2} & hexagonal & $\Gamma$-SOC & 0.23 & 0.39 & quadratic & OAI & / & Stable \\\hline
\ch{AlSnTe} & 72 & \twodtqclink{1.3.192} & 0.09 & \ctwodblink{2AlSnTe-1} & hexagonal & $\Gamma$-SOC & 0.31 & 0.37 & quadratic & SEBR & / & Stable \\\hline
\ch{CdGa2Se4} & 72 & \twodtqclink{4.3.10} & 0.02 & \ctwodblink{1CdGa2Se4-1} & hexagonal & $\Gamma$-SOC & 0.12 & 0.28 & quadratic & OOAI & / & Stable \\\hline

\hline
\end{longtable}

    \begin{longtable}{c|c|c|c|c|c|c|c|c|c|c|c|c}
    \caption{\label{semimetal_hexagonal-GM-SOC_unstable} Computationally unstable materials with valley type: hexagonal-$\Gamma$-SOC.}\\
    \centering
    Formula & LG & ID & Gap & Database & Lattice & valley & SOC gap & Twist score & Dispersion & Topology & Bulk & Mat. type\\
    \endfirsthead
    \hline
    \ch{Cu2S} & 80 & \twodtqclink{1.4.325} & 0.00 & \ctwodblink{1SCu2-1} & hexagonal & $\Gamma$-SOC & 0.14 & 0.81 & quadratic & SEBR & Yes & Unstable \\\hline
\ch{Ag2Se} & 80 & \twodtqclink{1.4.314} & 0.00 & \ctwodblink{1SeAg2-1} & hexagonal & $\Gamma$-SOC & 0.06 & 0.79 & quadratic & SEBR & / & Unstable \\\hline
\ch{Ag2S} & 80 & \twodtqclink{1.4.313} & 0.00 & \ctwodblink{1SAg2-1} & hexagonal & $\Gamma$-SOC & 0.21 & 0.79 & quadratic & SEBR & / & Unstable \\\hline
\ch{CrOS} & 69 & \twodtqclink{6.4.547} & 0.00 & \ctwodblink{1CrOS-1} & hexagonal & $\Gamma$-SOC & 0.05 & 0.48 & quadratic & LCEBR & / & Unstable \\\hline
\ch{WTe} & 78 & \twodtqclink{3.4.217} & 0.02 & \ctwodblink{2TeW-5} & hexagonal & $\Gamma$-SOC & 0.25 & 0.39 & quadratic & OAI & / & Unstable \\\hline

\hline
\end{longtable}

\subsubsection{$K$}

    \begin{longtable}{c|c|c|c|c|c|c|c|c|c|c|c|c}
    \caption{\label{semimetal_hexagonal-K_exp} Experimental materials with valley type: hexagonal-$K$.}\\
    \centering
    Formula & LG & ID & Gap & Database & Lattice & valley & SOC gap & Twist score & Dispersion & Topology & Bulk & Mat. type\\
    \endfirsthead
    \hline
    \ch{C} & 80 & \twodtqclink{5.1.3} & 0.00 & \ctwodblink{2C-1} & hexagonal & $K$ & 0.00 & 0.90 & linear & AccidentalFermi & Yes & \href{https://doi.org/10.1126/science.1102896}{Exp.M.Exfo} \\\hline
\ch{Si} & 72 & \twodtqclink{5.1.2} & 0.00 & \ctwodblink{2Si-1} & hexagonal & $K$ & 0.00 & 0.80 & linear & SEBR & Yes & \href{https://onlinelibrary.wiley.com/doi/full/10.1002/adma.201903013}{Exp.W.Exfo} \\\hline
\ch{Ge} & 72 & \twodtqclink{1.1.15} & 0.02 & \ctwodblink{2Ge-1} & hexagonal & $K$ & 0.02 & 0.68 & linear & SEBR & Yes & \href{https://pubs.rsc.org/en/content/articlehtml/2020/tc/d0tc03892j}{Exp.W.Exfo} \\\hline

\hline
\end{longtable}

    \begin{longtable}{c|c|c|c|c|c|c|c|c|c|c|c|c}
    \caption{\label{semimetal_hexagonal-K_unstable} Computationally unstable materials with valley type: hexagonal-$K$.}\\
    \centering
    Formula & LG & ID & Gap & Database & Lattice & valley & SOC gap & Twist score & Dispersion & Topology & Bulk & Mat. type\\
    \endfirsthead
    \hline
    \ch{FN} & 72 & \twodtqclink{1.4.171} & 0.02 & \ctwodblink{2FN-1} & hexagonal & $K$ & 0.02 & 0.70 & linear & SEBR & / & Unstable \\\hline
\ch{HN} & 72 & \twodtqclink{1.4.177} & 0.01 & \ctwodblink{2HN-1} & hexagonal & $K$ & 0.01 & 0.66 & linear & SEBR & / & Unstable \\\hline
\ch{ZrCl} & 72 & \twodtqclink{1.4.162} & 0.04 & \ctwodblink{2ClZr-4} & hexagonal & $K$ & 0.04 & 0.64 & linear & SEBR & / & Unstable \\\hline
\ch{ScSe} & 72 & \twodtqclink{1.4.228} & 0.00 & \ctwodblink{2ScSe-4} & hexagonal & $K$ & 0.01 & 0.59 & linear & SEBR & / & Unstable \\\hline
\ch{ScS} & 72 & \twodtqclink{1.4.223} & 0.00 & \ctwodblink{2SSc-5} & hexagonal & $K$ & 0.01 & 0.59 & linear & SEBR & / & Unstable \\\hline
\ch{ScTe} & 72 & \twodtqclink{1.4.229} & 0.00 & \ctwodblink{2ScTe-5} & hexagonal & $K$ & 0.00 & 0.57 & linear & SEBR & / & Unstable \\\hline
\ch{VO} & 72 & \twodtqclink{1.4.207} & 0.02 & \ctwodblink{2OV-1} & hexagonal & $K$ & 0.02 & 0.55 & linear & SEBR & / & Unstable \\\hline
\ch{ZrBr} & 72 & \twodtqclink{1.4.143} & 0.04 & \ctwodblink{2BrZr-4} & hexagonal & $K$ & 0.04 & 0.53 & linear & SEBR & / & Unstable \\\hline
\ch{TiCl} & 72 & \twodtqclink{1.4.159} & 0.01 & \ctwodblink{2ClTi-2} & hexagonal & $K$ & 0.01 & 0.51 & linear & SEBR & / & Unstable \\\hline

\hline
\end{longtable}

\subsubsection{$K$-SOC}

    \begin{longtable}{c|c|c|c|c|c|c|c|c|c|c|c|c}
    \caption{\label{semimetal_hexagonal-K-SOC_exp} Experimental materials with valley type: hexagonal-$K$-SOC.}\\
    \centering
    Formula & LG & ID & Gap & Database & Lattice & valley & SOC gap & Twist score & Dispersion & Topology & Bulk & Mat. type\\
    \endfirsthead
    \hline
    \ch{Sn} & 72 & \twodtqclink{1.1.17} & 0.07 & \ctwodblink{2Sn-1} & hexagonal & $K$-SOC & 0.07 & 0.66 & linear & SEBR & Yes & \href{https://www.sciencedirect.com/science/article/pii/S1005030223000415}{Exp.W.Exfo} \\\hline

\hline
\end{longtable}

    \begin{longtable}{c|c|c|c|c|c|c|c|c|c|c|c|c}
    \caption{\label{semimetal_hexagonal-K-SOC_exfo} Computationally exfoliable materials with valley type: hexagonal-$K$-SOC.}\\
    \centering
    Formula & LG & ID & Gap & Database & Lattice & valley & SOC gap & Twist score & Dispersion & Topology & Bulk & Mat. type\\
    \endfirsthead
    \hline
    \ch{HgPt2Se3} & 72 & \twodtqclink{1.2.121} & 0.13 & \mctwodlink{Pt2HgSe3} & hexagonal & $K$-SOC & 0.17 & 0.45 & linear & SEBR & Yes & Comp.Exfo \\\hline

\hline
\end{longtable}

    \begin{longtable}{c|c|c|c|c|c|c|c|c|c|c|c|c}
    \caption{\label{semimetal_hexagonal-K-SOC_stable} Computationally stable materials with valley type: hexagonal-$K$-SOC.}\\
    \centering
    Formula & LG & ID & Gap & Database & Lattice & valley & SOC gap & Twist score & Dispersion & Topology & Bulk & Mat. type\\
    \endfirsthead
    \hline
    \ch{Ta2STe2} & 72 & \twodtqclink{1.3.321} & 0.15 & \ctwodblink{1STa2Te2-1} & hexagonal & $K$-SOC & 0.15 & 0.66 & linear & SEBR & / & Stable \\\hline
\ch{AsCl} & 72 & \twodtqclink{1.3.193} & 0.19 & \ctwodblink{2AsCl-1} & hexagonal & $K$-SOC & 0.19 & 0.65 & linear & SEBR & / & Stable \\\hline
\ch{CdPt2Se3} & 72 & \twodtqclink{1.3.249} & 0.04 & \ctwodblink{2CdPt2Se3-1} & hexagonal & $K$-SOC & 0.16 & 0.47 & linear & SEBR & / & Stable \\\hline
\ch{CdPt2Te3} & 72 & \twodtqclink{1.3.250} & 0.05 & \ctwodblink{2CdPt2Te3-1} & hexagonal & $K$-SOC & 0.18 & 0.36 & linear & SEBR & / & Stable \\\hline

\hline
\end{longtable}

    \begin{longtable}{c|c|c|c|c|c|c|c|c|c|c|c|c}
    \caption{\label{semimetal_hexagonal-K-SOC_unstable} Computationally unstable materials with valley type: hexagonal-$K$-SOC.}\\
    \centering
    Formula & LG & ID & Gap & Database & Lattice & valley & SOC gap & Twist score & Dispersion & Topology & Bulk & Mat. type\\
    \endfirsthead
    \hline
    \ch{HgCl} & 72 & \twodtqclink{1.4.154} & 0.12 & \ctwodblink{2ClHg-2} & hexagonal & $K$-SOC & 0.12 & 0.75 & linear & SEBR & / & Unstable \\\hline
\ch{HgBr} & 72 & \twodtqclink{1.4.136} & 0.18 & \ctwodblink{2BrHg-2} & hexagonal & $K$-SOC & 0.18 & 0.73 & linear & SEBR & / & Unstable \\\hline
\ch{InS} & 72 & \twodtqclink{1.4.191} & 0.00 & \ctwodblink{2InS-5} & hexagonal & $K$-SOC & 0.17 & 0.67 & linear & SEBR & / & Unstable \\\hline
\ch{IrSe} & 72 & \twodtqclink{1.4.195} & 0.14 & \ctwodblink{2IrSe-5} & hexagonal & $K$-SOC & 0.14 & 0.46 & linear & SEBR & / & Unstable \\\hline

\hline
\end{longtable}

\subsubsection{nHSP}

    \begin{longtable}{c|c|c|c|c|c|c|c|c|c|c|c|c}
    \caption{\label{semimetal_hexagonal-nHSP_stable} Computationally stable materials with valley type: hexagonal-nHSP.}\\
    \centering
    Formula & LG & ID & Gap & Database & Lattice & valley & SOC gap & Twist score & Dispersion & Topology & Bulk & Mat. type\\
    \endfirsthead
    \hline
    \ch{TiH2} & 78 & \twodtqclink{1.3.383} & 0.00 & \ctwodblink{1TiH2-1} & hexagonal & nHSP & 0.01 & 0.69 & linear & NLC & / & Stable \\\hline
\ch{IrPS3} & 71 & \twodtqclink{1.3.186} & 0.00 & \ctwodblink{2IrPS3-1} & hexagonal & nHSP & 0.01 & 0.59 & linear & SEBR & / & Stable \\\hline
\ch{RhPSe3} & 71 & \twodtqclink{1.3.188} & 0.02 & \ctwodblink{2PRhSe3-1} & hexagonal & nHSP & 0.02 & 0.55 & linear & SEBR & / & Stable \\\hline

\hline
\end{longtable}

\subsubsection{nHSP-SOC}

    \begin{longtable}{c|c|c|c|c|c|c|c|c|c|c|c|c}
    \caption{\label{semimetal_hexagonal-nHSP-SOC_stable} Computationally stable materials with valley type: hexagonal-nHSP-SOC.}\\
    \centering
    Formula & LG & ID & Gap & Database & Lattice & valley & SOC gap & Twist score & Dispersion & Topology & Bulk & Mat. type\\
    \endfirsthead
    \hline
    \ch{ZrH2} & 78 & \twodtqclink{1.3.385} & 0.00 & \ctwodblink{1ZrH2-1} & hexagonal & nHSP-SOC & 0.07 & 0.68 & linear & NLC & / & Stable \\\hline
\ch{ZrTe} & 72 & \twodtqclink{1.3.333} & 0.11 & \ctwodblink{2TeZr-5} & hexagonal & nHSP-SOC & 0.11 & 0.43 & linear & SEBR & / & Stable \\\hline
\ch{AuTe} & 78 & \twodtqclink{1.3.341} & 0.02 & \ctwodblink{2AuTe-3} & hexagonal & nHSP-SOC & 0.21 & 0.29 & linear & NLC & / & Stable \\\hline

\hline
\end{longtable}

    \begin{longtable}{c|c|c|c|c|c|c|c|c|c|c|c|c}
    \caption{\label{semimetal_hexagonal-nHSP-SOC_unstable} Computationally unstable materials with valley type: hexagonal-nHSP-SOC.}\\
    \centering
    Formula & LG & ID & Gap & Database & Lattice & valley & SOC gap & Twist score & Dispersion & Topology & Bulk & Mat. type\\
    \endfirsthead
    \hline
    \ch{MoO} & 78 & \twodtqclink{1.4.277} & 0.00 & \ctwodblink{2MoO-1} & hexagonal & nHSP-SOC & 0.19 & 0.48 & linear & NLC & / & Unstable \\\hline
\ch{HfSe} & 72 & \twodtqclink{1.4.183} & 0.05 & \ctwodblink{2HfSe-3} & hexagonal & nHSP-SOC & 0.05 & 0.45 & linear & SEBR & / & Unstable \\\hline

\hline
\end{longtable}

\subsection{Square lattice}

\subsubsection{$\Gamma$-SOC}

    \begin{longtable}{c|c|c|c|c|c|c|c|c|c|c|c|c}
    \caption{\label{semimetal_square-GM-SOC_unstable} Computationally unstable materials with valley type: square-$\Gamma$-SOC.}\\
    \centering
    Formula & LG & ID & Gap & Database & Lattice & valley & SOC gap & Twist score & Dispersion & Topology & Bulk & Mat. type\\
    \endfirsthead
    \hline
    \ch{PtS} & 64 & \twodtqclink{1.4.118} & 0.01 & \ctwodblink{2PtS-6} & square & $\Gamma$-SOC & 0.38 & 0.33 & quadratic & SEBR & / & Unstable \\\hline

\hline
\end{longtable}

\subsubsection{nHSP-SOC}

    \begin{longtable}{c|c|c|c|c|c|c|c|c|c|c|c|c}
    \caption{\label{semimetal_square-nHSP-SOC_unstable} Computationally unstable materials with valley type: square-nHSP-SOC.}\\
    \centering
    Formula & LG & ID & Gap & Database & Lattice & valley & SOC gap & Twist score & Dispersion & Topology & Bulk & Mat. type\\
    \endfirsthead
    \hline
    \ch{SnS} & 61 & \twodtqclink{1.4.63} & 0.03 & \ctwodblink{S2Sn2-b8e2ee0faa87} & square & nHSP-SOC & 0.22 & 0.78 & linear & SEBR & / & Unstable \\\hline
\ch{RuCl2} & 59 & \twodtqclink{6.4.373} & 0.00 & \ctwodblink{1RuCl2-2} & square & nHSP-SOC & 0.13 & 0.33 & linear & LCEBR & / & Unstable \\\hline

\hline
\end{longtable}

\subsection{Rectangular lattice}

\subsubsection{HSP}

    \begin{longtable}{c|c|c|c|c|c|c|c|c|c|c|c|c}
    \caption{\label{semimetal_rectangular-HSP_unstable} Computationally unstable materials with valley type: rectangular-HSP.}\\
    \centering
    Formula & LG & ID & Gap & Database & Lattice & valley & SOC gap & Twist score & Dispersion & Topology & Bulk & Mat. type\\
    \endfirsthead
    \hline
    \ch{TaF} & 18 & \twodtqclink{1.4.14} & 0.00 & \ctwodblink{F2Ta2-3472bad93efb} & rectangular & $\Gamma$ & 0.04 & 0.45 & linear & SEBR & / & Unstable \\\hline

\hline
\end{longtable}

\subsubsection{nHSP}

    \begin{longtable}{c|c|c|c|c|c|c|c|c|c|c|c|c}
    \caption{\label{semimetal_rectangular-nHSP_exfo} Computationally exfoliable materials with valley type: rectangular-nHSP.}\\
    \centering
    Formula & LG & ID & Gap & Database & Lattice & valley & SOC gap & Twist score & Dispersion & Topology & Bulk & Mat. type\\
    \endfirsthead
    \hline
    \ch{Hg3S2} & 18 & \twodtqclink{1.2.29} & 0.05 & \mctwodlink{Hg3S2} & rectangular & nHSP & 0.05 & 0.44 & linear & SEBR & Yes & Comp.Exfo \\\hline

\hline
\end{longtable}

    \begin{longtable}{c|c|c|c|c|c|c|c|c|c|c|c|c}
    \caption{\label{semimetal_rectangular-nHSP_stable} Computationally stable materials with valley type: rectangular-nHSP.}\\
    \centering
    Formula & LG & ID & Gap & Database & Lattice & valley & SOC gap & Twist score & Dispersion & Topology & Bulk & Mat. type\\
    \endfirsthead
    \hline
    \ch{RhFS} & 46 & \twodtqclink{1.3.81} & 0.01 & \ctwodblink{2FRhS-2} & rectangular & nHSP & 0.05 & 0.31 & linear & SEBR & / & Stable \\\hline

\hline
\end{longtable}

\subsubsection{nHSP-SOC}

    \begin{longtable}{c|c|c|c|c|c|c|c|c|c|c|c|c}
    \caption{\label{semimetal_rectangular-nHSP-SOC_exp} Experimental materials with valley type: rectangular-nHSP-SOC.}\\
    \centering
    Formula & LG & ID & Gap & Database & Lattice & valley & SOC gap & Twist score & Dispersion & Topology & Bulk & Mat. type\\
    \endfirsthead
    \hline
    \ch{WS2} & 15 & \twodtqclink{1.3.24} & 0.04 & \ctwodblink{2WS2-1} & rectangular & nHSP-SOC & 0.15 & 0.29 & linear & NLC & Yes & \href{https://doi.org/10.1063/1.5091997}{Exp.W.Exfo} \\\hline
\ch{MoS2} & 15 & \twodtqclink{1.1.1} & 0.05 & \ctwodblink{2MoS2-1} & rectangular & nHSP-SOC & 0.05 & 0.28 & linear & NLC & Yes & \href{https://www.nature.com/articles/s41467-017-00640-2}{Exp.M.Exfo} \\\hline

\hline
\end{longtable}

    \begin{longtable}{c|c|c|c|c|c|c|c|c|c|c|c|c}
    \caption{\label{semimetal_rectangular-nHSP-SOC_exfo} Computationally exfoliable materials with valley type: rectangular-nHSP-SOC.}\\
    \centering
    Formula & LG & ID & Gap & Database & Lattice & valley & SOC gap & Twist score & Dispersion & Topology & Bulk & Mat. type\\
    \endfirsthead
    \hline
    \ch{Hg2O} & 10 & \twodtqclink{6.2.307} & 0.07 & \mctwodlink{Hg2O} & rectangular & nHSP-SOC & 0.07 & 0.63 & linear & LCEBR & Yes & Comp.Exfo \\\hline
\ch{Hg2O} & 17 & \twodtqclink{1.2.25} & 0.08 & \mctwodlink{Hg2O} & rectangular & nHSP-SOC & 0.08 & 0.33 & linear & NLC & Yes & Comp.Exfo \\\hline

\hline
\end{longtable}

    \begin{longtable}{c|c|c|c|c|c|c|c|c|c|c|c|c}
    \caption{\label{semimetal_rectangular-nHSP-SOC_stable} Computationally stable materials with valley type: rectangular-nHSP-SOC.}\\
    \centering
    Formula & LG & ID & Gap & Database & Lattice & valley & SOC gap & Twist score & Dispersion & Topology & Bulk & Mat. type\\
    \endfirsthead
    \hline
    \ch{AgGaSeTe} & 13 & \twodtqclink{6.3.520} & 0.10 & \ctwodblink{1AgGaSeTe-1} & rectangular & nHSP-SOC & 0.11 & 0.64 & linear & LCEBR & / & Stable \\\hline
\ch{Ta2Se3} & 18 & \twodtqclink{1.3.50} & 0.03 & \ctwodblink{2Ta2Se3-1} & rectangular & nHSP-SOC & 0.11 & 0.36 & linear & SEBR & / & Stable \\\hline

\hline
\end{longtable}

    \begin{longtable}{c|c|c|c|c|c|c|c|c|c|c|c|c}
    \caption{\label{semimetal_rectangular-nHSP-SOC_unstable} Computationally unstable materials with valley type: rectangular-nHSP-SOC.}\\
    \centering
    Formula & LG & ID & Gap & Database & Lattice & valley & SOC gap & Twist score & Dispersion & Topology & Bulk & Mat. type\\
    \endfirsthead
    \hline
    \ch{CrS2} & 15 & \twodtqclink{1.4.5} & 0.00 & \ctwodblink{Cr2S4-45828f7c9596} & rectangular & nHSP-SOC & 0.18 & 0.29 & linear & NLC & / & Unstable \\\hline

\hline
\end{longtable}
\clearpage

\section{Catalogue of twistable insulators}\label{app:sec:table_twistable_insulator}

\subsection{Summary of results}

\begin{table}[hbp]
\centering
\scriptsize
\begin{tabular}{c|c|c|c|c|c|c|c}
\hline\hline
Lattice & Valley & \# of materials & Experimental & Comp. exfoliated & Comp. stable & Comp. unstable & Total \# \\ \hline
\multirow{8}{*}{Hexagonal} & $\Gamma$ & 414 & 28 & 28 & 298 & 60 & \multirow{8}{*}{1072} \\ \cline{2-7}
 & $\Gamma$-SOC & 167 & 13 & 10 & 124 & 20 &  \\ \cline{2-7}
 & $K$ & 115 & 7 & 8 & 75 & 25 &  \\ \cline{2-7}
 & $K$-SOC & 64 & 7 & 2 & 43 & 12 &  \\ \cline{2-7}
 & $M$ & 94 & 10 & 10 & 54 & 20 &  \\ \cline{2-7}
 & $M$-SOC & 6 & 1 & 0 & 3 & 2 &  \\ \cline{2-7}
 & nHSP & 115 & 10 & 8 & 79 & 18 &  \\ \cline{2-7}
 & nHSP-SOC & 97 & 1 & 3 & 65 & 28 &  \\ \hline\multirow{8}{*}{Square} & $\Gamma$ & 81 & 0 & 23 & 45 & 13 & \multirow{8}{*}{167} \\ \cline{2-7}
 & $\Gamma$-SOC & 17 & 0 & 3 & 14 & 0 &  \\ \cline{2-7}
 & $M$ & 25 & 0 & 2 & 20 & 3 &  \\ \cline{2-7}
 & $M$-SOC & 4 & 0 & 0 & 3 & 1 &  \\ \cline{2-7}
 & $X$ & 15 & 0 & 2 & 8 & 5 &  \\ \cline{2-7}
 & $X$-SOC & 1 & 0 & 0 & 0 & 1 &  \\ \cline{2-7}
 & nHSP & 13 & 0 & 1 & 3 & 9 &  \\ \cline{2-7}
 & nHSP-SOC & 11 & 0 & 0 & 7 & 4 &  \\ \hline\multirow{4}{*}{Rectangular} & HSP & 475 & 4 & 96 & 290 & 85 & \multirow{4}{*}{710} \\ \cline{2-7}
 & HSP-SOC & 34 & 0 & 0 & 17 & 17 &  \\ \cline{2-7}
 & nHSP & 139 & 3 & 13 & 98 & 25 &  \\ \cline{2-7}
 & nHSP-SOC & 62 & 2 & 4 & 38 & 18 &  \\ \hline\multirow{4}{*}{Oblique} & HSP & 19 & 0 & 6 & 10 & 3 & \multirow{4}{*}{38} \\ \cline{2-7}
 & HSP-SOC & 4 & 0 & 0 & 3 & 1 &  \\ \cline{2-7}
 & nHSP & 7 & 0 & 4 & 3 & 0 &  \\ \cline{2-7}
 & nHSP-SOC & 8 & 0 & 1 & 3 & 4 &  \\ \hline
\hline
\end{tabular}
\caption{\label{app:table:statistic_insulator} Statistics of twistable insulator, classified into Bravais lattice, valley, and strength of SOC. 
When the SOC splitting near the twisting point is strong, the valley is labeled as ``valley-SOC'' type. 
The materials are further grouped into four types, \ie, experimental, computationally (Comp.) exfoliable, computationally stable, and computationally unstable. }
\end{table}

\subsection{Hexagonal lattice}

\subsubsection{$\Gamma$}

    \begin{longtable}{c|c|c|c|c|c|c|c|c|c|c|c}
    \caption{\label{app:table_insulator_hexagonal-GM_exp} Experimental materials with valley type: hexagonal-$\Gamma$.}\\
    \centering
    Formula & LG & ID & Gap & Database & Lattice & VBM valley & CBM valley & Twist score & Topology & Bulk & Mat. type\\
    \endfirsthead
    \hline
    \ch{GaN} & 78 & \twodtqclink{6.1.78} & 1.82 & \ctwodblink{1GaN-1} & hexagonal & $K$ & $\Gamma$ & 0.70 & LCEBR & Yes & Exp. \\\hline
\ch{GeH} & 72 & \twodtqclink{3.1.18} & 0.90 & \ctwodblink{2GeH-1} & hexagonal & / & $\Gamma$ & 0.66 & OAI & Yes & \href{https://doi.org/10.1021/nn4009406}{Exp.M.Exfo} \\\hline
\ch{PbI2} & 72 & \twodtqclink{6.1.54} & 1.87 & \ctwodblink{1PbI2-1} & hexagonal & / & $\Gamma$ & 0.65 & LCEBR & Yes & \href{https://doi.org/10.1016/j.matlet.2016.01.034}{Exp.M.Exfo} \\\hline
\ch{InSe} & 78 & \twodtqclink{3.1.37} & 1.40 & \ctwodblink{2InSe-1} & hexagonal & nHSP & $\Gamma$ & 0.64 & OAI & Yes & \href{https://onlinelibrary.wiley.com/doi/10.1002/adma.201302616}{Exp.M.Exfo} \\\hline
\ch{BiClTe} & 69 & \twodtqclink{6.3.1963} & 0.95 & \ctwodblink{1BiClTe-2} & hexagonal & $\Gamma$-SOC & $\Gamma$ & 0.63 & LCEBR & Yes & \href{https://analyticalsciencejournals.onlinelibrary.wiley.com/doi/full/10.1002/jrs.5253}{Exp.M.Exfo} \\\hline
\ch{BiClTe} & 69 & \twodtqclink{6.1.11} & 0.94 & \ctwodblink{1BiClTe-1} & hexagonal & / & $\Gamma$ & 0.63 & LCEBR & Yes & \href{https://analyticalsciencejournals.onlinelibrary.wiley.com/doi/full/10.1002/jrs.5253}{Exp.M.Exfo} \\\hline
\ch{BiClTe} & 69 & \twodtqclink{6.1.12} & 0.94 & \mctwodlink{BiTeCl} & hexagonal & $\Gamma$-SOC & $\Gamma$ & 0.63 & LCEBR & Yes & \href{https://analyticalsciencejournals.onlinelibrary.wiley.com/doi/full/10.1002/jrs.5253}{Exp.M.Exfo} \\\hline
\ch{Bi} & 72 & \twodtqclink{1.1.10} & 0.53 & \mctwodlink{Bi} & hexagonal & $\Gamma$-SOC & $\Gamma$ & 0.60 & SEBR & Yes & Exp. \\
\hline

\hline
\end{longtable}

    \begin{longtable}{c|c|c|c|c|c|c|c|c|c|c|c}
    \caption{\label{app:table_insulator_hexagonal-GM_exfo} Computationally exfoliable materials with valley type: hexagonal-$\Gamma$.}\\
    \centering
    Formula & LG & ID & Gap & Database & Lattice & VBM valley & CBM valley & Twist score & Topology & Bulk & Mat. type\\
    \endfirsthead
    \hline
    \ch{PbTe} & 69 & \twodtqclink{6.2.1129} & 1.14 & \mctwodlink{PbTe} & hexagonal & $\Gamma$-SOC & $\Gamma$ & 0.69 & LCEBR & Yes & Comp.Exfo \\\hline
\ch{Tl2S} & 72 & \twodtqclink{6.2.1302} & 1.27 & \mctwodlink{Tl2S} & hexagonal & $\Gamma$ & $M$ & 0.65 & LCEBR & Yes & Comp.Exfo \\\hline
\ch{NaSnP} & 69 & \twodtqclink{6.2.1128} & 1.07 & \mctwodlink{NaSnP} & hexagonal & nHSP-SOC & $\Gamma$ & 0.65 & LCEBR & Yes & Comp.Exfo \\\hline
\ch{TbBrH2} & 72 & \twodtqclink{6.2.1255} & 1.58 & \mctwodlink{Tb2H4Br2} & hexagonal & $\Gamma$ & / & 0.64 & LCEBR & Yes & Comp.Exfo \\\hline
\ch{Li3As} & 78 & \twodtqclink{6.2.1312} & 0.80 & \mctwodlink{Li3As} & hexagonal & / & $\Gamma$ & 0.63 & LCEBR & Yes & Comp.Exfo \\\hline
\ch{Li4SrP2} & 72 & \twodtqclink{6.2.1277} & 1.28 & \mctwodlink{Li4SrP2} & hexagonal & / & $\Gamma$ & 0.62 & LCEBR & Yes & Comp.Exfo \\\hline
\ch{GaSe} & 69 & \twodtqclink{6.2.1106} & 0.88 & \mctwodlink{Ga2Se2} & hexagonal & $K$-SOC & $\Gamma$ & 0.59 & LCEBR & Yes & Comp.Exfo \\\hline
\ch{K2SnAs2S6} & 66 & \twodtqclink{6.2.1064} & 1.12 & \mctwodlink{K2SnAs2S6} & hexagonal & / & $\Gamma$ & 0.59 & LCEBR & Yes & Comp.Exfo \\
\hline

\hline
\end{longtable}

    \begin{longtable}{c|c|c|c|c|c|c|c|c|c|c|c}
    \caption{\label{app:table_insulator_hexagonal-GM_stable} Computationally stable materials with valley type: hexagonal-$\Gamma$.}\\
    \centering
    Formula & LG & ID & Gap & Database & Lattice & VBM valley & CBM valley & Twist score & Topology & Bulk & Mat. type\\
    \endfirsthead
    \hline
    \ch{CuI} & 78 & \twodtqclink{6.3.2638} & 1.29 & \ctwodblink{1CuI-1} & hexagonal & / & $\Gamma$ & 0.75 & LCEBR & / & Stable \\\hline
\ch{CuBr} & 78 & \twodtqclink{6.3.2619} & 1.02 & \ctwodblink{1BrCu-1} & hexagonal & / & $\Gamma$ & 0.75 & LCEBR & Yes & Stable \\\hline
\ch{CuCl} & 78 & \twodtqclink{6.3.2637} & 1.09 & \ctwodblink{1ClCu-1} & hexagonal & / & $\Gamma$ & 0.75 & LCEBR & / & Stable \\\hline
\ch{AgI} & 78 & \twodtqclink{6.3.2600} & 1.55 & \ctwodblink{1AgI-1} & hexagonal & $\Gamma$-SOC & $\Gamma$ & 0.74 & LCEBR & / & Stable \\\hline
\ch{AgBr} & 78 & \twodtqclink{6.3.2598} & 1.58 & \ctwodblink{1AgBr-1} & hexagonal & / & $\Gamma$ & 0.74 & LCEBR & / & Stable \\\hline
\ch{AgCl} & 78 & \twodtqclink{6.3.2599} & 1.68 & \ctwodblink{1AgCl-1} & hexagonal & $\Gamma$-SOC & $\Gamma$ & 0.72 & LCEBR & / & Stable \\\hline
\ch{PbO2} & 72 & \twodtqclink{6.3.2544} & 1.31 & \ctwodblink{1PbO2-1} & hexagonal & / & $\Gamma$ & 0.71 & LCEBR & / & Stable \\\hline
\ch{WO2} & 78 & \twodtqclink{3.3.520} & 1.34 & \ctwodblink{1WO2-1} & hexagonal & $\Gamma$ & $K$ & 0.71 & OAI & / & Stable \\
\hline

\hline
\end{longtable}

    \begin{longtable}{c|c|c|c|c|c|c|c|c|c|c|c}
    \caption{\label{app:table_insulator_hexagonal-GM_unstable} Computationally unstable materials with valley type: hexagonal-$\Gamma$.}\\
    \centering
    Formula & LG & ID & Gap & Database & Lattice & VBM valley & CBM valley & Twist score & Topology & Bulk & Mat. type\\
    \endfirsthead
    \hline
    \ch{AlAs} & 78 & \twodtqclink{6.4.691} & 1.24 & \ctwodblink{1AlAs-1} & hexagonal & $K$ & $\Gamma$ & 0.75 & LCEBR & / & Unstable \\\hline
\ch{CdS} & 78 & \twodtqclink{6.4.702} & 1.63 & \ctwodblink{1CdS-1} & hexagonal & / & $\Gamma$ & 0.73 & LCEBR & / & Unstable \\\hline
\ch{SrO} & 78 & \twodtqclink{6.4.725} & 1.75 & \ctwodblink{1OSr-1} & hexagonal & $K$-SOC & $\Gamma$ & 0.71 & LCEBR & / & Unstable \\\hline
\ch{GeO2} & 78 & \twodtqclink{6.4.712} & 1.39 & \ctwodblink{1GeO2-3} & hexagonal & / & $\Gamma$ & 0.71 & LCEBR & / & Unstable \\\hline
\ch{CdSe} & 69 & \twodtqclink{6.4.528} & 1.11 & \ctwodblink{1CdSe-1} & hexagonal & $\Gamma$-SOC & $\Gamma$ & 0.69 & LCEBR & / & Unstable \\\hline
\ch{CdTe} & 69 & \twodtqclink{6.4.529} & 1.01 & \ctwodblink{1CdTe-1} & hexagonal & $\Gamma$-SOC & $\Gamma$ & 0.69 & LCEBR & / & Unstable \\\hline
\ch{CdO} & 78 & \twodtqclink{6.4.701} & 0.76 & \ctwodblink{1CdO-1} & hexagonal & / & $\Gamma$ & 0.68 & LCEBR & / & Unstable \\\hline
\ch{GeH2O2} & 72 & \twodtqclink{6.4.655} & 1.09 & \ctwodblink{1GeH2O2-1} & hexagonal & $K$ & $\Gamma$ & 0.68 & LCEBR & / & Unstable \\
\hline

\hline
\end{longtable}

\subsubsection{$\Gamma$-SOC}

    \begin{longtable}{c|c|c|c|c|c|c|c|c|c|c|c}
    \caption{\label{app:table_insulator_hexagonal-GM-SOC_exp} Experimental materials with valley type: hexagonal-$\Gamma$-SOC.}\\
    \centering
    Formula & LG & ID & Gap & Database & Lattice & VBM valley & CBM valley & Twist score & Topology & Bulk & Mat. type\\
    \endfirsthead
    \hline
    \ch{BiITe} & 69 & \twodtqclink{6.1.13} & 0.70 & \ctwodblink{1BiITe-1} & hexagonal & nHSP-SOC & $\Gamma$-SOC & 0.55 & LCEBR & Yes & \href{https://doi.org/10.1088/2053-1583/aac652}{Exp.M.Exfo} \\\hline
\ch{Sb} & 72 & \twodtqclink{3.1.25} & 1.01 & \ctwodblink{2Sb-1} & hexagonal & $\Gamma$-SOC & nHSP & 0.54 & OAI & Yes & \href{https://pubs.rsc.org/en/content/articlelanding/2019/ta/c9ta06072c}{Exp.W.Exfo} \\\hline
\ch{As} & 72 & \twodtqclink{3.1.11} & 1.48 & \ctwodblink{2As-1} & hexagonal & $\Gamma$-SOC & nHSP & 0.50 & OAI & Yes & \href{https://onlinelibrary.wiley.com/doi/full/10.1002/adfm.201807004}{Exp.W.Exfo} \\\hline
\ch{PtSe2} & 72 & \twodtqclink{6.1.58} & 1.18 & \ctwodblink{1PtSe2-1} & hexagonal & $\Gamma$-SOC & nHSP & 0.48 & LCEBR & Yes & \href{https://www.nature.com/articles/s41467-020-16266-w}{Exp.M.Exfo} \\\hline
\ch{BiClTe} & 69 & \twodtqclink{6.3.1963} & 0.95 & \ctwodblink{1BiClTe-2} & hexagonal & $\Gamma$-SOC & $\Gamma$ & 0.44 & LCEBR & Yes & \href{https://analyticalsciencejournals.onlinelibrary.wiley.com/doi/full/10.1002/jrs.5253}{Exp.M.Exfo} \\\hline
\ch{BiBrTe} & 69 & \twodtqclink{6.1.10} & 0.92 & \ctwodblink{1BiBrTe-2} & hexagonal & $\Gamma$-SOC & $\Gamma$-SOC & 0.41 & LCEBR & Yes & \href{https://www.nature.com/articles/s41467-023-44439-w}{Exp.M.Exfo} \\\hline
\ch{SnSe2} & 72 & \twodtqclink{6.1.71} & 0.76 & \ctwodblink{1SnSe2-1} & hexagonal & $\Gamma$-SOC & $M$ & 0.39 & LCEBR & Yes & \href{https://www.nature.com/articles/s41467-020-16266-w}{Exp.M.Exfo} \\\hline
\ch{BiClTe} & 69 & \twodtqclink{6.1.12} & 0.94 & \mctwodlink{BiTeCl} & hexagonal & $\Gamma$-SOC & $\Gamma$ & 0.38 & LCEBR & Yes & \href{https://analyticalsciencejournals.onlinelibrary.wiley.com/doi/full/10.1002/jrs.5253}{Exp.M.Exfo} \\
\hline

\hline
\end{longtable}

    \begin{longtable}{c|c|c|c|c|c|c|c|c|c|c|c}
    \caption{\label{app:table_insulator_hexagonal-GM-SOC_exfo} Computationally exfoliable materials with valley type: hexagonal-$\Gamma$-SOC.}\\
    \centering
    Formula & LG & ID & Gap & Database & Lattice & VBM valley & CBM valley & Twist score & Topology & Bulk & Mat. type\\
    \endfirsthead
    \hline
    \ch{GeTe} & 69 & \twodtqclink{6.2.1108} & 1.25 & \mctwodlink{GeTe} & hexagonal & $\Gamma$-SOC & nHSP & 0.50 & LCEBR & Yes & Comp.Exfo \\\hline
\ch{PbTe} & 69 & \twodtqclink{6.2.1129} & 1.14 & \mctwodlink{PbTe} & hexagonal & $\Gamma$-SOC & $\Gamma$ & 0.48 & LCEBR & Yes & Comp.Exfo \\\hline
\ch{AsSb} & 69 & \twodtqclink{6.2.1090} & 1.18 & \mctwodlink{SbAs} & hexagonal & $\Gamma$-SOC & nHSP-SOC & 0.46 & LCEBR & Yes & Comp.Exfo \\\hline
\ch{Bi2STe2} & 69 & \twodtqclink{6.2.1092} & 0.36 & \mctwodlink{Bi2Te2S-Cu5FeS4(hR5)} & hexagonal & nHSP-SOC & $\Gamma$-SOC & 0.36 & LCEBR & Yes & Comp.Exfo \\\hline
\ch{AlLiTe2} & 69 & \twodtqclink{6.2.1084} & 0.67 & \mctwodlink{LiAlTe2} & hexagonal & $\Gamma$-SOC & $\Gamma$-SOC & 0.36 & LCEBR & Yes & Comp.Exfo \\\hline
\ch{BiCuP2Se6} & 67 & \twodtqclink{3.2.159} & 0.65 & \mctwodlink{CuBiP2Se6} & hexagonal & / & $\Gamma$-SOC & 0.34 & OAI & Yes & Comp.Exfo \\\hline
\ch{ZrBrN} & 69 & \twodtqclink{6.2.1099} & 2.59 & \mctwodlink{ZrNBr-BiTeI} & hexagonal & $\Gamma$-SOC & $K$-SOC & 0.26 & LCEBR & Yes & Comp.Exfo \\\hline
\ch{Y2CI2} & 72 & \twodtqclink{6.2.1201} & 0.24 & \mctwodlink{Y2CI2} & hexagonal & $\Gamma$-SOC & $M$ & 0.26 & LCEBR & Yes & Comp.Exfo \\\hline
\ch{TmI3} & 71 & \twodtqclink{6.2.1159} & 2.75 & \mctwodlink{Tm2I6} & hexagonal & $\Gamma$-SOC & / & 0.24 & LCEBR & Yes & Comp.Exfo \\\hline
\ch{LuI3} & 71 & \twodtqclink{6.2.1154} & 2.84 & \mctwodlink{Lu2I6} & hexagonal & $\Gamma$-SOC & / & 0.24 & LCEBR & Yes & Comp.Exfo \\\hline

\hline
\end{longtable}

    \begin{longtable}{c|c|c|c|c|c|c|c|c|c|c|c}
    \caption{\label{app:table_insulator_hexagonal-GM-SOC_stable} Computationally stable materials with valley type: hexagonal-$\Gamma$-SOC.}\\
    \centering
    Formula & LG & ID & Gap & Database & Lattice & VBM valley & CBM valley & Twist score & Topology & Bulk & Mat. type\\
    \endfirsthead
    \hline
    \ch{GaAs} & 69 & \twodtqclink{6.3.1940} & 1.07 & \ctwodblink{1AsGa-1} & hexagonal & $K$ & $\Gamma$-SOC & 0.69 & LCEBR & / & Stable \\\hline
\ch{BiBrS} & 69 & \twodtqclink{6.3.1954} & 1.23 & \ctwodblink{1BiBrS-1} & hexagonal & nHSP-SOC & $\Gamma$-SOC & 0.65 & LCEBR & / & Stable \\\hline
\ch{BiBrSe} & 69 & \twodtqclink{6.3.1957} & 1.38 & \ctwodblink{1BiBrSe-2} & hexagonal & nHSP-SOC & $\Gamma$-SOC & 0.65 & LCEBR & / & Stable \\\hline
\ch{BiBrSe} & 69 & \twodtqclink{6.3.1956} & 1.03 & \ctwodblink{1BiBrSe-1} & hexagonal & nHSP-SOC & $\Gamma$-SOC & 0.65 & LCEBR & / & Stable \\\hline
\ch{BiIS} & 69 & \twodtqclink{6.3.1969} & 1.14 & \ctwodblink{1BiIS-1} & hexagonal & / & $\Gamma$-SOC & 0.65 & LCEBR & / & Stable \\\hline
\ch{BrSSb} & 69 & \twodtqclink{6.3.2007} & 1.23 & \ctwodblink{1BrSSb-1} & hexagonal & nHSP-SOC & $\Gamma$-SOC & 0.65 & LCEBR & / & Stable \\\hline
\ch{BrSbSe} & 69 & \twodtqclink{6.3.2009} & 1.07 & \ctwodblink{1BrSbSe-1} & hexagonal & nHSP-SOC & $\Gamma$-SOC & 0.65 & LCEBR & / & Stable \\\hline
\ch{ClSSb} & 69 & \twodtqclink{6.3.2038} & 1.33 & \ctwodblink{1ClSSb-1} & hexagonal & nHSP & $\Gamma$-SOC & 0.65 & LCEBR & / & Stable \\
\hline

\hline
\end{longtable}

    \begin{longtable}{c|c|c|c|c|c|c|c|c|c|c|c}
    \caption{\label{app:table_insulator_hexagonal-GM-SOC_unstable} Computationally unstable materials with valley type: hexagonal-$\Gamma$-SOC.}\\
    \centering
    Formula & LG & ID & Gap & Database & Lattice & VBM valley & CBM valley & Twist score & Topology & Bulk & Mat. type\\
    \endfirsthead
    \hline
    \ch{FOSb} & 69 & \twodtqclink{6.4.558} & 1.32 & \ctwodblink{1FOSb-1} & hexagonal & $K$ & $\Gamma$-SOC & 0.65 & LCEBR & / & Unstable \\\hline
\ch{HgBrFS} & 69 & \twodtqclink{6.4.503} & 1.01 & \ctwodblink{1BrFHgS-1} & hexagonal & / & $\Gamma$-SOC & 0.62 & LCEBR & / & Unstable \\\hline
\ch{AlSb} & 69 & \twodtqclink{6.4.478} & 1.45 & \ctwodblink{1AlSb-1} & hexagonal & $K$ & $\Gamma$-SOC & 0.59 & LCEBR & / & Unstable \\\hline
\ch{BiFO} & 69 & \twodtqclink{6.4.492} & 1.85 & \ctwodblink{1BiFO-1} & hexagonal & $K$ & $\Gamma$-SOC & 0.59 & LCEBR & / & Unstable \\\hline
\ch{HgFIS} & 69 & \twodtqclink{6.4.554} & 0.98 & \ctwodblink{1FHgIS-1} & hexagonal & / & $\Gamma$-SOC & 0.57 & LCEBR & / & Unstable \\\hline
\ch{CdFIO} & 69 & \twodtqclink{6.4.522} & 1.36 & \ctwodblink{1CdFIO-1} & hexagonal & $\Gamma$-SOC & $\Gamma$ & 0.54 & LCEBR & / & Unstable \\\hline
\ch{ZnI2} & 78 & \twodtqclink{6.4.720} & 0.93 & \ctwodblink{1ZnI2-3} & hexagonal & $\Gamma$-SOC & $M$ & 0.53 & LCEBR & / & Unstable \\\hline
\ch{CdTe} & 69 & \twodtqclink{6.4.529} & 1.01 & \ctwodblink{1CdTe-1} & hexagonal & $\Gamma$-SOC & $\Gamma$ & 0.52 & LCEBR & / & Unstable \\
\hline

\hline
\end{longtable}

\subsubsection{$K$}

    \begin{longtable}{c|c|c|c|c|c|c|c|c|c|c|c}
    \caption{\label{app:table_insulator_hexagonal-K_exp} Experimental materials with valley type: hexagonal-$K$.}\\
    \centering
    Formula & LG & ID & Gap & Database & Lattice & VBM valley & CBM valley & Twist score & Topology & Bulk & Mat. type\\
    \endfirsthead
    \hline
    \ch{ZrBrN} & 72 & \twodtqclink{6.3.2323} & 1.62 & \ctwodblink{2BrNZr-1} & hexagonal & nHSP & $K$ & 0.65 & LCEBR & Yes & \href{https://doi.org/10.1016/j.chemphys.2024.112487}{Exp.W.Exfo} \\\hline
\ch{ZrClN} & 72 & \twodtqclink{6.3.2408} & 1.91 & \ctwodblink{2ClNZr-1} & hexagonal & $\Gamma$ & $K$ & 0.58 & LCEBR & Yes & \href{https://doi.org/10.1002/smll.202107490}{Exp.M.Exfo} \\\hline
\ch{MoS2} & 78 & \twodtqclink{3.1.39} & 1.60 & \ctwodblink{1MoS2-1} & hexagonal & $K$-SOC & $K$ & 0.57 & OAI & Yes & \href{https://onlinelibrary.wiley.com/doi/full/10.1002/advs.202301243}{Exp.M.Exfo} \\\hline
\ch{MoSe2} & 78 & \twodtqclink{3.1.41} & 1.34 & \ctwodblink{1MoSe2-1} & hexagonal & $K$-SOC & $K$ & 0.57 & OAI & Yes & \href{https://doi.org/10.1364/OE.21.004908}{Exp.M.Exfo} \\\hline
\ch{MoTe2} & 78 & \twodtqclink{3.1.43} & 0.96 & \ctwodblink{1MoTe2-1} & hexagonal & $K$-SOC & $K$ & 0.54 & OAI & Yes & \href{https://doi.org/10.1038/nature24043}{Exp.M.Exfo} \\\hline
\ch{GaN} & 78 & \twodtqclink{6.1.78} & 1.82 & \ctwodblink{1GaN-1} & hexagonal & $K$ & $\Gamma$ & 0.52 & LCEBR & Yes & Exp. \\\hline
\ch{MoSSe} & 69 & \twodtqclink{3.1.10} & 1.48 & \ctwodblink{1MoSSe-1} & hexagonal & $K$-SOC & $K$ & 0.51 & OAI & Yes & \href{https://www.nature.com/articles/nnano.2017.100}{Exp.Substr} \\\hline

\hline
\end{longtable}

    \begin{longtable}{c|c|c|c|c|c|c|c|c|c|c|c}
    \caption{\label{app:table_insulator_hexagonal-K_exfo} Computationally exfoliable materials with valley type: hexagonal-$K$.}\\
    \centering
    Formula & LG & ID & Gap & Database & Lattice & VBM valley & CBM valley & Twist score & Topology & Bulk & Mat. type\\
    \endfirsthead
    \hline
    \ch{PbO6Se2} & 72 & \twodtqclink{6.2.1287} & 0.49 & \mctwodlink{PbSe2O6} & hexagonal & / & $K$ & 0.49 & LCEBR & Yes & Comp.Exfo \\\hline
\ch{SnPS3} & 66 & \twodtqclink{3.2.152} & 1.59 & \mctwodlink{SnPS3} & hexagonal & / & $K$ & 0.49 & OAI & Yes & Comp.Exfo \\\hline
\ch{AgErP2Se6} & 67 & \twodtqclink{3.2.153} & 1.50 & \mctwodlink{AgErP2Se6} & hexagonal & / & $K$ & 0.37 & OAI & Yes & Comp.Exfo \\\hline
\ch{AgTmP2Se6} & 67 & \twodtqclink{3.2.158} & 1.50 & \mctwodlink{AgTmP2Se6} & hexagonal & / & $K$ & 0.36 & OAI & Yes & Comp.Exfo \\\hline
\ch{YFSe} & 72 & \twodtqclink{6.2.1225} & 2.57 & \mctwodlink{Y2Se2F2} & hexagonal & / & $K$ & 0.33 & LCEBR & Yes & Comp.Exfo \\\hline
\ch{PrIS} & 72 & \twodtqclink{6.2.1268} & 2.47 & \mctwodlink{Pr2S2I2} & hexagonal & nHSP & $K$ & 0.28 & LCEBR & Yes & Comp.Exfo \\\hline
\ch{NdIS} & 72 & \twodtqclink{6.2.1266} & 2.47 & \mctwodlink{Nd2S2I2} & hexagonal & nHSP & $K$ & 0.28 & LCEBR & Yes & Comp.Exfo \\\hline
\ch{GdIS} & 72 & \twodtqclink{6.2.1238} & 2.57 & \mctwodlink{Gd2S2I2} & hexagonal & nHSP & $K$ & 0.26 & LCEBR & Yes & Comp.Exfo \\\hline

\hline
\end{longtable}

    \begin{longtable}{c|c|c|c|c|c|c|c|c|c|c|c}
    \caption{\label{app:table_insulator_hexagonal-K_stable} Computationally stable materials with valley type: hexagonal-$K$.}\\
    \centering
    Formula & LG & ID & Gap & Database & Lattice & VBM valley & CBM valley & Twist score & Topology & Bulk & Mat. type\\
    \endfirsthead
    \hline
    \ch{BP} & 78 & \twodtqclink{6.3.2606} & 0.91 & \ctwodblink{1BP-1} & hexagonal & $K$ & $K$ & 0.72 & LCEBR & / & Stable \\\hline
\ch{TiF2} & 78 & \twodtqclink{3.3.500} & 1.28 & \ctwodblink{1TiF2-1} & hexagonal & $K$ & / & 0.71 & OAI & / & Stable \\\hline
\ch{ZrCl2} & 78 & \twodtqclink{3.3.495} & 0.99 & \ctwodblink{1ZrCl2-1} & hexagonal & $K$ & nHSP & 0.70 & OAI & Yes & Stable \\\hline
\ch{InBr} & 72 & \twodtqclink{6.3.2321} & 1.34 & \ctwodblink{2BrIn-3} & hexagonal & nHSP & $K$ & 0.69 & LCEBR & / & Stable \\\hline
\ch{InI} & 72 & \twodtqclink{6.3.2500} & 1.01 & \ctwodblink{2IIn-2} & hexagonal & nHSP & $K$ & 0.69 & LCEBR & / & Stable \\\hline
\ch{AlO} & 78 & \twodtqclink{3.3.475} & 1.32 & \ctwodblink{2AlO-1} & hexagonal & $K$ & $M$ & 0.69 & OAI & / & Stable \\\hline
\ch{TiCl2} & 78 & \twodtqclink{3.3.494} & 0.90 & \ctwodblink{1TiCl2-1} & hexagonal & $K$ & / & 0.68 & OAI & / & Stable \\\hline
\ch{InCl} & 72 & \twodtqclink{6.3.2406} & 1.55 & \ctwodblink{2ClIn-4} & hexagonal & / & $K$ & 0.68 & LCEBR & / & Stable \\
\hline

\hline
\end{longtable}

    \begin{longtable}{c|c|c|c|c|c|c|c|c|c|c|c}
    \caption{\label{app:table_insulator_hexagonal-K_unstable} Computationally unstable materials with valley type: hexagonal-$K$.}\\
    \centering
    Formula & LG & ID & Gap & Database & Lattice & VBM valley & CBM valley & Twist score & Topology & Bulk & Mat. type\\
    \endfirsthead
    \hline
    \ch{AlAs} & 78 & \twodtqclink{6.4.691} & 1.24 & \ctwodblink{1AlAs-1} & hexagonal & $K$ & $\Gamma$ & 0.75 & LCEBR & / & Unstable \\\hline
\ch{HfF2} & 78 & \twodtqclink{3.4.181} & 1.14 & \ctwodblink{1HfF2-1} & hexagonal & $K$ & $\Gamma$ & 0.71 & OAI & / & Unstable \\\hline
\ch{ZrF2} & 78 & \twodtqclink{3.4.184} & 1.52 & \ctwodblink{1ZrF2-1} & hexagonal & $K$ & / & 0.71 & OAI & / & Unstable \\\hline
\ch{AlO} & 72 & \twodtqclink{3.4.124} & 1.18 & \ctwodblink{2AlO-2} & hexagonal & $K$ & $M$ & 0.69 & OAI & / & Unstable \\\hline
\ch{SnF2O2} & 72 & \twodtqclink{6.4.639} & 1.08 & \ctwodblink{1SnF2O2-1} & hexagonal & $K$ & $\Gamma$ & 0.68 & LCEBR & / & Unstable \\\hline
\ch{F2GeO2} & 72 & \twodtqclink{6.4.634} & 1.09 & \ctwodblink{1GeF2O2-1} & hexagonal & $K$ & $M$ & 0.68 & LCEBR & / & Unstable \\\hline
\ch{GeH2O2} & 72 & \twodtqclink{6.4.655} & 1.09 & \ctwodblink{1GeH2O2-1} & hexagonal & $K$ & $\Gamma$ & 0.68 & LCEBR & / & Unstable \\\hline
\ch{F2Si} & 72 & \twodtqclink{6.4.646} & 1.78 & \ctwodblink{1SiF2-2} & hexagonal & $K$ & $\Gamma$ & 0.66 & LCEBR & / & Unstable \\
\hline

\hline
\end{longtable}

\subsubsection{$K$-SOC}

    \begin{longtable}{c|c|c|c|c|c|c|c|c|c|c|c}
    \caption{\label{app:table_insulator_hexagonal-K-SOC_exp} Experimental materials with valley type: hexagonal-$K$-SOC.}\\
    \centering
    Formula & LG & ID & Gap & Database & Lattice & VBM valley & CBM valley & Twist score & Topology & Bulk & Mat. type\\
    \endfirsthead
    \hline
    \ch{WSe2} & 78 & \twodtqclink{3.1.46} & 1.26 & \ctwodblink{1WSe2-1} & hexagonal & $K$-SOC & $K$-SOC & 0.71 & OAI & Yes & \href{https://journals.aps.org/prl/abstract/10.1103/PhysRevLett.113.026803}{Exp.M.Exfo} \\\hline
\ch{WS2} & 78 & \twodtqclink{3.1.45} & 1.55 & \ctwodblink{1WS2-1} & hexagonal & $K$-SOC & $K$-SOC & 0.70 & OAI & Yes & \href{https://www.science.org/doi/10.1126/science.aao3503}{Exp.M.Exfo} \\\hline
\ch{MoS2} & 78 & \twodtqclink{3.1.39} & 1.60 & \ctwodblink{1MoS2-1} & hexagonal & $K$-SOC & $K$ & 0.68 & OAI & Yes & \href{https://onlinelibrary.wiley.com/doi/full/10.1002/advs.202301243}{Exp.M.Exfo} \\\hline
\ch{MoSe2} & 78 & \twodtqclink{3.1.41} & 1.34 & \ctwodblink{1MoSe2-1} & hexagonal & $K$-SOC & $K$ & 0.64 & OAI & Yes & \href{https://doi.org/10.1364/OE.21.004908}{Exp.M.Exfo} \\\hline
\ch{WTe2} & 78 & \twodtqclink{3.1.47} & 0.75 & \ctwodblink{1WTe2-1} & hexagonal & $K$-SOC & $K$-SOC & 0.61 & OAI & Yes & \href{https://pubs.acs.org/doi/10.1021/acs.inorgchem.0c02049}{Exp.W.Exfo} \\\hline
\ch{MoTe2} & 78 & \twodtqclink{3.1.43} & 0.96 & \ctwodblink{1MoTe2-1} & hexagonal & $K$-SOC & $K$ & 0.60 & OAI & Yes & \href{https://doi.org/10.1038/nature24043}{Exp.M.Exfo} \\\hline
\ch{MoSSe} & 69 & \twodtqclink{3.1.10} & 1.48 & \ctwodblink{1MoSSe-1} & hexagonal & $K$-SOC & $K$ & 0.59 & OAI & Yes & \href{https://www.nature.com/articles/nnano.2017.100}{Exp.Substr} \\\hline

\hline
\end{longtable}

    \begin{longtable}{c|c|c|c|c|c|c|c|c|c|c|c}
    \caption{\label{app:table_insulator_hexagonal-K-SOC_exfo} Computationally exfoliable materials with valley type: hexagonal-$K$-SOC.}\\
    \centering
    Formula & LG & ID & Gap & Database & Lattice & VBM valley & CBM valley & Twist score & Topology & Bulk & Mat. type\\
    \endfirsthead
    \hline
    \ch{GaSe} & 69 & \twodtqclink{6.2.1106} & 0.88 & \mctwodlink{Ga2Se2} & hexagonal & $K$-SOC & $\Gamma$ & 0.45 & LCEBR & Yes & Comp.Exfo \\\hline
\ch{ZrBrN} & 69 & \twodtqclink{6.2.1099} & 2.59 & \mctwodlink{ZrNBr-BiTeI} & hexagonal & $\Gamma$-SOC & $K$-SOC & 0.41 & LCEBR & Yes & Comp.Exfo \\\hline

\hline
\end{longtable}

    \begin{longtable}{c|c|c|c|c|c|c|c|c|c|c|c}
    \caption{\label{app:table_insulator_hexagonal-K-SOC_stable} Computationally stable materials with valley type: hexagonal-$K$-SOC.}\\
    \centering
    Formula & LG & ID & Gap & Database & Lattice & VBM valley & CBM valley & Twist score & Topology & Bulk & Mat. type\\
    \endfirsthead
    \hline
    \ch{HfCl2} & 78 & \twodtqclink{3.3.493} & 0.89 & \ctwodblink{1HfCl2-1} & hexagonal & $K$-SOC & nHSP-SOC & 0.67 & OAI & / & Stable \\\hline
\ch{CrS2} & 78 & \twodtqclink{3.3.497} & 0.90 & \ctwodblink{1CrS2-1} & hexagonal & $K$-SOC & $K$ & 0.66 & OAI & / & Stable \\\hline
\ch{ZrBr2} & 78 & \twodtqclink{3.3.483} & 0.83 & \ctwodblink{1ZrBr2-1} & hexagonal & $K$-SOC & nHSP & 0.65 & OAI & / & Stable \\\hline
\ch{WSSe} & 69 & \twodtqclink{3.3.292} & 1.42 & \ctwodblink{1SSeW-1} & hexagonal & $K$-SOC & $K$-SOC & 0.65 & OAI & / & Stable \\\hline
\ch{TiBr2} & 78 & \twodtqclink{3.3.482} & 0.76 & \ctwodblink{1TiBr2-1} & hexagonal & $K$-SOC & / & 0.63 & OAI & / & Stable \\\hline
\ch{WSeTe} & 69 & \twodtqclink{3.3.294} & 1.06 & \ctwodblink{1SeTeW-1} & hexagonal & $K$-SOC & $K$-SOC & 0.63 & OAI & / & Stable \\\hline
\ch{HfBr2} & 78 & \twodtqclink{3.3.481} & 0.72 & \ctwodblink{1HfBr2-1} & hexagonal & $K$-SOC & nHSP-SOC & 0.62 & OAI & / & Stable \\\hline
\ch{ZrBrCl} & 69 & \twodtqclink{3.3.267} & 0.91 & \ctwodblink{1BrClZr-1} & hexagonal & $K$-SOC & nHSP & 0.62 & OAI & / & Stable \\
\hline

\hline
\end{longtable}

    \begin{longtable}{c|c|c|c|c|c|c|c|c|c|c|c}
    \caption{\label{app:table_insulator_hexagonal-K-SOC_unstable} Computationally unstable materials with valley type: hexagonal-$K$-SOC.}\\
    \centering
    Formula & LG & ID & Gap & Database & Lattice & VBM valley & CBM valley & Twist score & Topology & Bulk & Mat. type\\
    \endfirsthead
    \hline
    \ch{I2Si} & 78 & \twodtqclink{6.4.719} & 1.38 & \ctwodblink{1SiI2-2} & hexagonal & nHSP-SOC & $K$-SOC & 0.65 & LCEBR & / & Unstable \\\hline
\ch{HfBrF} & 69 & \twodtqclink{3.4.88} & 1.09 & \ctwodblink{1BrFHf-1} & hexagonal & $K$-SOC & nHSP-SOC & 0.65 & OAI & / & Unstable \\\hline
\ch{HfClF} & 69 & \twodtqclink{3.4.99} & 1.18 & \ctwodblink{1ClFHf-1} & hexagonal & $K$-SOC & nHSP-SOC & 0.65 & OAI & / & Unstable \\\hline
\ch{ZrFI} & 69 & \twodtqclink{3.4.109} & 1.07 & \ctwodblink{1FIZr-1} & hexagonal & $K$-SOC & / & 0.65 & OAI & / & Unstable \\\hline
\ch{HfFI} & 69 & \twodtqclink{3.4.107} & 0.96 & \ctwodblink{1FHfI-1} & hexagonal & $K$-SOC & nHSP-SOC & 0.63 & OAI & / & Unstable \\\hline
\ch{SrO} & 78 & \twodtqclink{6.4.725} & 1.75 & \ctwodblink{1OSr-1} & hexagonal & $K$-SOC & $\Gamma$ & 0.63 & LCEBR & / & Unstable \\\hline
\ch{Br2Si} & 78 & \twodtqclink{6.4.697} & 1.98 & \ctwodblink{1SiBr2-2} & hexagonal & nHSP-SOC & $K$-SOC & 0.59 & LCEBR & / & Unstable \\\hline
\ch{ClPTe} & 69 & \twodtqclink{6.4.544} & 0.94 & \ctwodblink{1ClPTe-1} & hexagonal & nHSP-SOC & $K$-SOC & 0.57 & LCEBR & / & Unstable \\
\hline

\hline
\end{longtable}

\subsubsection{$M$}

    \begin{longtable}{c|c|c|c|c|c|c|c|c|c|c|c}
    \caption{\label{app:table_insulator_hexagonal-M_exp} Experimental materials with valley type: hexagonal-$M$.}\\
    \centering
    Formula & LG & ID & Gap & Database & Lattice & VBM valley & CBM valley & Twist score & Topology & Bulk & Mat. type\\
    \endfirsthead
    \hline
    \ch{SnS2} & 72 & \twodtqclink{6.1.75} & 1.58 & \ctwodblink{SnS2-42a44e4e7298} & hexagonal & / & $M$ & 0.70 & LCEBR & Yes & \href{https://pubs.acs.org/doi/10.1021/acsaelm.2c01010}{Exp.M.Exfo} \\\hline
\ch{HfS2} & 72 & \twodtqclink{6.1.43} & 1.24 & \ctwodblink{1HfS2-1} & hexagonal & / & $M$ & 0.64 & LCEBR & Yes & \href{https://doi.org/10.1002/adma.201503864}{Exp.M.Exfo} \\\hline
\ch{SnSe2} & 72 & \twodtqclink{6.1.71} & 0.76 & \ctwodblink{1SnSe2-1} & hexagonal & $\Gamma$-SOC & $M$ & 0.63 & LCEBR & Yes & \href{https://www.nature.com/articles/s41467-020-16266-w}{Exp.M.Exfo} \\\hline
\ch{CdI2} & 72 & \twodtqclink{6.1.38} & 2.18 & \ctwodblink{1CdI2-2} & hexagonal & $\Gamma$-SOC & $M$ & 0.60 & LCEBR & Yes & \href{https://pubs.acs.org/doi/10.1021/acsnano.7b01507}{Exp.Substr} \\\hline
\ch{C3N} & 80 & \twodtqclink{3.1.48} & 0.39 & \ctwodblink{2NC3-1} & hexagonal & $M$ & / & 0.59 & OAI & Yes & Exp. \\\hline
\ch{HSi} & 72 & \twodtqclink{3.1.19} & 2.18 & \ctwodblink{2HSi-1} & hexagonal & / & $M$ & 0.57 & OAI & Yes & Exp. \\\hline
\ch{ZrS2} & 72 & \twodtqclink{6.1.61} & 1.16 & \ctwodblink{1ZrS2-1} & hexagonal & / & $M$ & 0.55 & LCEBR & Yes & \href{https://onlinelibrary.wiley.com/doi/full/10.1002/smll.202205763}{Exp.M.Exfo} \\\hline
\ch{HfSe2} & 72 & \twodtqclink{6.1.44} & 0.45 & \ctwodblink{1HfSe2-1} & hexagonal & $\Gamma$-SOC & $M$ & 0.49 & LCEBR & Yes & \href{https://doi.org/10.1063/1.4917458}{Exp.M.Exfo} \\
\hline

\hline
\end{longtable}

    \begin{longtable}{c|c|c|c|c|c|c|c|c|c|c|c}
    \caption{\label{app:table_insulator_hexagonal-M_exfo} Computationally exfoliable materials with valley type: hexagonal-$M$.}\\
    \centering
    Formula & LG & ID & Gap & Database & Lattice & VBM valley & CBM valley & Twist score & Topology & Bulk & Mat. type\\
    \endfirsthead
    \hline
    \ch{Tl2S} & 72 & \twodtqclink{6.2.1302} & 1.27 & \mctwodlink{Tl2S} & hexagonal & $\Gamma$ & $M$ & 0.71 & LCEBR & Yes & Comp.Exfo \\\hline
\ch{Lu2CCl2} & 72 & \twodtqclink{6.2.1199} & 1.04 & \mctwodlink{CLu2Cl2} & hexagonal & / & $M$ & 0.68 & LCEBR & Yes & Comp.Exfo \\\hline
\ch{BC3} & 80 & \twodtqclink{3.2.206} & 0.63 & \mctwodlink{B2C6} & hexagonal & / & $M$ & 0.67 & OAI & Yes & Comp.Exfo \\\hline
\ch{Y2Br2C} & 72 & \twodtqclink{6.2.1198} & 0.81 & \mctwodlink{Y2CBr2} & hexagonal & / & $M$ & 0.62 & LCEBR & Yes & Comp.Exfo \\\hline
\ch{Tl2O} & 72 & \twodtqclink{6.2.1288} & 0.93 & \mctwodlink{Tl2O} & hexagonal & $M$ & $M$ & 0.59 & LCEBR & Yes & Comp.Exfo \\\hline
\ch{HgNa4P2} & 72 & \twodtqclink{6.2.1261} & 1.15 & \mctwodlink{Na4HgP2} & hexagonal & / & $M$ & 0.57 & LCEBR & Yes & Comp.Exfo \\\hline
\ch{Ga2S3} & 72 & \twodtqclink{6.2.1237} & 0.52 & \mctwodlink{Ga2S3} & hexagonal & / & $M$ & 0.52 & LCEBR & Yes & Comp.Exfo \\\hline
\ch{La2GeI2} & 72 & \twodtqclink{6.2.1240} & 0.30 & \mctwodlink{La2GeI2} & hexagonal & / & $M$ & 0.45 & LCEBR & Yes & Comp.Exfo \\\hline
\ch{Y2CI2} & 72 & \twodtqclink{6.2.1201} & 0.24 & \mctwodlink{Y2CI2} & hexagonal & $\Gamma$-SOC & $M$ & 0.44 & LCEBR & Yes & Comp.Exfo \\\hline
\ch{KPt2Se3} & 72 & \twodtqclink{6.2.1276} & 1.11 & \mctwodlink{KPt2Se3} & hexagonal & / & $M$ & 0.42 & LCEBR & Yes & Comp.Exfo \\\hline

\hline
\end{longtable}%

\subsubsection{$M$-SOC}

    \begin{longtable}{c|c|c|c|c|c|c|c|c|c|c|c}
    \caption{\label{app:table_insulator_hexagonal-M-SOC_exp} Experimental materials with valley type: hexagonal-$M$-SOC.}\\
    \centering
    Formula & LG & ID & Gap & Database & Lattice & VBM valley & CBM valley & Twist score & Topology & Bulk & Mat. type\\
    \endfirsthead
    \hline
    \ch{GaTe} & 78 & \twodtqclink{3.1.35} & 1.29 & \ctwodblink{2GaTe-1} & hexagonal & / & $M$-SOC & 0.58 & OAI & Yes & \href{https://doi.org/10.1039/C6CP01963C}{Exp.M.Exfo} \\\hline

\hline
\end{longtable}%

\subsubsection{nHSP}

    %

\subsubsection{nHSP-SOC}

    %

\subsection{Square lattice}

\subsubsection{$\Gamma$}

    %

\subsubsection{$\Gamma$-SOC}

    %

\subsubsection{$M$}

    %

\subsubsection{$M$-SOC}

    %

\subsubsection{$X$}

    %

\subsubsection{$X$-SOC}

    %

\subsubsection{nHSP}

    %

\subsubsection{nHSP-SOC}

    %

\subsection{Rectangular lattice}

\subsubsection{HSP}

    %

\subsubsection{HSP-SOC}

    %

\subsubsection{nHSP}

    %

\subsubsection{nHSP-SOC}

    %

\subsection{Oblique lattice}

\subsubsection{HSP}

    %

\subsubsection{HSP-SOC}

    %

\subsubsection{nHSP}

    %

\subsubsection{nHSP-SOC}

    %

\section{Band Structures of Twistable Materials}\label{app:sec:bandplot_intro}
In the following two sections, we present the monolayer band structures of promising twistable semimetals and insulators. The materials included are selected based on the following criteria: (i) they are either experimentally reported in monolayer or few-layer form, computationally exfoliable, or computationally stable with corresponding bulk materials; (ii) a cutoff twist score of 0.25 is applied for semimetals, and 0.4 for insulators; and (iii) a maximum of 10 materials are included for each class.

The complete catalog of twistable materials, along with their detailed electronic properties, can be accessed at \webtwoDTQC.

\onecolumngrid
\cleardoublepage 
\includepdf[pages={{},-}]{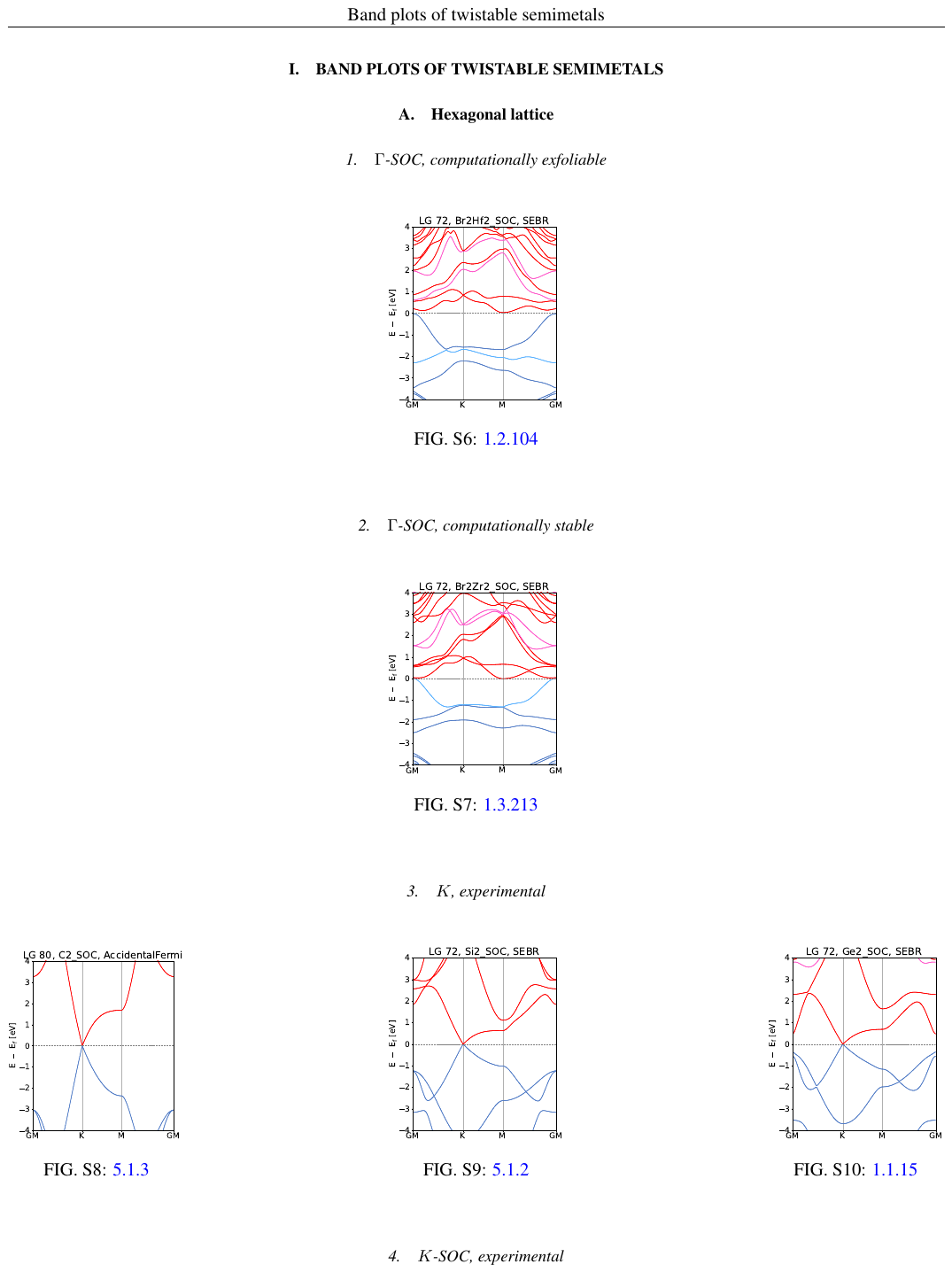}

\cleardoublepage 

\end{document}